\documentclass[10pt,twocolumn]{article}
\usepackage[top=2cm, bottom=2cm, left=2cm, right=2cm]{geometry}
\usepackage{times}  %
\usepackage[sort&compress,numbers]{natbib}  %
\usepackage[hyphens]{url}
\usepackage{flushend}
\usepackage[hyphens]{url}  %

\usepackage[pdftex]{graphicx}
\usepackage{amsmath,amsthm,amssymb,graphicx}
\usepackage{algorithm,algpseudocode,verbatim,color}
\usepackage{array}
\usepackage[skip=0pt,justification=centering]{subcaption}
\usepackage[hyphens]{url}
\usepackage[all]{xy}
\usepackage[labelfont=bf]{caption}
\usepackage{multirow}
\usepackage{makecell}
\usepackage{pifont}%
\usepackage{longtable}
\usepackage[T1]{fontenc}
\usepackage{paralist}
\usepackage{xspace}
\usepackage{booktabs}
\usepackage[bookmarks=true, bookmarksnumbered=true, colorlinks=true, linkcolor=linkcol, citecolor=citecol, urlcolor=urlcol, hypertexnames=true]{hyperref}

\let\OLDthebibliography\thebibliography
\renewcommand\thebibliography[1]{
  \OLDthebibliography{#1}
  \setlength{\parskip}{0pt}
  \setlength{\itemsep}{0pt plus 0.2ex}
}

\usepackage[marginal,hang,stable]{footmisc}
\renewcommand{\footnoterule}{
	\kern -3pt
	\hrule width 1in
	\kern 2pt
}

\usepackage{kantlipsum}
\usepackage[table, usenames, dvipsnames]{xcolor}
\usepackage[breakable, theorems, skins]{tcolorbox}
\tcbset{enhanced}

\usepackage[raggedright,compact]{titlesec}
\titlespacing*{\section}{0pt}{*3}{3pt}
\titlespacing*{\subsection}{0pt}{*2}{2pt}
\titlespacing*{\subsubsection}{0pt}{*2}{2pt}
  
\definecolor{darkgreen}{RGB}{0, 100, 0}
\definecolor{linkcol}{rgb}{0.3,0,0}
\definecolor{citecol}{rgb}{0.3,0,0}
\definecolor{urlcol}{rgb}{0.3,0,0}
\definecolor{vlightgray}{gray}{0.925}
\definecolor{teal}{rgb}{0.0, 0.5, 0.5}
\definecolor{pos}{HTML}{40B0A6}
\definecolor{neg}{HTML}{E1BE6A}
\definecolor{semipos}{HTML}{F0E442}

\newcommand{\descr}[1]{\smallskip\noindent\textbf{#1}}

\usepackage{mathtools}
\newlength{\myeqskip}  
\AtBeginDocument{%
    \setlength\abovedisplayskip{\myeqskip}%
    \setlength\belowdisplayskip{\myeqskip}%
    \setlength\abovedisplayshortskip{\myeqskip-\baselineskip}%
    \setlength\belowdisplayshortskip{\myeqskip}}

\usepackage{enumitem}

\usepackage{fourier}

\makeatletter
\renewenvironment{thebibliography}[1]
  {\section*{\refname}
   \small
   \begin{list}{\@biblabel{\@arabic\c@enumiv}}%
     {\setlength{\itemsep}{4pt}%
      \usecounter{enumiv}}}%
  {\end{list}}
\makeatother

\usepackage{microtype}

\newtheorem{theorem}{Theorem}

\newtheorem{assumption}{Assumption}

\newif\ifcomment
\commenttrue
\commentfalse
\ifcomment
	\newcommand{\SC}[1]{\textcolor{blue}{\textbf{SC:} #1}}
	\newcommand{\YH}[1]{\textcolor{green}{\textbf{YH:} #1}}
	\newcommand{\GB}[1]{\textcolor{red}{\textbf{GB:} #1}}
	\newcommand{\edc}[1]{\textcolor{purpe}{\textbf{EDC:} #1}}
\else
	\newcommand\edc[1]{}
	\newcommand{\SC}[1]{}
	\newcommand{\YH}[1]{}
	\newcommand{\GB}[1]{}
\fi

\newcommand{\mywidth}{0.18}

\begin{document}

\sloppy

\title{\bf PROTEAN: Federated Intrusion Detection in Non-IID Environments through Prototype-Based Knowledge Sharing\thanks{Published in the Proceedings of the 30th European Symposium on Research in Computer Security (ESORICS 2025).}}

\author{Sara Chennoufi$^1$, Yufei Han$^2$, Gregory Blanc$^1$, Emiliano De Cristofaro$^3$, Christophe Kiennert$^1$\\[1ex]
\normalsize $^1$SAMOVAR, Télécom SudParis, Institut Polytechnique de Paris\\[-0.5ex]
\normalsize $^2$PIRAT, INRIA Rennes,\\[-0.5ex]
\normalsize $^3$University of California, Riverside}
\date{}

\maketitle              %

\begin{abstract} 
In distributed networks, participants often face diverse and fast-evolving cyberattacks.
This makes techniques based on Federated Learning (FL) a promising mitigation strategy.
By only exchanging model updates, FL participants can collaboratively build detection models without revealing sensitive information, e.g., network structures or security postures. 
However, the effectiveness of FL solutions is often hindered by significant data heterogeneity, as attack patterns often differ drastically across organizations due to varying security policies. 
To address these challenges, we introduce PROTEAN, a Prototype Learning-based framework geared to facilitate collaborative and privacy-preserving intrusion detection. 
PROTEAN enables accurate detection in environments with highly non-IID attack distributions and promotes direct knowledge sharing by exchanging class prototypes of different attack types among participants. 
This allows organizations to better understand attack techniques not present in their data collections.
We instantiate PROTEAN on two cyber intrusion datasets collected from IIoT and 5G-connected participants and evaluate its performance in terms of utility and privacy, demonstrating its effectiveness in addressing data heterogeneity while improving cyber attack understanding in federated intrusion detection systems (IDSs).  
\end{abstract}

\section{Introduction} 
Cyber intrusions vary widely, challenging detection efforts. Signature-based methods lack adaptability, while Machine Learning (ML)-based IDSs trained on local data struggle with unseen threats. This highlights the need for collaborative intelligence \cite{lavaur2022evolution}, though data sharing raises serious privacy concerns.
FL has emerged as a privacy-friendly alternative for distributed ML-based IDS without disclosing local training data~\cite{Naseri2022CCS}. 
Each participating organization trains the model locally and shares only model updates to jointly aggregate a global model. %

In the practice of distributed IDS, FL faces the notable non-IID challenge raised by heterogeneous data distributions of different end-point devices. 
In distributed IDS applications, certain types of cyber intrusions may be specific to a subset of participants while being rare or completely absent in others, resulting in highly imbalanced distributions of attack data across participants. For example, while DDoS attacks are common across various types of organizations, entities that primarily host database servers are more frequently targeted by SQL injection or brute-force attacks aimed at gaining unauthorized database access. Conversely, botnet attacks, such as those involving Mirai, are more likely to affect organizations with IoT infrastructure. 
Although prior research has proposed aggregation techniques to mitigate the effects of non-IID data in FL (e.g., \cite{li2020federated, wang2020federated, pmlr-v139-li21h, karimireddy2020scaffold}), these methods primarily address heterogeneous training data distribution caused by varying class proportions across participants and are less effective when the class imbalance arises from the absence or rarity of certain classes in local datasets.
Specifically, they enhance the overall performance of the globally aggregated model \cite{li2020federated,wang2020federated,karimireddy2020scaffold}, or adapt the global model to heterogeneous local data distributions, i.e., personalization of the global model \cite{pmlr-v139-li21h}. 
Nevertheless, they overlook the class-specific accuracy for rarely appearing classes that can only be seen by a few participants.

Moreover, FL shares only global model parameters, which do not provide a comprehensive understanding of the threat across participants. 
This is particularly evident in non-IID scenarios.
When shared back with local participants, the globally aggregated model fails to reveal characteristics of attacks that are not locally observed by the client. %
Understanding the attack class distribution is crucial to grasping the broader threat landscape beyond merely achieving accurate detection and is needed to make ML-based detection more explainable.

As a result, we set out to achieve two main objectives.
\textit{First}, jointly trained detection models of a distributed IDS should reach high accuracy in the presence of highly skewed distributions of different cyber attack types. 
Specifically, we consider a cross-silo scenario where attacks are distributed in a highly imbalanced manner, aiming to
improve detection performance over rarely appearing attack types.
\textit{Second}, beyond reaching accurate detection results, cyber attack knowledge sharing should also boost the explainability of the detection model, helping reach a comprehensive understanding of the distribution of cyber attack types. 

To do so, our work introduces PROTEAN, a collaborative intrusion detection mechanism. 
Unlike standard FL, PROTEAN draws from prototype learning~\cite{10.5555/3294996.3295163,tan2022fedproto}. 
It operates by sharing model parameters for aggregation alongside class-specific prototypes of attack behaviors encountered by local participants during federated training. 
PROTEAN enforces a two-fold alignment process: 
\begin{itemize}
\item \textit{(Model parameter)} Detection model parameters trained locally by different participants should converge to a unified global model. 
\item \textit{(Class prototype)} PROTEAN requires class prototypes of the same attack type generated by different participants to be as similar as possible.
\end{itemize}
This improves training convergence by minimizing discrepancies between locally trained models, thereby enhancing detection capabilities in environments where attacks are heterogeneously distributed across participants.
Additionally, sharing class-specific prototypes enables participants to generalize their understanding of various cyber attack types.
For example, a participant with limited exposure to a rare attack type can learn to identify and classify it by receiving the corresponding prototype during training.
This not only accelerates convergence in heterogeneous and imbalanced attack class distributions but also promotes the transfer of attack knowledge among participants.

Overall, our work aims to answer the following research questions:\smallskip
\begin{enumerate}
    \item Can PROTEAN improve the FL-based IDS detection performance under heterogeneous attack data distributions with highly imbalanced cyber attack types? 
    \item Can attack knowledge sharing via sharing class-specific prototypes help detect rare or unique cyber attacks on local participants? 
    \item Can sharing attack class prototypes and model updates allow attackers to perform reconstruction attacks \cite{zhu2019deep,tan2024defending} to recover training data, triggering privacy risk of the FL IDS?     
\end{enumerate}

We theoretically prove the convergence of PROTEAN and provide a complexity analysis to demonstrate the applicability of PROTEAN for distributed IDS training. 
We also show the superior performance of PROTEAN compared to prior FL-based IDS methods in heterogeneous environments on two IoT and 5G intrusion detection datasets: {\em 5G-NIDD}~\cite{samarakoon20225g} and {\em X-IIoTID}~\cite{mpb6-py55-21}. 
In extremely imbalanced data distributions, PROTEAN improves on the state-of-the-art (SOTA) FL-based IDS method~\cite{Naseri2022CCS} with F1 scores 23\% higher on the X-IIoTID dataset and 5\% on the 5G-NIDD dataset. 
Finally, we audit the privacy leakage risks using a reconstruction attack with an adversary observing the shared prototypes either on a compromised local participant or via a semi-honest global server.
This follows previous insights from Zhu et al.~\cite{zhu2019deep,tan2024defending}, which has proved effective in reconstructing training data in FL. 
We experimentally show that the attack fails to recover meaningful feature values of the original data from the local participants. We also evaluate the impact of differential privacy (DP) on utility and privacy preservation.

\section{Related Work}
\label{section: related work}

\descr{FL with non-IID data.} 
A few aggregation techniques have been proposed to address non-IIDness and heterogeneous data distributions in FL. 
Some use optimization techniques to avoid model divergence: e.g., FedProx~\cite{li2020federated} incorporates a proximal term to keep the local model close to the global one;
FedOpt~\cite{asad2020fedopt} adapts classic optimizers like Adam and SGD-Momentum, while FedNova~\cite{wang2021novel} normalizes contributions from participants, and SCAFFOLD~\cite{karimireddy2020scaffold} adds control variables to account for the data drift over local participants.
These robust aggregation methods focus on smoothing the federated optimization process facing the heterogeneous local data distributions, improving the convergence speed of FL methods.

The other line of non-IID FL methods relies on personalizing techniques. 
For instance, FedAlt /FedSim~\cite{pillutla2022federated} aggregate parts of the model while keeping other parts private. 
Ditto~\cite{pmlr-v139-li21h} simultaneously tunes the locally adapted models and the global aggregated model, forcing the consistency between the local and global models. 
By transferring the knowledge from other participants while keeping parts frozen, transfer learning helps adapt the model even when data distributions are different. 
Additionally, Model-Contrastive Federated Learning (Moon)~\cite{li2021model}, and Model-Agnostic Meta-Learning (MAML)~\cite{fallah2020personalized} have been used in the context of FL heterogeneity. 
The personalized FL algorithms are dedicated to adapting the global model to the local data distributions of different participants. 
However, previous efforts to improve non-IID federated training often prioritize evaluating the overall performance of the aggregated global model. Nevertheless, they overlook the specific detection performance for classes that appear rarely, existing only in the training data of a few participants. 
Consequently, commonly used global metrics, heavily influenced by the majority classes, can mask poor performance on these minority classes, leading to an inaccurate assessment of accuracy for underrepresented classes.

\descr{FL-based IDSs.} Similar FL techniques have also been applied to IDSs with heterogeneous data ~\cite{fan2020iotdefender, abosata2023customised}. 
The evaluation only focuses on eliminating specific attacks for some participants while maintaining an IID distribution for other classes. 
Federated clustering ~\cite{briggs2020federated} is also applied for non-IID cases. But it largely ignores the class-specific detection performance and does not show how well participants benefit from knowledge sharing on new classes.

\descr{Knowledge Sharing in FL.} 
A major challenge of vanilla FL is the degradation of model performance due to the aggregation of local models trained on highly skewed class distributions of local training datasets. This results in a global model that struggles to generalize effectively. 
Local models often overfit to their specific data distributions, limiting the effectiveness of global knowledge sharing. 
Pei et al.~\cite{pei2022knowledge} propose integrating transfer learning into the FL framework to enhance knowledge sharing. However, their approach primarily addresses the semi-supervised learning scenario and does not fully tackle the non-IID bottleneck.
On the contrary, Federated Prototype Learning (FPL)~\cite{Long2022FedPCL,tan2022fedproto} offers a more effective approach to knowledge sharing, particularly under non-IID data conditions. Nevertheless, as unveiled in our study, the two vanilla FPL algorithms ~\cite{Long2022FedPCL,tan2022fedproto} fail to reach accurate detection in the extreme heterogeneous environments, where some cyber attack types mostly appear on a few participants'  local training data, while barely exist in the others' datasets. The core reason is that the imbalanced distribution of the attack types across local participants leads to a significant distribution drift between the embedding space spanned by locally trained ML models, which makes the aggregation of local class prototypes in FPL diverge in the training process.

\section{Methodology}
\label{section: methodology}

\begin{figure*}[t]
    \centering
    \fbox{
   \includegraphics[width=0.95\linewidth]{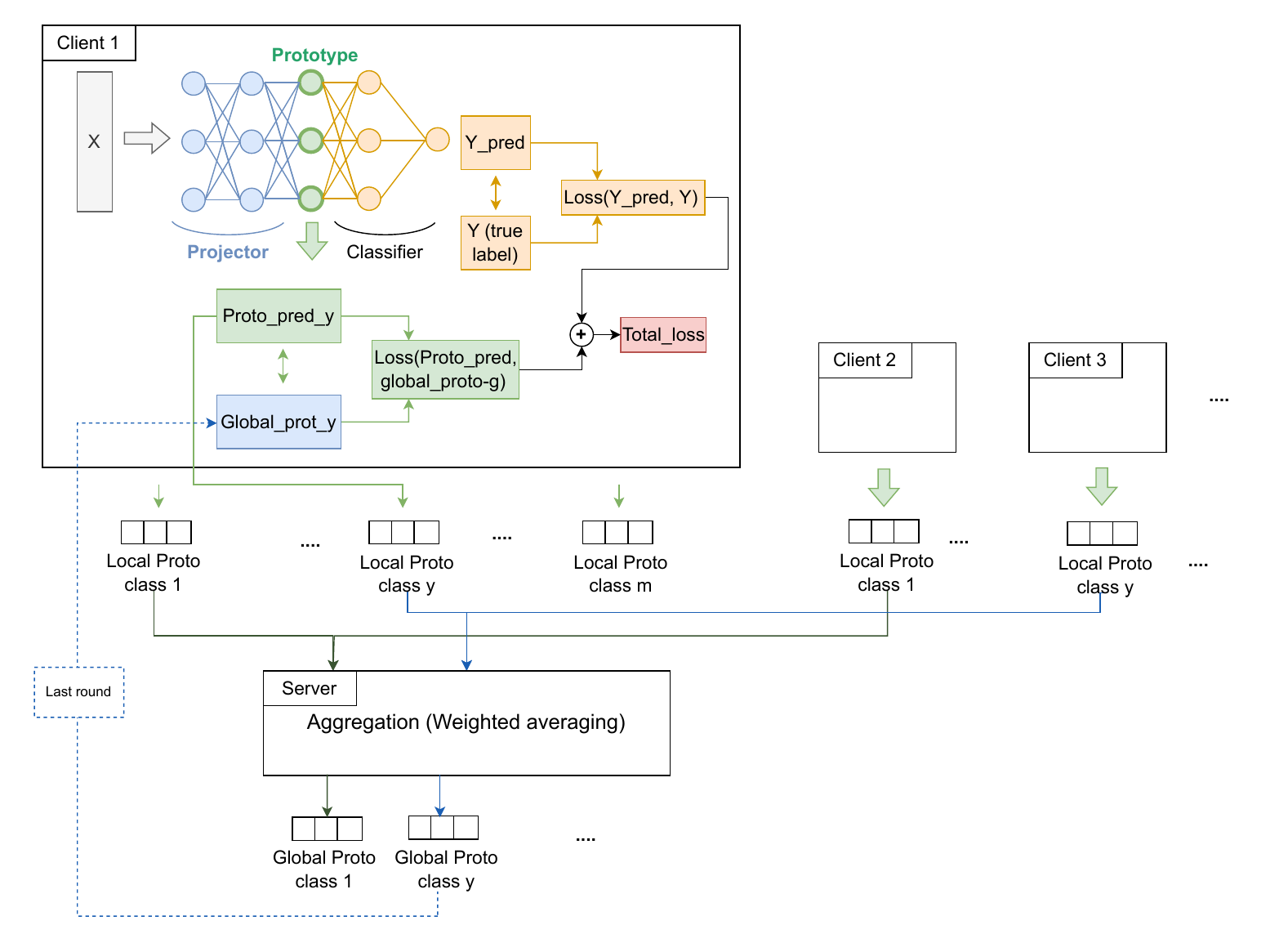}}
    \caption{PROTEAN training process. Each participant trains a local model and extracts prototypes that summarize observed attacks. These prototypes and models are sent to the server, which aggregates them and sends the aggregated results to the participants to refine their knowledge using both local and global insights.}
     
    \label{fig:fedproto}
\end{figure*}

\subsection{Notations}
Let \( M \) denote the number of participants and \( K \) the total number of cyber attack classes.
Each participant $i$ owns data $D_i$; $d_i\subseteq D_i$ is a batch. For any class $j$, $N_{i,j}$ designates the number of samples of that class on participant $i$, and $N_i = \sum_{j=1}^{K}N_{i,j}$ the total number of samples on that participant.  %
Let $f_{\omega}(x)\rightarrow{y}$ be the locally trained detection model with $\omega$ the parameters of $f$. 
We denote $x$ and $y$ as the features of a cyber attack and the corresponding predicted attack type, respectively. 
To clarify, we further decompose the detection model as $y = f_{\omega}(x) = c(\phi(x))$, 
where $c$ and $\phi$ are the classification head and embedding layers of the detection model, respectively. 
For instance, for a ten-layered convolutional neural network (CNN), we use the first eight layers as the embedding function and the remaining two layers as the classification head. 
An input $x$ is first fed into the embedding function, then the classification head $c$ is applied to the generated embedding $\phi(x)$ to produce the attack type prediction. 
At round $t$, the model parameters trained by participant $i$ are denoted as $\omega^{t}_i$.
The logit output on $x_i$ from the locally trained detection model with respect to class $j$ is denoted as $p_{i,j} = f(\omega^{t}_i; x_i)$. 
Each participant also builds a local prototype $C_{i,j}^{t}$ for every class $j$ by averaging the embeddings of its class $j$ samples; the server aggregates those local prototypes into the global prototype $\bar{C}_j^{t}$. The dimensionality of a prototype is denoted as \( d \), and the number of model parameters as \( m \).
Let $L_S$ denote the cross-entropy loss, $L_R$ the prototype alignment loss. $\lambda$ and $\mu$ are hyperparameter scalars to balance the loss terms in the optimization objective.

\subsection{Description of PROTEAN} 
The workflow of PROTEAN is outlined in Algorithm \ref{alg: fedproto_modified} and illustrated in Figure~\ref{fig:fedproto}. 
Its training process is established based on the FL protocol. 
Each participant trains the parameters $\omega$ of the detection model locally using privately owned training samples. 
Simultaneously, each participant $i$ also generates the class prototype vector $C_{i,j}$, 
as given in Equation~\ref{eq:protovec}: %
\begin{equation}\label{eq:protovec}
\small
C_{i,j} = \frac{1}{N_{i,j}} \sum_{x_{i,j} \in \text{class}\,\,\,{j}}\phi(x_{i,j})
\end{equation}
These locally computed class prototypes encapsulate the essential characteristics of each attack type observed by the participant, with one prototype vector per attack type. 
At each federated training round, each participant transmits the locally generated class prototypes $\{C_{i,j}\}$, as well as the parameters $\omega$ of the local detection model, to a central server.

\begin{algorithm}[t]
\caption{PROTEAN}
\small
    \label{alg: fedproto_modified}
\begin{algorithmic}[1]
        \State \textbf{Input:} $D_i, \omega_i, i = 1, \dots, M$
        \State Initialize the global prototypes $\{ \bar{C}^{0}_{j} \}, j = 1, \dots, K$ for all $K$ classes
        \State \textbf{Server executes:}
        \For{each round $t = 1, 2, \dots$}
            \For{each participant $i$ in parallel}
                \State $\{C^{t}_{i,j}\}, \omega^{t}_i \leftarrow \text{LocalUpdate}(i, \bar{C}^{t-1}_{j}, \omega^{t-1})$
            \EndFor
            \State Aggregate the local model parameters $\omega^{t}$ on the central server (Equation~\ref{eq: proto_agreg})
            \State Update the global prototypes by aggregating the local prototypes $\{C^{t}_{i,j}\}$ (Equation~\ref{eq: proto_agreg})
            \State Share back the global prototypes $\{ \bar{C}^{t}_{j} \}$ and the global model parameters $\omega^{t}$ to local clients
        \EndFor
        \State \textbf{Function LocalUpdate($i, \bar{C}^{t-1}_i, \omega^{t-1}$)}
        \For{each local epoch}
            \For{each batch $(x_i, y_i) \in d_i$}
                \State Compute local prototypes using the embedding layers of the detection model
                \State Optimize the learning objective in Equation~\ref{eq:optimization_objective} to update local model parameters $\omega^{t}_i$
            \EndFor
        \EndFor
        \State \textbf{return} $\{C^{t}_{i,j}\}$, $\omega^{t}_i$

\end{algorithmic}
\end{algorithm}

As given in Equation~\ref{eq: proto_agreg}, the server then aggregates these local class prototypes. %
Each $\bar{C}_j$ represents the shared knowledge of the attack class $j$ observed by different participants. 
The global prototypes and the aggregated detection model parameters are then redistributed to the participants, allowing each one to retrain its local detection model using both its local data and the globally shared prototypes.
This iterative process enables each participant to align both the detection model parameters and the class prototypes, ensuring that the shared knowledge remains representative of the data distribution across all participants.
During inference, the global detection model and prototypes are used by each local participant to detect attack classes on their device. 
The global class prototypes and model parameters are computed on the central server as: %
\begin{equation}
    \label{eq: proto_agreg}
    \small
    \omega^{t} = \frac{1}{M} \sum\limits_{i=1}^{M} \omega^{t}_i, \quad
    \bar{C}^{t}_j = \frac{1}{M} \sum\limits_{i=1}^{M} C^{t}_{i,j}
\end{equation}

The objective of local training, as shown in Equation~\ref{eq:optimization_objective}, is to optimize each participant's local classification performance and align the local prototypes with the global prototypes.
This learning objective incorporates three items. 
First, it minimizes the cross-entropy loss $\mathcal{L}_S$, encouraging accurate classification of cyber intrusions.
Second, minimizing the prototype alignment loss $\mathcal{L}_R$ ensures that the local prototypes \( C_{i,j} \) generated by participant \( i \) for each class \( j \) are closely aligned with the corresponding global prototypes \( \bar{C}_j \).
We denote with $\|C_{i,j}^{t}-\bar{C}_j^{t-1}\|$ the $L^2$ distance between the two prototypes, with a lower distance indicating that the local prototype $C_{i,j}$ is better aligned to the global prototype $\bar{C}_{j}$. 
Enforcing the alignment between the local and global class prototypes in the embedding space acts as a regularization, requiring locally trained models to provide similar class prototypes. This prototype alignment term hence encourages the convergence of the federated model training across different participants. 
Furthermore, the third term enforces model parameter alignment between different participants. It penalizes the locally trained parameters $\omega^{t}_i$ of the detection model if they deviate significantly from the global model parameters derived in the last round $ \omega^{t-1}$. Intuitively, with the model parameter alignment, locally tuned model parameters should stay close to the globally aggregated model parameters. It suppresses the bias from the distribution drift of local training datasets hosted by different participants. Jointly applying the prototype alignment (the second term) and model alignment (the third term) thereby boosts the convergence of the training process in the extreme heterogeneous training data distribution \cite{li2020federated}: %

\begin{equation}
    \label{eq:optimization_objective}
    \small
\begin{split}
   \small
   \omega^{*t}_i = & \underset{\omega^{t}_i}{\arg\min}\,\,\sum_{(x_i ,y_i) \sim d_i} \mathcal{L}_S(f(\omega^{t}_i; x_i), y_i) +   \lambda \sum_{j=1}^{K} \mathcal{L}_R(\bar{C}^{t-1}_j, C^{t}_{i,j}) + \\
   & + \frac{\mu}{2} \|\omega^{t}_i - \omega^{t-1}\|^2~~\text{s.t.}\\
   & \mathcal{L}_S(f(\omega^{t}_i; x_i), y_i) =    - \frac{1}{N_{i}}\sum_{(x_i,y_i)\sim d_i} \sum_{j=1}^{K} y_{i,j}\log(p_{i,j}), \\
   & \mathcal{L}_R(\bar{C}^{t-1}_j, C^{t}_{i,j}) =  \|C^{t}_{i,j} - \bar{C}^{t-1}_{j}\|^2
\end{split}
\end{equation}

where $x_{i}$ and $y_{i}$ are the features and attack class labels of the training sample hosted by the participant $i$. %
 $\bar{C}^{t-1}_j$ is the global prototype of the attack class $j$ obtained in the $t-1$-th round and shared back to the local participants at the beginning of the $t$-th federated training round. During inference, the global model classifies new instances by assigning the label of the global prototype with the smallest $L^2$ distance to the instance's embedding. 
 To guarantee the applicability of PROTEAN, we perform the theoretical reasoning of the asymptotic convergence of PROTEAN. Due to the space limit, we provide the convergence proof  in Appendix~\ref{app:proof}. 

\subsection{Direct Knowledge Sharing to Detect Rare Cyber Attacks}
Introducing the prototypes does not only aim to improve federated model training with non-IID training data but also facilitates direct knowledge sharing about cyber attacks between the participating organizations. 
In the context of PROTEAN, a global attack class prototype $\bar{C}_j$ serves as a representation of typical features associated with a specific attack type $j$. 
Note that one attack has multiple local prototypes $C_{i,j}$. These class prototypes reflect the diverse modes of data distribution for this attack type.  
This multi-prototype approach enables direct knowledge sharing by allowing various local participants -- each equipped with excessively non-IID cyber attack data -- to exchange these refined representations of different attack types.

Consider, for instance, an FL network deployed with various organizations as the local participants. Each organization monitors its own network traffic to raise alerts to malicious activity. 
Rather than exchanging raw network data, which may include confidential information, the organizations can access the global cyber attack class prototypes generated by the PROTEAN framework to estimate the distribution of various types of cyber attacks.  
These attack class-specific prototypes are high-level summaries that encapsulate the collective insights and characteristics of the respective attacks as observed across all participants. 
These participants collaborate by sending their local class prototypes to a central server. 
For example, if one participant generates a prototype $C_{i,{DDoS}}$ that captures specific traffic patterns indicative of this attack, this can be aggregated with prototypes from other participants. %
The result is a {\em global prototype} $\bar{C}_{DDoS}$ that reflects a comprehensive view of how such attacks manifest across diverse networks. By continuously updating their local prototypes in response to insights gained from the global prototype, the participants collectively strengthen the overall detection capabilities. %

Once the global prototype is established, each participant receives it, enabling them to enhance their local models. If a participant has only limited or even no training samples for a specific attack, e.g., DDoS Attack, the global prototype of this attack aggregated on the central server can be used as a valuable reference for the participant. It can help him to learn the representative profiles of this rarely appearing or previously unseen attack within the embedding space of the globally learned detection model. 
In other words, local participants can rely on this global prototype to identify potential DDoS attacks in future traffic flows. By aggregating and sharing these global prototypes of attacks across all participants, local participants gain the ability to perform few-shot or zero-shot \cite{8413121} learning on rare or previously unseen attack types, enhancing IDS capabilities even in the presence of heterogeneous data distributions.

\subsection{Comparison with FPL in Extreme Heterogeneous Environments}

For a local participant, the detection model trained with FPL \cite{tan2022fedproto} struggles when tested on the rarely or unseen attack classes due to the lack of optimization for these classes during training. If a participant's embedding model has never been tuned with training samples from such classes, the local training process will introduce a statistical bias into the detection model, which prevents the model from learning meaningful representations of these classes. Consequently, during inference, the embedding layer of this locally trained model generates noisy local prototypes for these unseen or rarely appearing classes. It results in poor alignment with the corresponding global prototypes formed by the other participants who have encountered these classes. 
This misalignment results in diverged prototype representations between different participants, which in turn causes incorrect detection output.

PROTEAN addresses this issue by simultaneously aggregating both local prototypes and model parameters. This design reduces the discrepancies between the embedding spaces of locally trained models, ensuring that the embeddings of locally generated prototypes align more closely with the global prototype representations of the same class. This alignment enhances the convergence of prototype generation and improves consistency across locally trained models. In addition, PROTEAN introduces a regularization term (the final term in Equation~\ref{eq:optimization_objective}) in the loss function, dragging the parameters of local models close to those of the globally aggregated model. Benefitting from the design, we show that PROTEAN can reach 35\% higher classification accuracy than the SOTA FPL algorithm in extreme heterogeneous data distributions. %

\subsection{Computational and Communication Complexity of PROTEAN}
Prototype dimensionality \( d \), is much smaller than the number of model parameters \( m \). Thus, communication and computational cost caused by prototype transfer is negligible.

The per-round {\bf communication} cost for classic FL (e.g., FedAvg) is \( 2 M m \) as each participant transmits its model parameters to the server, which then sends the aggregated model back to the participants.
In contrast, PROTEAN incurs a per-round communication cost of \( 2 M (m + dK) \), where the additional term \( dK \) accounts for the transmission of prototype data, consisting of one \( d \)-dimensional vector per class across \( K \) classes.

The additional {\bf computational} cost introduced by PROTEAN compared to classical FL primarily stems from three operations: 
\begin{inparaenum}[(i)]
\item aggregating participant prototypes by class on the server side, 
\item aggregating prototypes over all batches on the participant side, and 
\item computing the prototype loss on the participant side, too. 
\end{inparaenum}
Since these operations involve only small vectors, their cost is minimal relative to the full model parameter updates.

On the server side, aggregating model parameters requires \( O(mM) \) operations for classic FL. PROTEAN also aggregates participant prototypes, adding a computational cost of \( O(MKd) \). However, since \( Kd \ll m \), the overall server-side computational complexity remains \( O(mM) \).
On the participant side, let \( e \) denote the number of epochs, \( s \) the total number of training instances, and \( b \) the batch size. For each batch, the forward pass requires \( O(bm) \) operations, and the loss computation requires \( O(b(K + d) + m) \) operations for PROTEAN (due to prototype-related computations). The backward pass requires approximately \( O(bm) \) operations. As a result, the per-batch complexity remains dominated by \( O(bm) \), leading to an overall per-participant complexity of \( O(e s m) \) for all approaches. Although PROTEAN introduces an additional term \( O\left(\frac{K d s}{b}\right) \) for prototype processing, this term is negligible relative to \( O(e s m) \) because \( K d \ll m \).

\section{Evaluating the Knowledge Sharing in PROTEAN}
\label{section: evaluation}
In this section, we present an experimental evaluation of PROTEAN to assess its feasibility in terms of utility and knowledge sharing for rare attacks.

\subsection{Experimental Setup}

\noindent\textbf{Datasets.} We use benchmark datasets, namely X-IIoTID~\cite{mpb6-py55-21} and 5G-NIDD~\cite{samarakoon20225g}.
Both datasets capture real-world network attack data from distributed end-point devices, such as IoT devices and 5G end-devices. X-IIoTID is often used in the context of intrusion detection in Industrial Internet of Things (IIoT), a key use case in 5G studies. %
It includes both general and specific attack classes.
The general classes are: 
Normal, Weaponization, Ransom Denial of Service (RDOS), Reconnaissance, Lateral Movement, Exfiltration, Exploitation, Tampering, Command and Control (C\&C), and Crypto-ransomware. 
We remove Crypto-ransomware as it contains very few instances with valid values.

The 5G-NIDD dataset was collected from the 5G Test Network Finland (5GTN) at Oulu University, with additional testbed elements like Nokia Pico Base Stations, attacker nodes, and benign traffic-generating participants. 
It consists of live traffic 
using protocols like HTTP, HTTPS, SSH, and SFTP. 
The dataset covers various types of DoS attacks, such as ICMP Flood, UDP Flood, SYN Flood, HTTP Flood, and Slowrate DoS.
It also incorporates various port scanning techniques, including SYN Scan, TCP Connect Scan, and UDP Scan.

\descr{Non-IID Attack Data Distributions.}
We adopt the Dirichlet distribution model to simulate various non-IID levels of data across local participants~\cite{marfoq2021federated, li2022federated}. 
By changing the $\alpha$ hyperparameter of the Dirichlet distribution, we control the degree of heterogeneity across different local participants, with smaller values of $\alpha$ corresponding to less uniform distributions (non-IID-like behavior). 
We conduct our experiments using different $\alpha$ values, i.e., \{0.75, 0.5, 0.25\}. 
By focusing on lower $\alpha$ values and thus more heterogeneous distributions of attack data, we study how well knowledge sharing in PROTEAN can help detect unseen attacks in heterogeneous environments. 

\descr{Baselines.} We involve the Four SOTA FL methods applied in collaborative training of attack detection models, including Cerberus \cite{Naseri2022CCS}, MOON \cite{li2021model}, FedProx \cite{li2020federated}, and FedProto \cite{tan2022fedproto}. 
Cerberus applies the standard FedAvg aggregation. MOON and FedProx are non-IID FL methods, enforcing the consistency constraint over locally trained models to bridge the distribution gap between local participants. 
FedProto is the vanilla federated prototype learning method, which only aggregates local prototypes. In contrast, PROTEAN performs the aggregation of the locally trained class prototypes and detection model parameters, reaching the alignment of class prototypes and models. 
In the empirical study, these baselines are noted as \textit{Cerberus}, \textit{MOON-IDS}, \textit{FedProx-IDS}, and \textit{FPL-IDS}. 
In addition, we include a variant of PROTEAN, which aggregates the parameters of the embedding layers $\phi(*)$ of the detection model instead of all the parameters. 
We name this baseline as \textit{PROTEAN-embedding}. We compare PROTEAN and PROTEAN-embedding to verify the model aggregation strategy adopted in PROTEAN.

\descr{Detection Model.} We use a CNN implemented in Pytorch consisting of two convolutional layers (64 and 128 filters) with ReLU activations, each followed by max pooling and dropout (0.2 and 0.5). 
The flattened output is passed through a dense layer with 128 units, followed by the final classification layer. 
The architecture outputs both a class prediction via log-softmax and the global class prototype vectors. 
For the proximal term in Equation~\ref{eq:optimization_objective}, we use $\mu$ = 0.1. %

\descr{Evaluation Settings and Metrics.} 
PROTEAN targets cross-silo FL scenarios involving a limited number of participants \cite{liu2022privacy}, e.g., companies and ISPs. With this setting, we assume 10 participants in the following experiments.
For each dataset, we split the samples into 80\% for training and 20\% for testing, and distribute the former to the participants in a non-IID manner according to the Dirichlet distribution. 
We train PROTEAN with 10 rounds of federated training and three epochs of local training on each participant.
To evaluate the performance of the detection model over multiple classes, we use \textit{Macro accuracy}, \textit{Accuracy}, \textit{F1 score}, and \textit{Precision}.  
\SC{added for reviewer asking for training time}Experiments were done on a Linux Ubuntu 22.04.4 with 11th Gen Intel® Core™ i7-11800H processor (16 threads, 4.6 GHz max) and 31 GiB of RAM.

\begin{table*}[t]
    \centering
    \small
    \begin{tabular}{lrrrrrr}
    \toprule
    \multirow{2}{*}{\textbf{Algorithm}} & \multicolumn{3}{c}{\textbf{X-IIoTID}} & \multicolumn{3}{c}{\textbf{5G-NIDD}} \\ \cmidrule(lr){2-4} \cmidrule(lr){5-7}
    & $\alpha{=}$0.75 & $\alpha{=}$0.50 & $\alpha{=}$0.25 & $\alpha{=}$0.75 & $\alpha{=}$0.50 & $\alpha{=}$0.25 \\ \midrule
    Cerberus & 52.23 (5.19) & 56.52 (16.64) & 53.57 (4.81) & 66.05 (5.4) & 54.45 (2.22) & 47.40 (11.19) \\ 
    Moon-IDS & 41.1 (1.81) & 35.58 (3.59) & 27.17 (5.34) & 39.06 (2.46) & 31.55 (4.32) & 25.03 (7.5) \\
    FedProx-IDS & 89.72 (0.87) & 89.15 (3.75) & 87.29 (5.41) & 82.27(0.21) & 77.98 (6.71) & 71.67 (5.96) \\
    FPL-IDS & 64.57 (2.48) & 65.46 (3.06) & 57.32 (4.92) & 56.06 (2.89) & 51.08 (0.47) & 44.93 (8.04) \\ 
    PROTEAN-embedding & 90.88 (1.36) & 91.31 (1.37) & 92.2 (0.47) & 78.78(2.58) & 77.42 (0.13) & 81.52 (0.24) \\
    PROTEAN (Ours) & \textbf{92.67 (0.32)} & \textbf{93.62 (0.29)} & \textbf{93.43 (0.16)} & \textbf{85.83 (3.14)} & \textbf{82.50 (0.95)} & \textbf{79.97 (5.72)} \\ \bottomrule
    \end{tabular}
    \caption{Mean and standard deviation of the macro accuracy with varying $\alpha$ values on X-IIoTID and 5G-NIDD datasets. Results for each $\alpha$ are averaged over three runs with different random seeds.}
    \label{tab:macro_acc_combined}
\end{table*}

\subsection{Utility of PROTEAN in Heterogeneous Environments}  
We start by evaluating the
accuracy scores on each local participant and report the average over all 10 participants as the PROTEAN's global utility metrics in detecting different attacks. 
The empirical results can be found in Table~\ref{tab:macro_acc_combined}.
These results demonstrate that PROTEAN outperforms Cerberus, MOON, FedProx-IDS, and FPL-IDS  
across all $\alpha$ values and on both datasets. 
For instance, in a very heterogeneous environment with $\alpha=0.25$, on the X-IIoTID dataset, PROTEAN achieves 40\%, 66.45\%, 6.33\% and 36.3\% %
higher utility than the baselines.

PROTEAN outperformed the original FPL. Specifically, in the X-IIoTID dataset, where each participant has a skewed class distribution with $\alpha=0.5$, 5 out of 10 participants lack 2 to 4 classes in their training sets. When evaluated across all classes—including those unseen during training—FPL-IDS achieves a macro-averaged classification accuracy of 68.06\% vs.~92.24\% for  PROTEAN. Under an even more severely skewed distribution with $\alpha=0.25$, where 8 out of 10 participants miss between 1 and 5 classes, FPL-IDS's macro-averaged accuracy drops to 63.92\% vs.~93.64\% for PROTEAN. Similarly, for the 5G-NIDD dataset at $\alpha = 0.75$, 7 out of 10 users are missing 1 to 3 classes, with one user missing 5 classes. Under these conditions, FPL-IDS achieves a macro accuracy of 60.14\%, compared to 88.07\% for PROTEAN. At $\alpha = 0.5$, with 8 out of 10 users missing between 1 and 5 classes, FPL-IDS's macro accuracy further decreases to 51.63\% while PROTEAN's is 83.84\%. Finally, for $\alpha = 0.25$, FPL-IDS records a macro accuracy of 48.88\% while PROTEAN achieves 85.88\%. This demonstrates PROTEAN's effectiveness in leveraging prototype-based knowledge, enabling robust detection of unseen or rare classes, even in highly heterogeneous environments.

PROTEAN increasingly improves over the baselines with lower $\alpha$ values. 
This is due to the divergence of prototypes and model parameters locally tuned by participants during the training of Cerberus, FedProx-IDS, and FPL-IDS. It impedes effective aggregation and hinders the ability to generalize in highly heterogeneous distributions. 
FedProx-IDS achieves higher macro accuracy than Cerberus by mitigating model divergence across different participants, but still underperforms compared to PROTEAN due to its introduction of prototypes and model parameter alignment into the training process. Both alignment-driven learning objectives shrink the gap between locally tuned class prototypes. These prototypes, in turn, enhance accuracy over rarely appearing attack classes of each participant, thereby increasing the overall utility.

Furthermore, PROTEAN achieves higher macro-accuracy compared to PROTEAN-embedding.
This result highlights the importance of aggregating all parameters to improve classification performance, especially for infrequently occurring attack classes.
Theoretically, this full-parameter aggregation promotes greater consistency within the embedding space formed by the locally trained models.
Such a design enhances knowledge sharing by aggregating the local class prototypes, which in turn improves the accuracy for low-prevalence attack classes.

\begin{figure*}[t]
    \centering
            \includegraphics[width=0.6\linewidth]{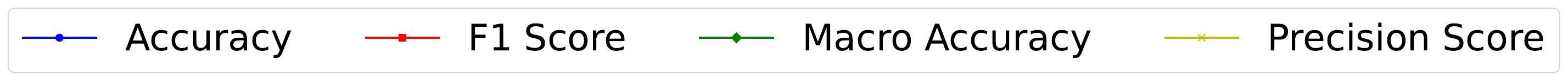}
        \centering
        \begin{subfigure}[t]{\linewidth}
        \includegraphics[width=\linewidth]{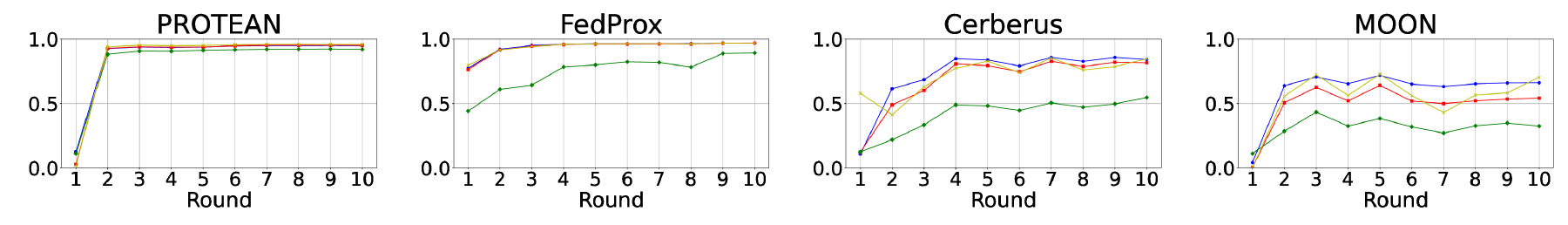}
         \caption{$\alpha=0.75$}
		\end{subfigure}
        \begin{subfigure}[t]{\linewidth}
        \includegraphics[width=\linewidth]{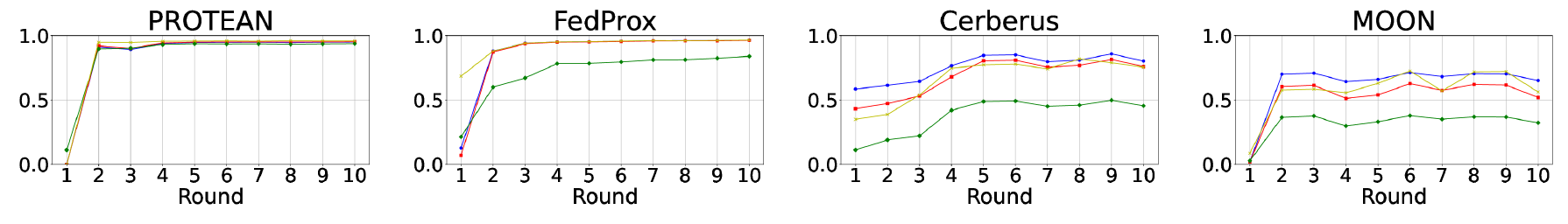}
         \caption{$\alpha=0.5$}
		\end{subfigure}
        \begin{subfigure}[t]{\linewidth}
        \includegraphics[width=\linewidth]{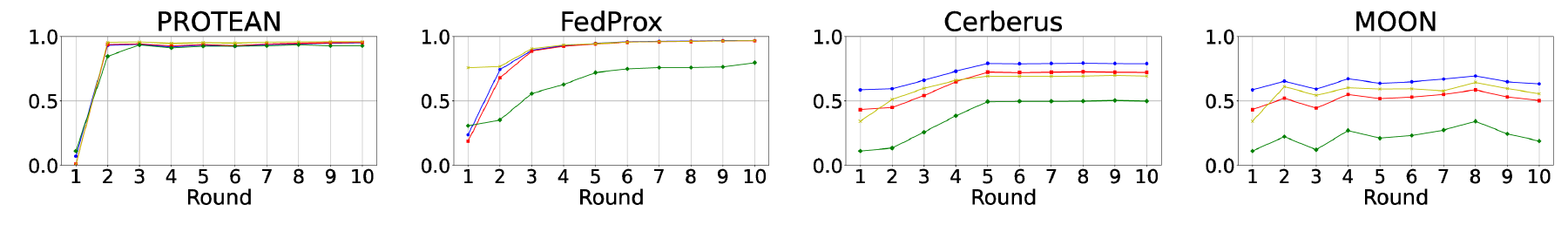}
         \caption{$\alpha=0.25$}
		\end{subfigure}

    \caption{Per-round accuracy, F1 score, macro accuracy, and precision of PROTEAN, FedProx-IDS, Cerberus, and MOON-IDS with $\alpha=$ 0.75, 0.5, and 0.25 on the X-IIoTID dataset.}
    \label{fig: metrics_xiiotid}
\end{figure*}

\descr{Convergence.} We also evaluate the convergence speed of PROTEAN in heterogeneous environments, via the evolution of accuracy, F1 score, macro accuracy, and precision scores for increasing FL rounds. 
We report the results in Figures~\ref{fig: metrics_xiiotid} and~\ref{fig: metrics_5Gnidd}, respectively, for the X-IIoTID and the 5G-NIDD datasets, with $\alpha$ valued as 0.75, 0.5, and 0.25. To ease presentation, we defer all the plots on the 5G-NIDD dataset to the Appendix. 
We omit FPL-IDS as its performance is significantly lower than the other detection models over all the rounds. 

For both datasets, PROTEAN outperforms Cerberus across all metrics; once again, this is primarily due to FedAvg's tendency to diverge in non-IID settings. 
PROTEAN also achieves higher macro accuracy than FedProx-IDS and exhibits faster convergence. 
This stems from sharing prototypes by class, enhancing detection accuracy for each class. 
In other words, by improving the accuracy of each class, PROTEAN effectively boosts the overall macro accuracy. 
\SC{added for reviewer asking for scaling}We also tested with a higher number of participants (20) and found out that PROTEAN maintained essentially the same effectiveness. For example, for $\alpha$=0.25 it attained 95.4\% accuracy 95.4\% F\textsubscript{1}, 91.7\% macro-accuracy and 95.5\% precision on X-IIoTID, and 97.1\% accuracy, 97.5\% F\textsubscript{1}, 83.7\% macro-accuracy and 98.0\% precision on 5G-NIDD.

Moreover, PROTEAN converges faster for both datasets in the second round, while FedProx-IDS converges in the fourth round for X-IIoTID and the fifth for 5G-NIDD, and Cerberus converges in the fifth round for both datasets. 
This reduces the number of communication rounds required to reach good performance, which is beneficial in FL settings where communication costs may be significant. %

\begin{figure*}[t]
    \centering
    \begin{subfigure}[t]{0.32\linewidth}  
    \includegraphics[width=\linewidth]{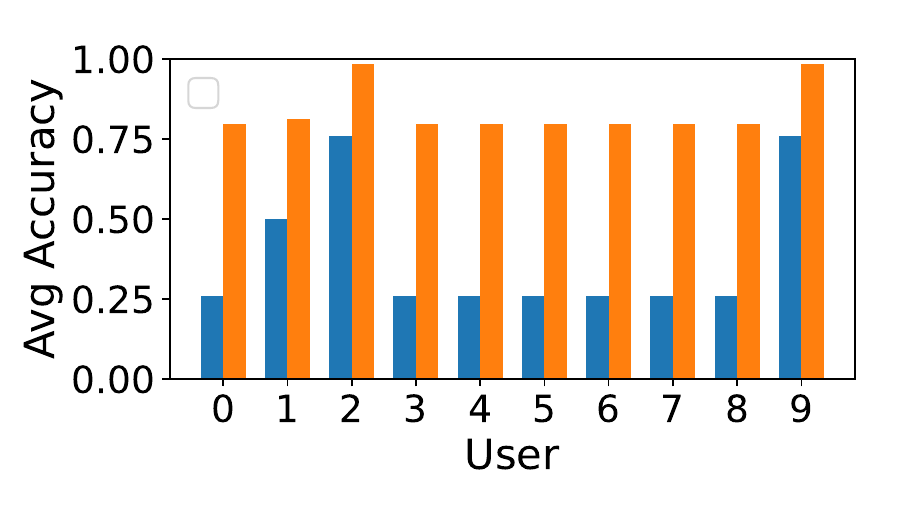}
	\caption{$\alpha{=}$0.75}
	\end{subfigure}
    \begin{subfigure}[t]{0.32\linewidth}  
    \includegraphics[width=\linewidth]{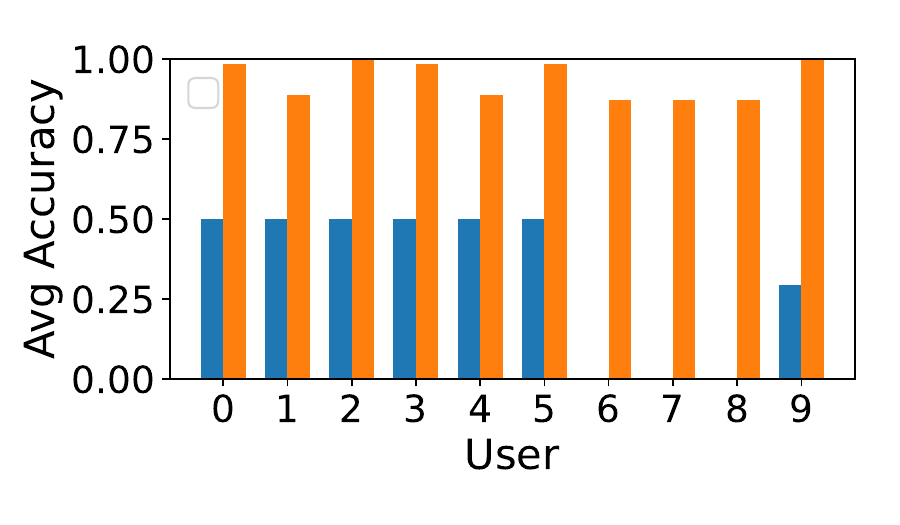}
	\caption{$\alpha=0.5$}
	\end{subfigure}
    \begin{subfigure}[t]{0.32\linewidth}  
    \includegraphics[width=\linewidth]{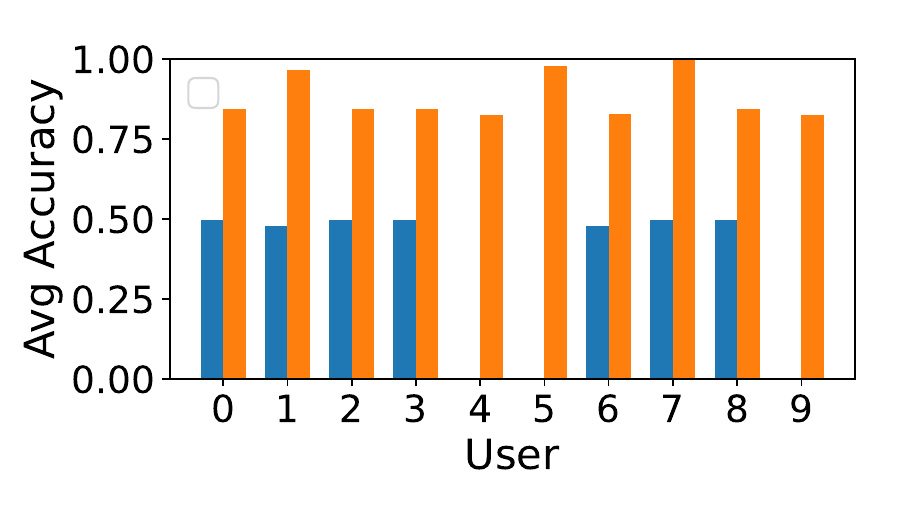}
	\caption{$\alpha{=}$0.25}
	\end{subfigure}
    \caption{Comparison of the averaged accuracy on two rare classes of each participant on the X-IIoTID dataset with different $\alpha$ values. Blue and orange bars represent Cerberus and PROTEAN, respectively.}
    \label{fig: ra_comparaisions_xiiotid}
\end{figure*}
\begin{figure*}[t]
    \centering

    \includegraphics[width=\mywidth\linewidth]{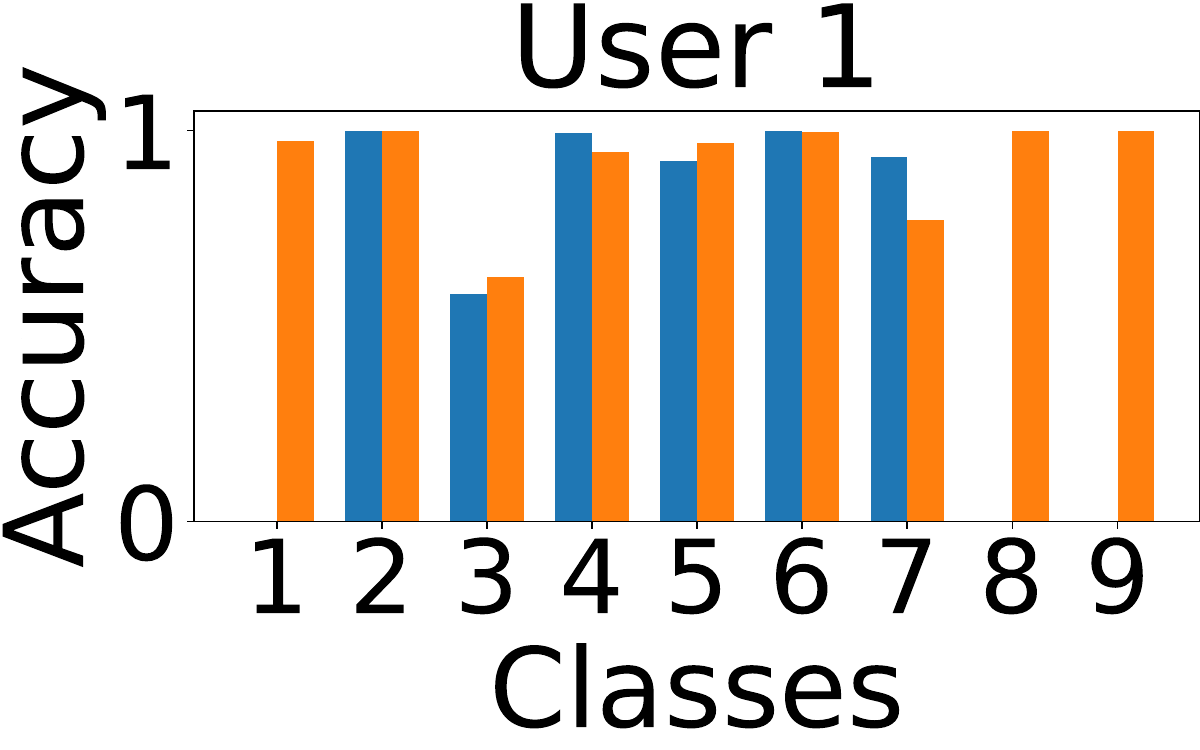}~%
    \includegraphics[width=\mywidth\linewidth]{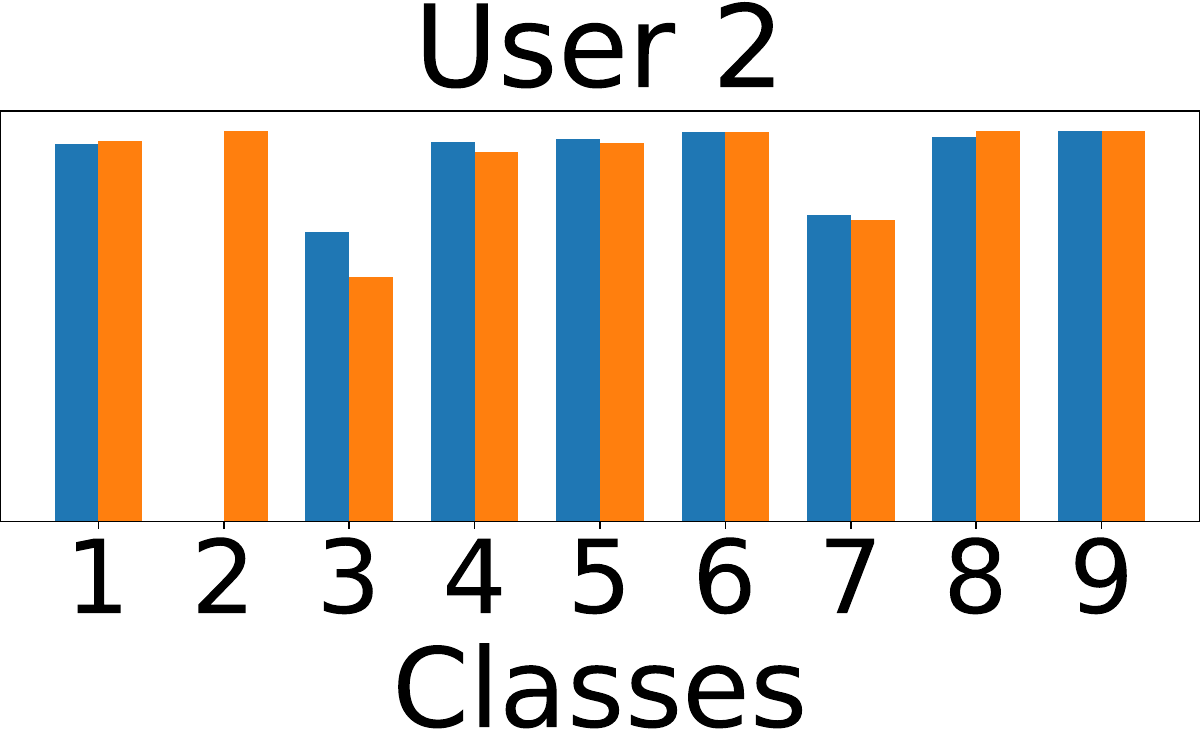}~%
    \includegraphics[width=\mywidth\linewidth]{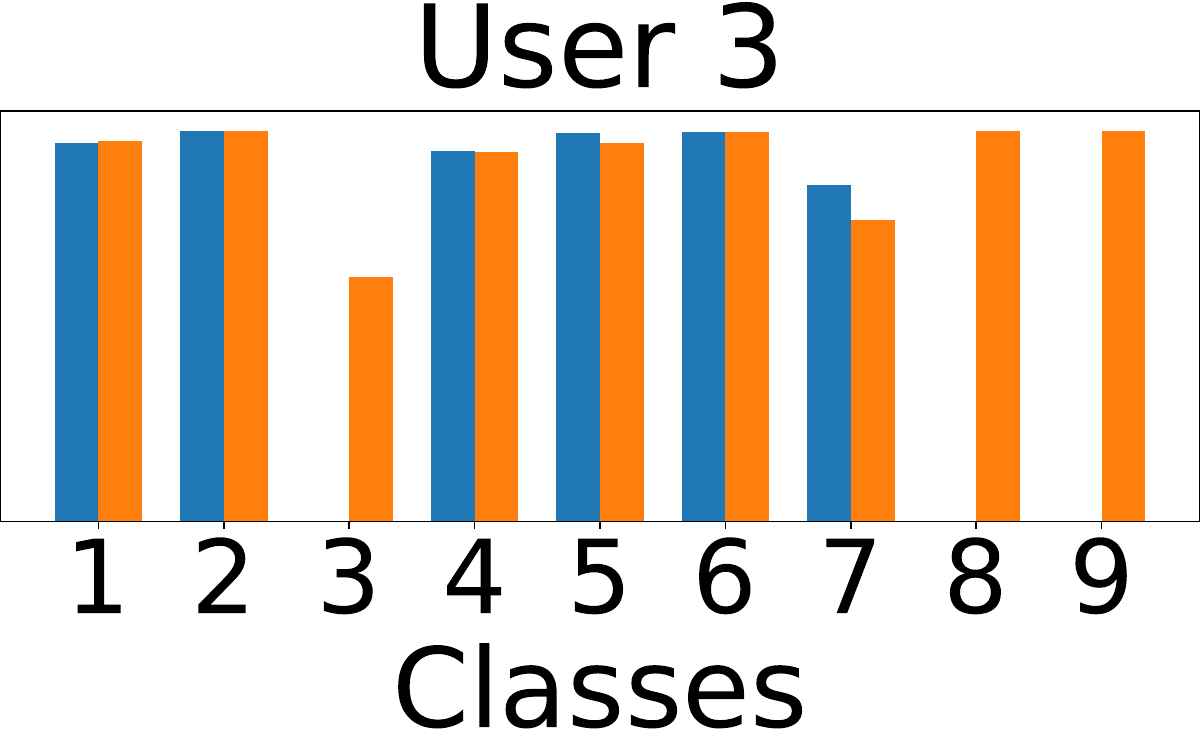}~%
    \includegraphics[width=\mywidth\linewidth]{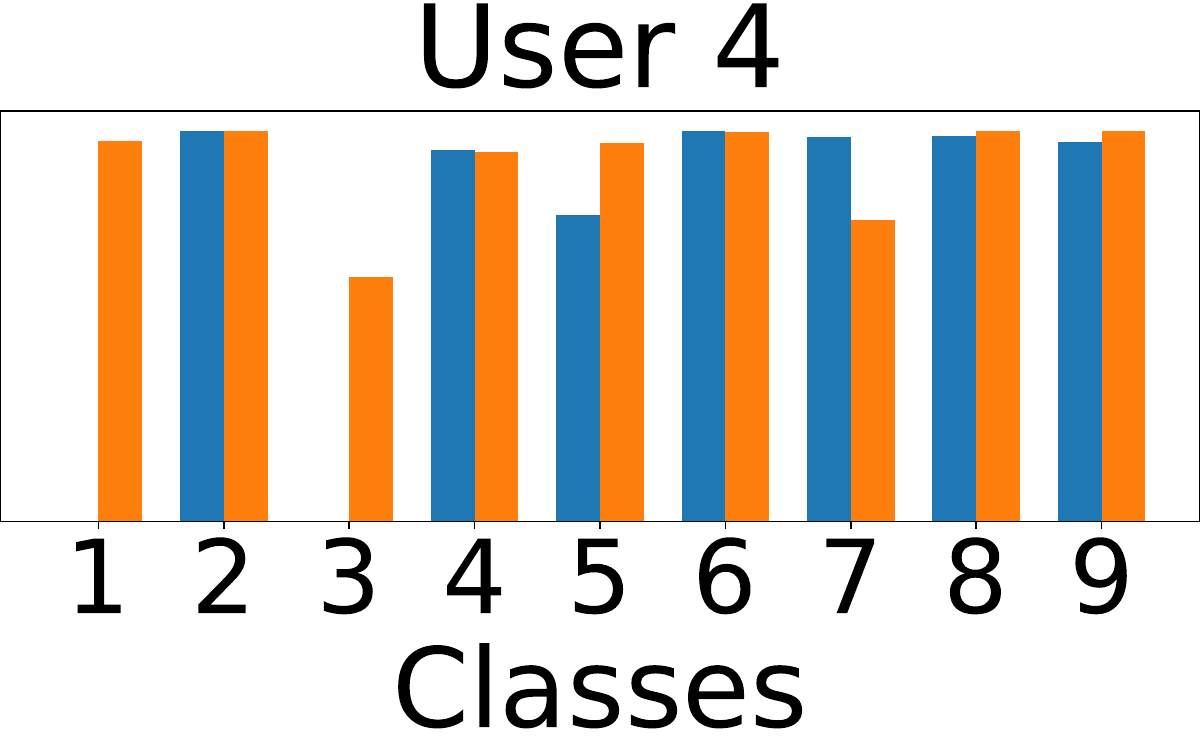}~%
    \includegraphics[width=\mywidth\linewidth]{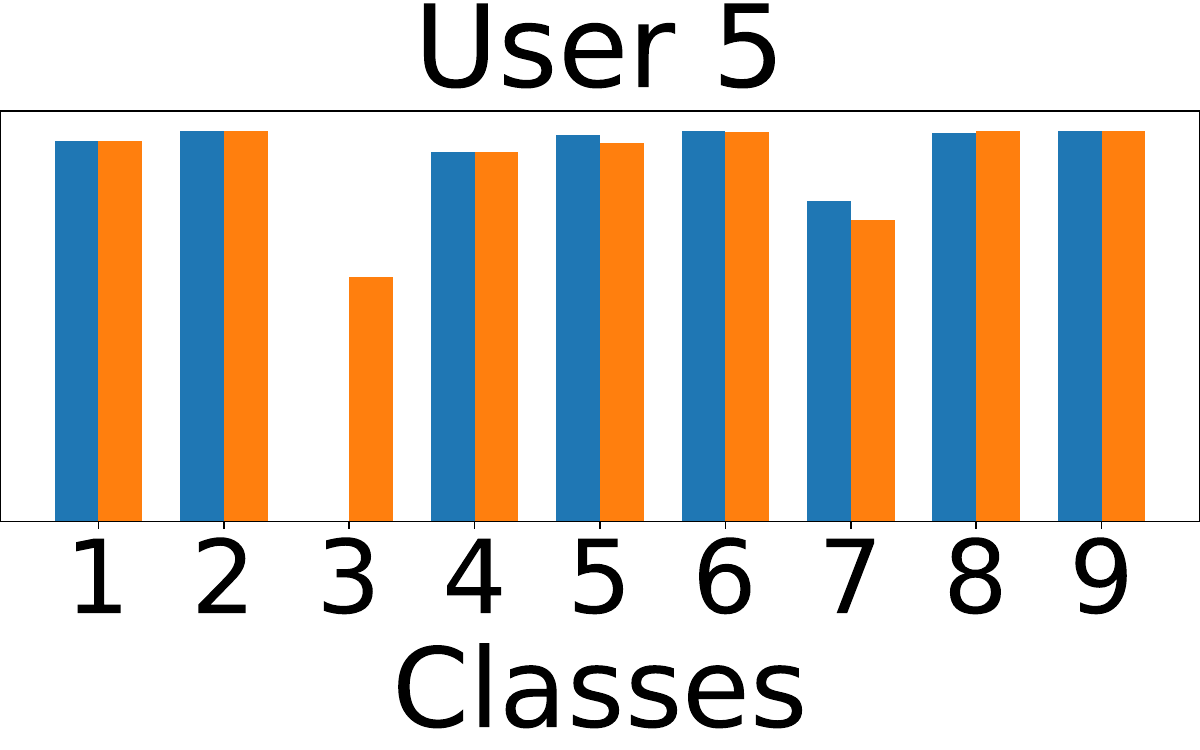}
    \includegraphics[width=\mywidth\linewidth]{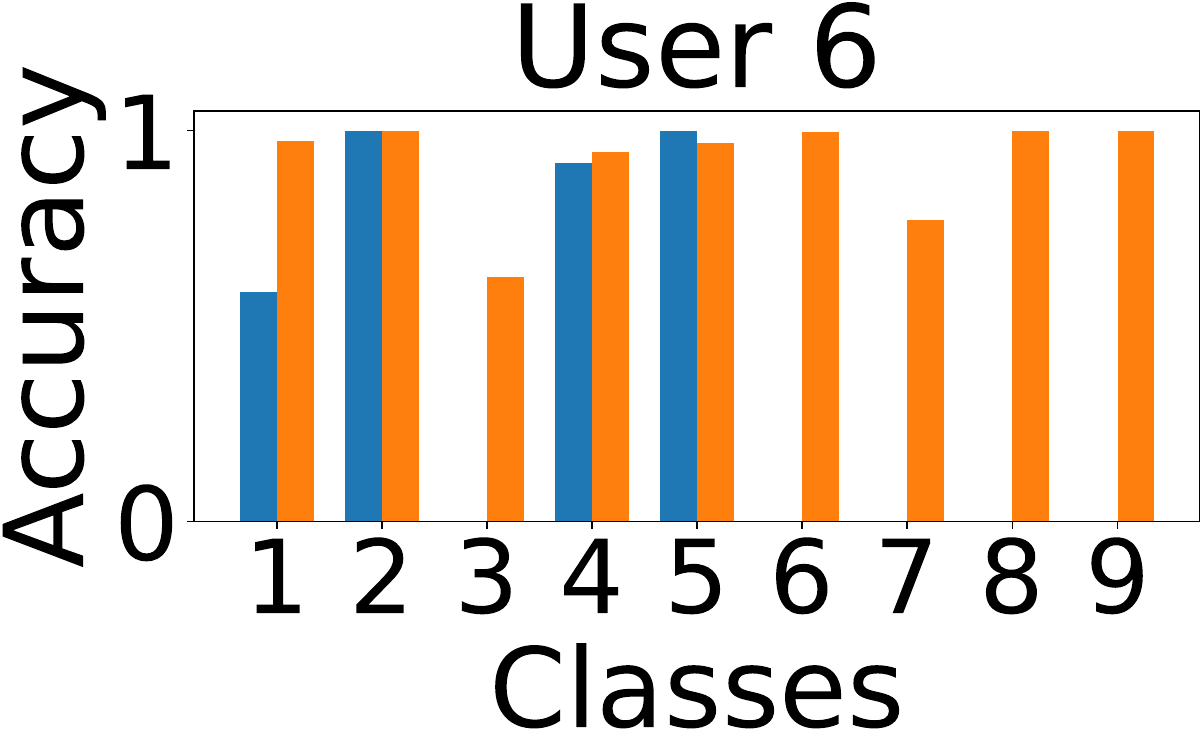}~%
    \includegraphics[width=\mywidth\linewidth]{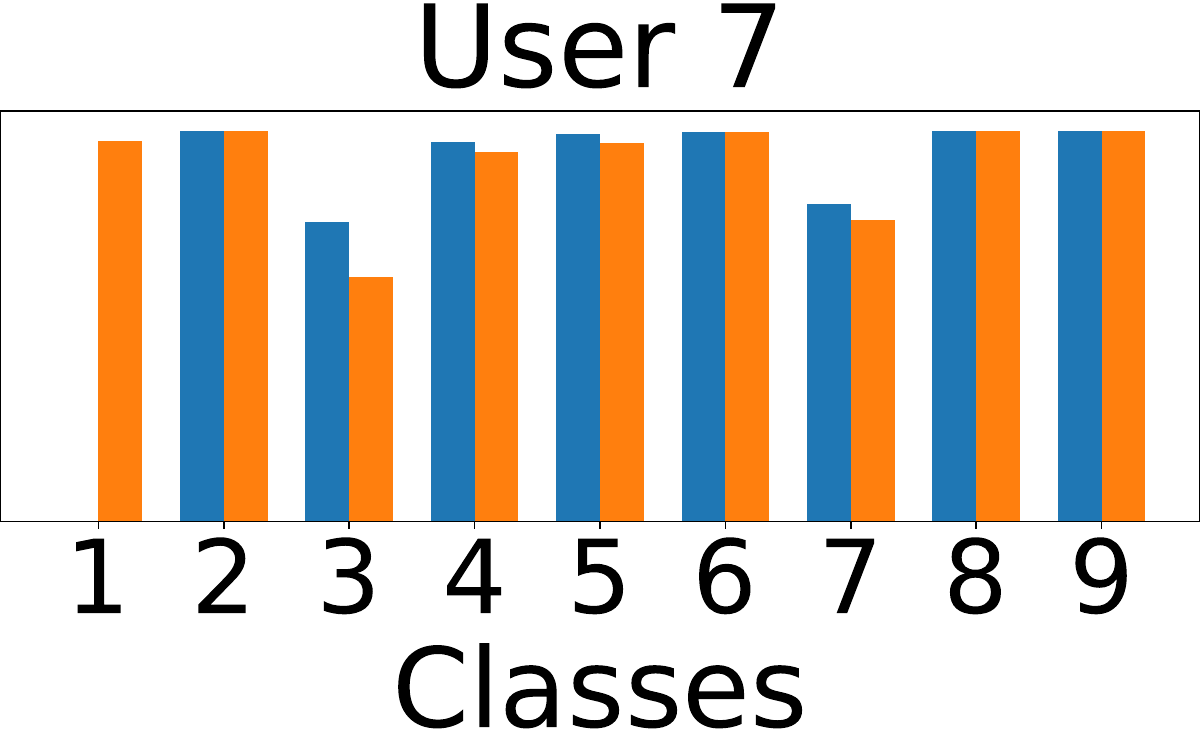}~%
    \begin{subfigure}[t]{\mywidth\linewidth}
        \includegraphics[width=\linewidth]{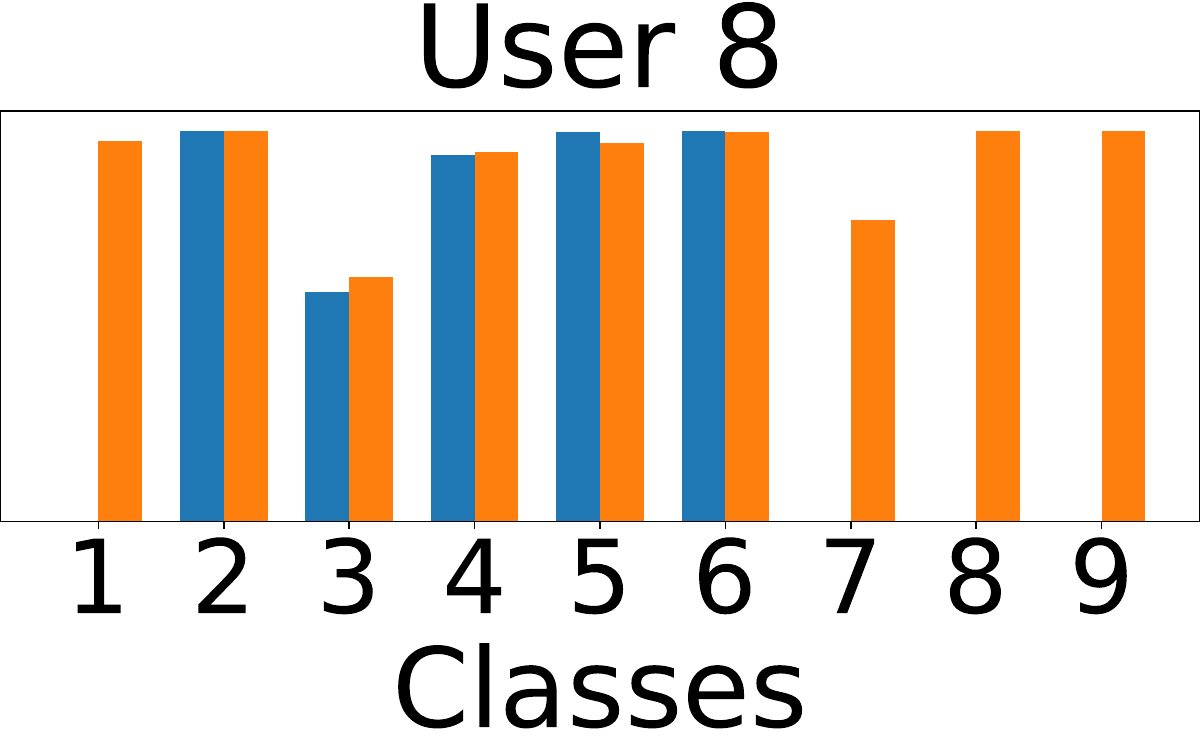}%
        \caption{$\alpha{=}$0.75}
    \end{subfigure} %
    ~\includegraphics[width=\mywidth\linewidth]{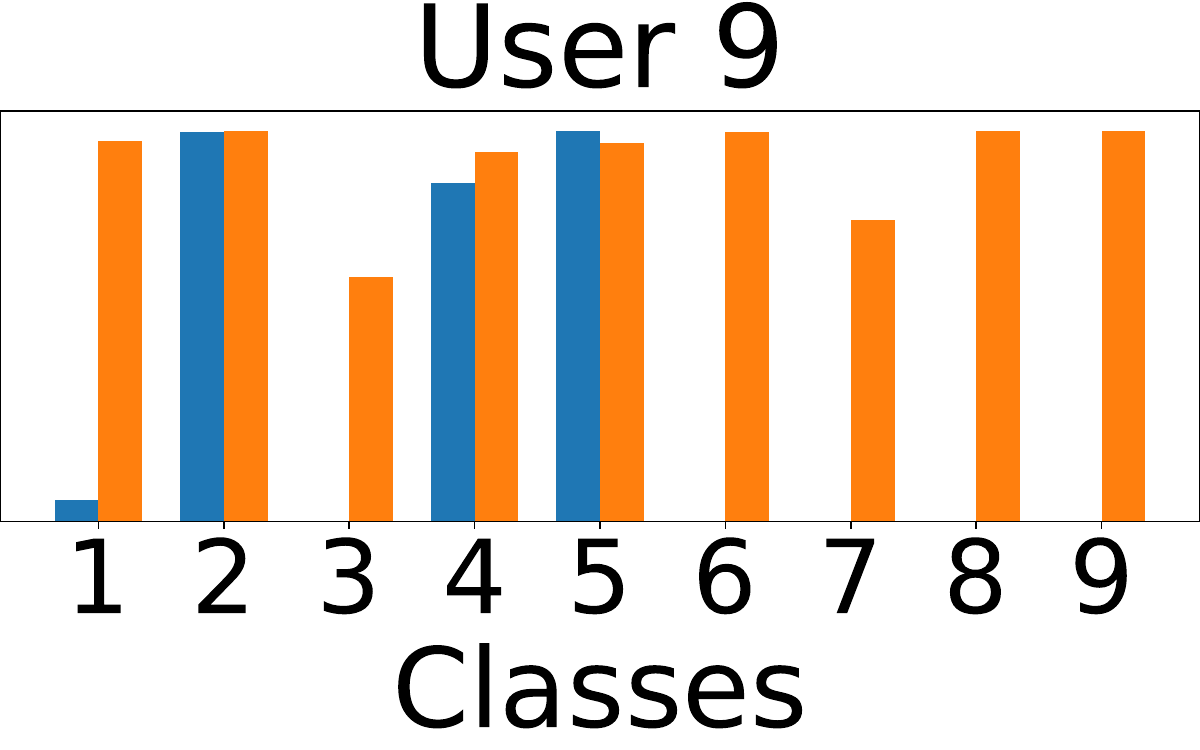}~ %
    \includegraphics[width=\mywidth\linewidth]{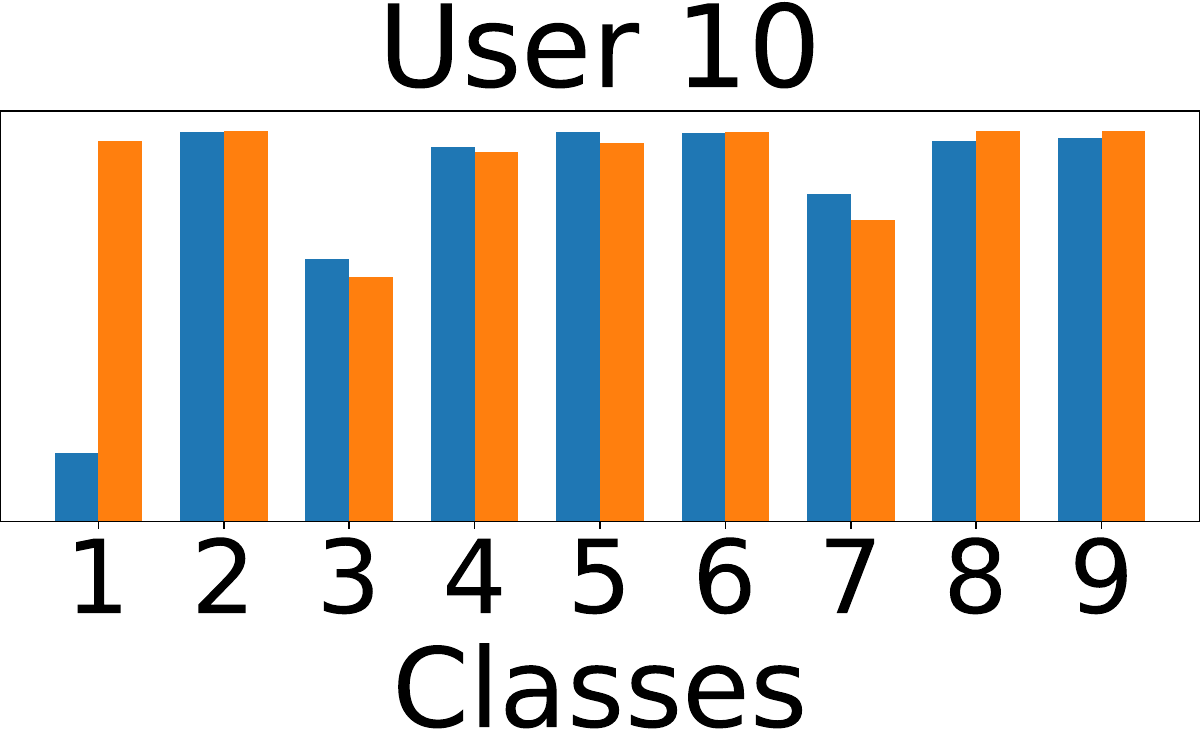}
    \includegraphics[width=\mywidth\linewidth]{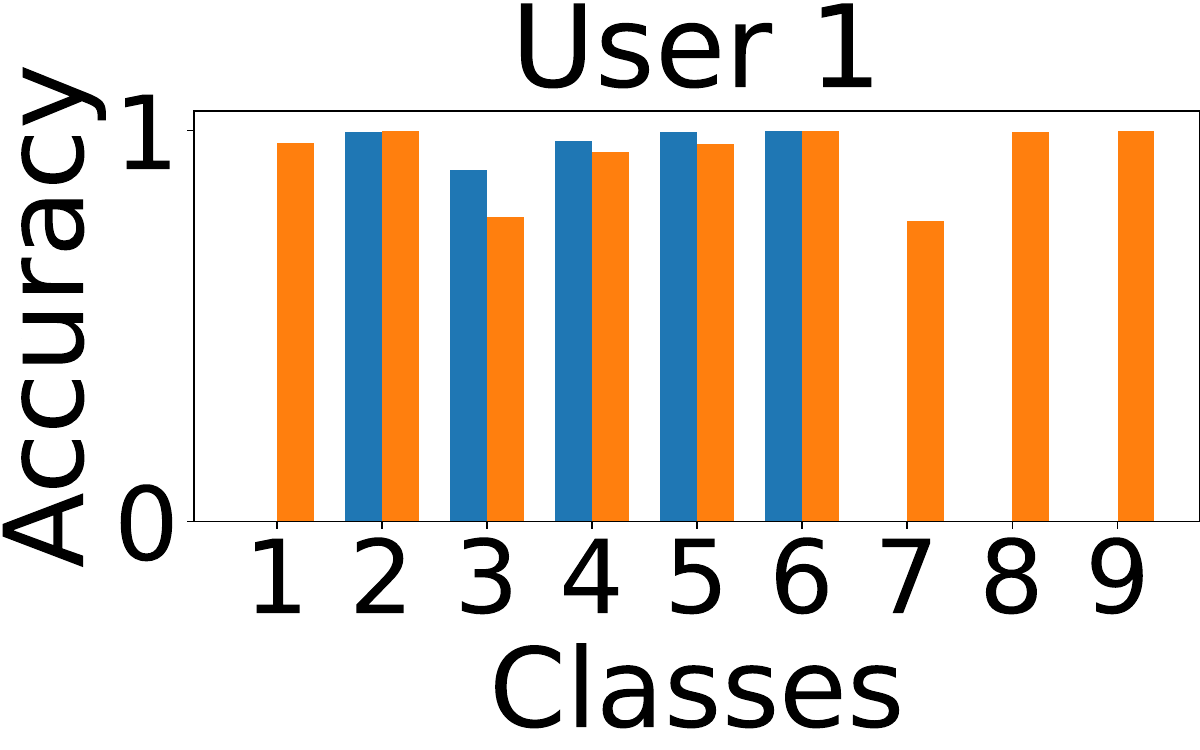}~%
    \includegraphics[width=\mywidth\linewidth]{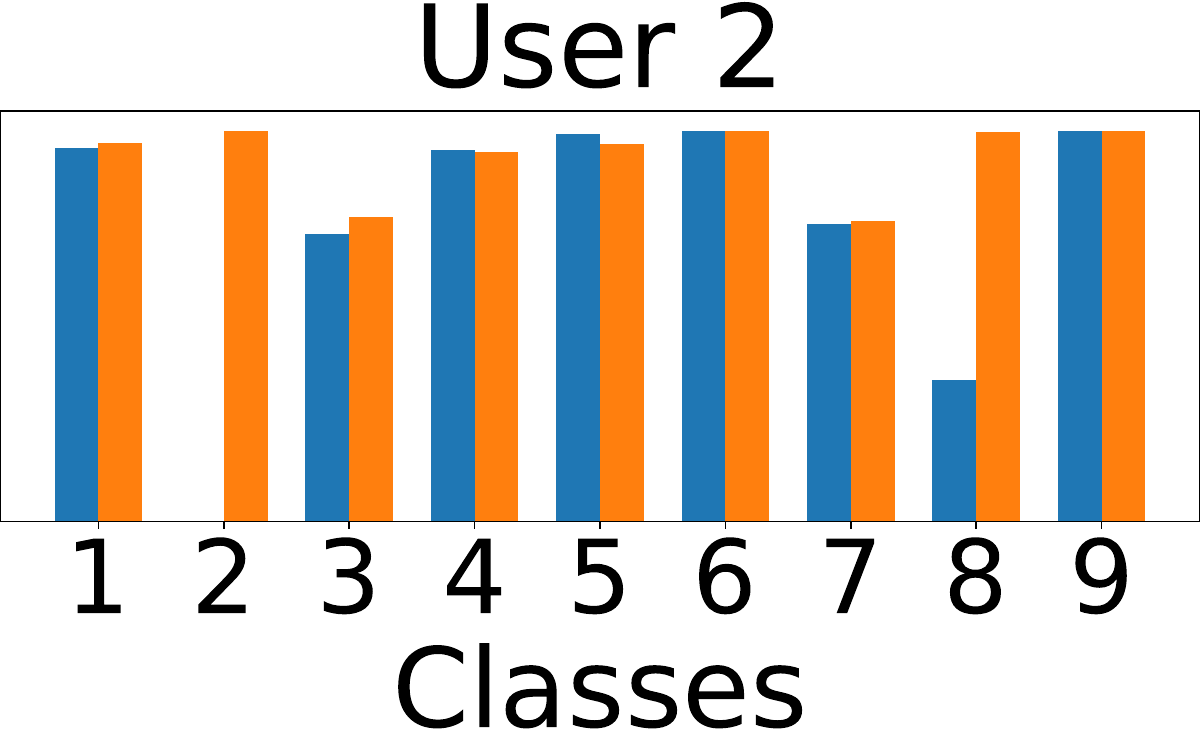}~%
    \includegraphics[width=\mywidth\linewidth]{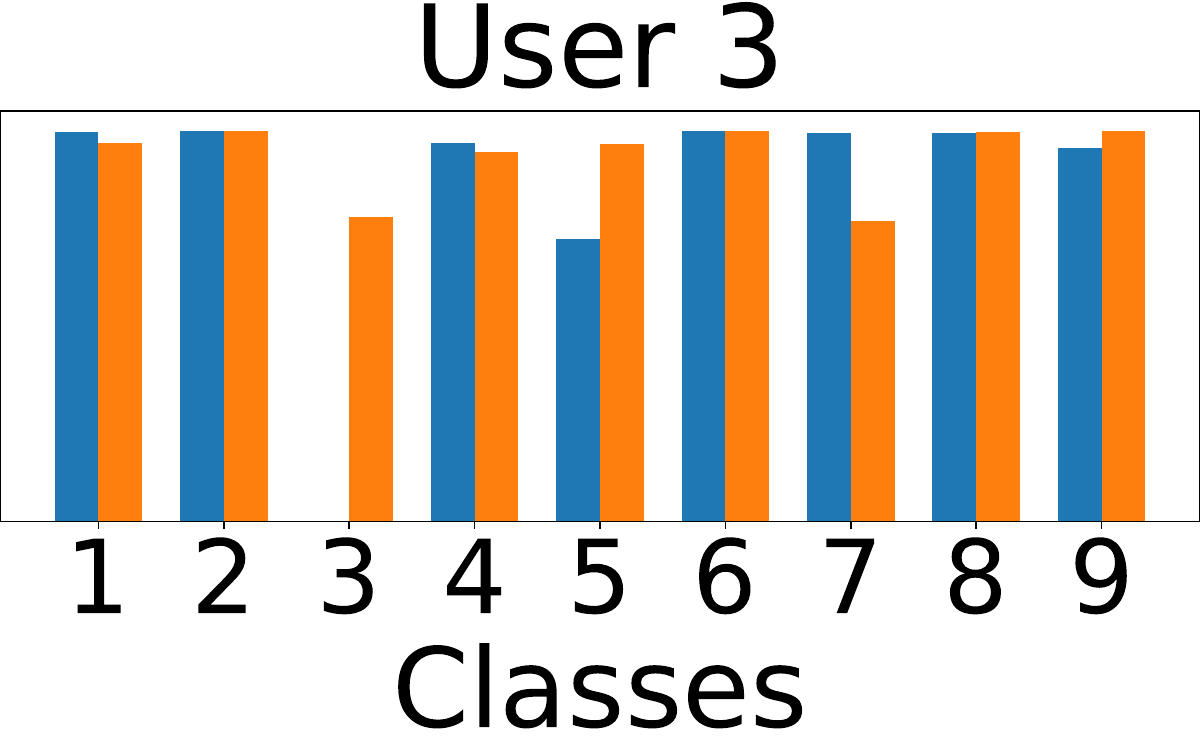}~%
    \includegraphics[width=\mywidth\linewidth]{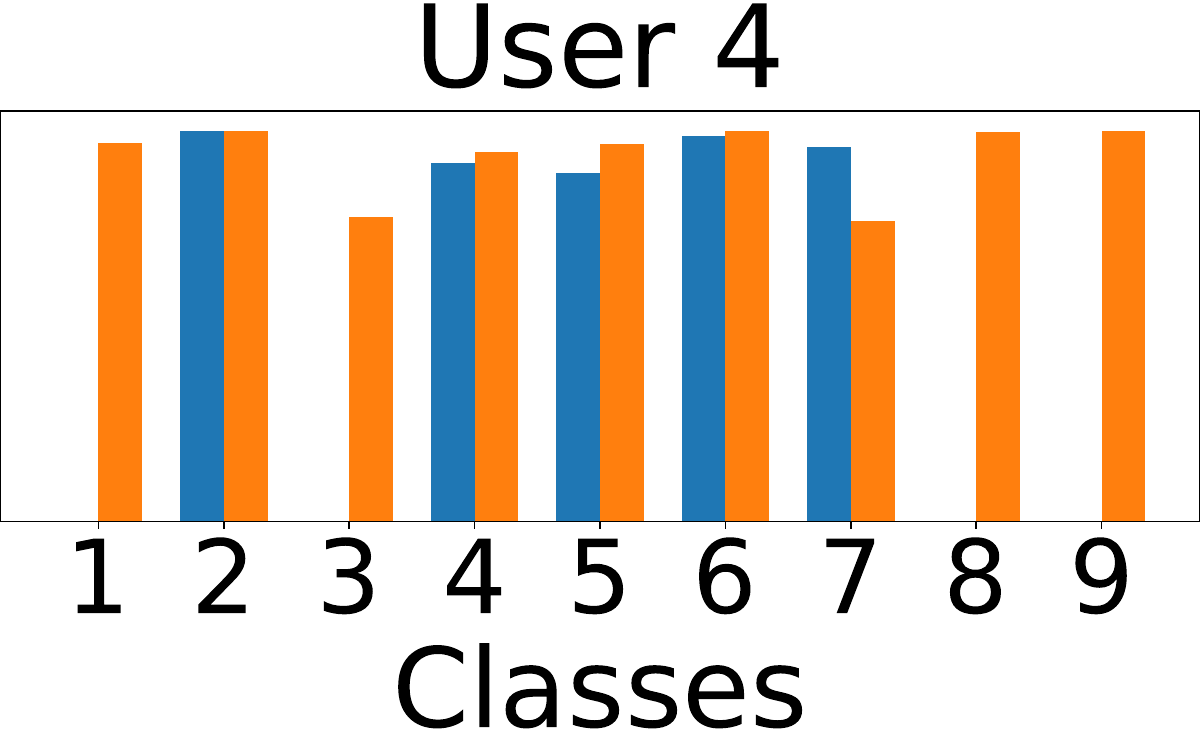}~%
    \includegraphics[width=\mywidth\linewidth]{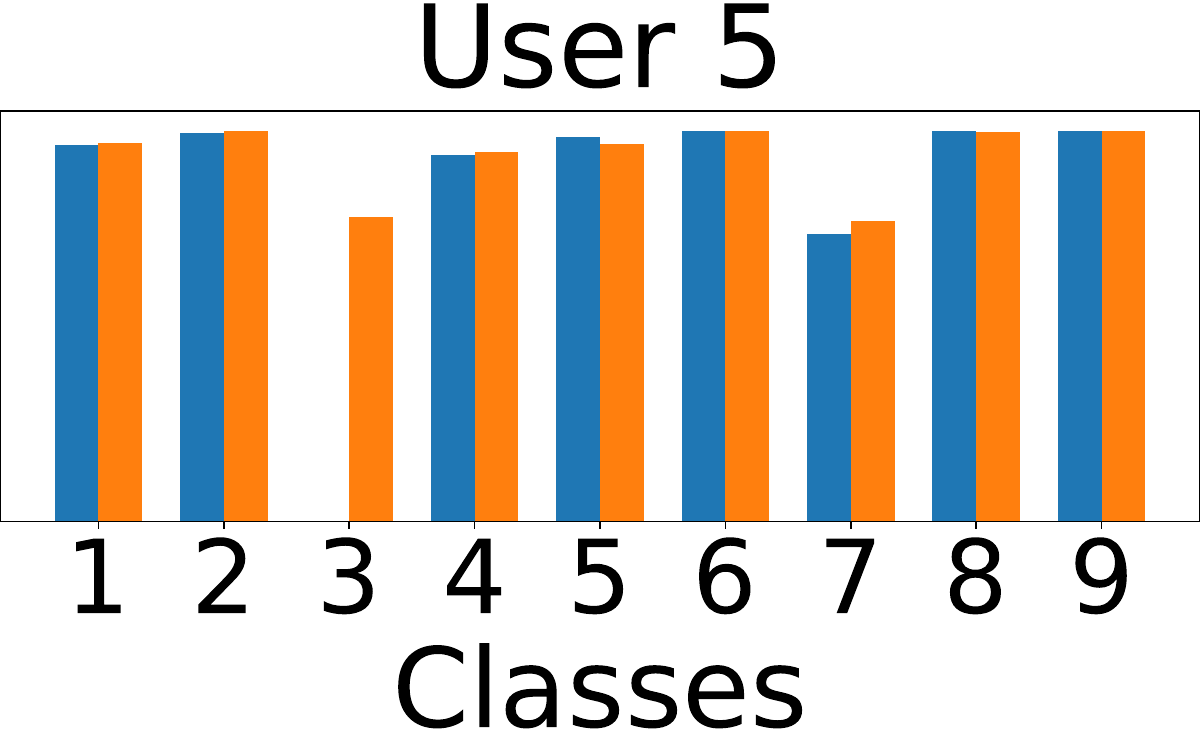}
    \includegraphics[width=\mywidth\linewidth]{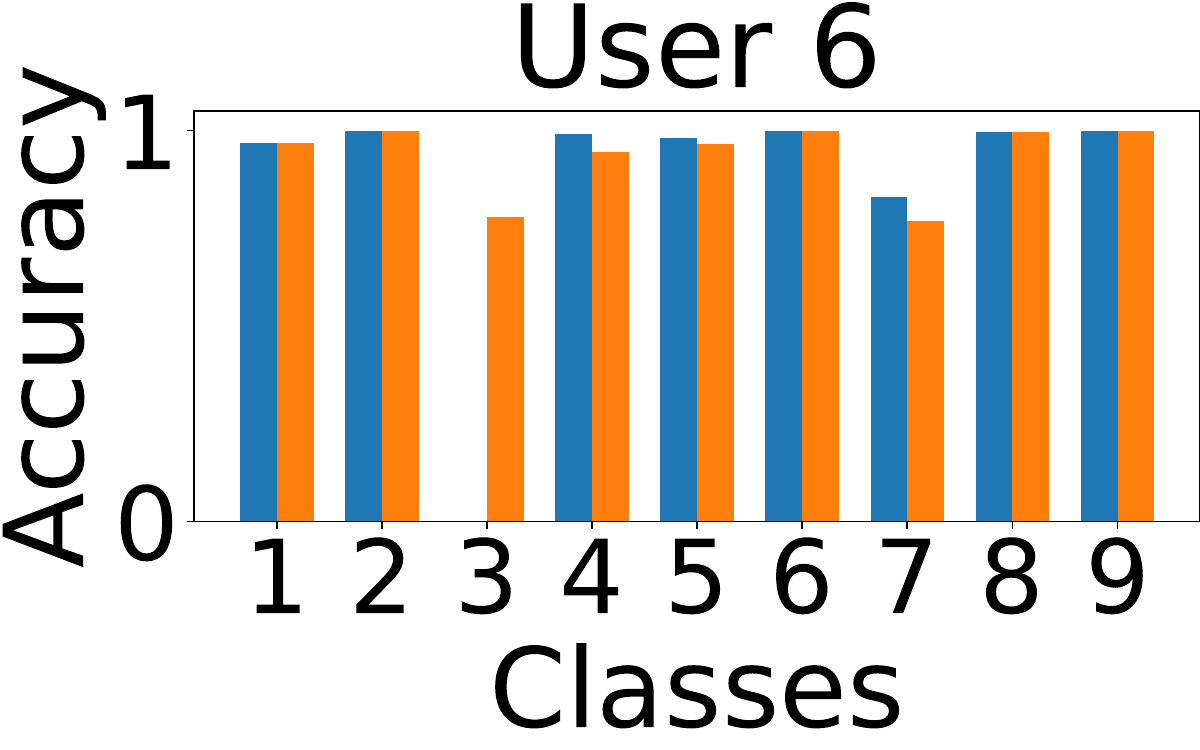}~%
    \includegraphics[width=\mywidth\linewidth]{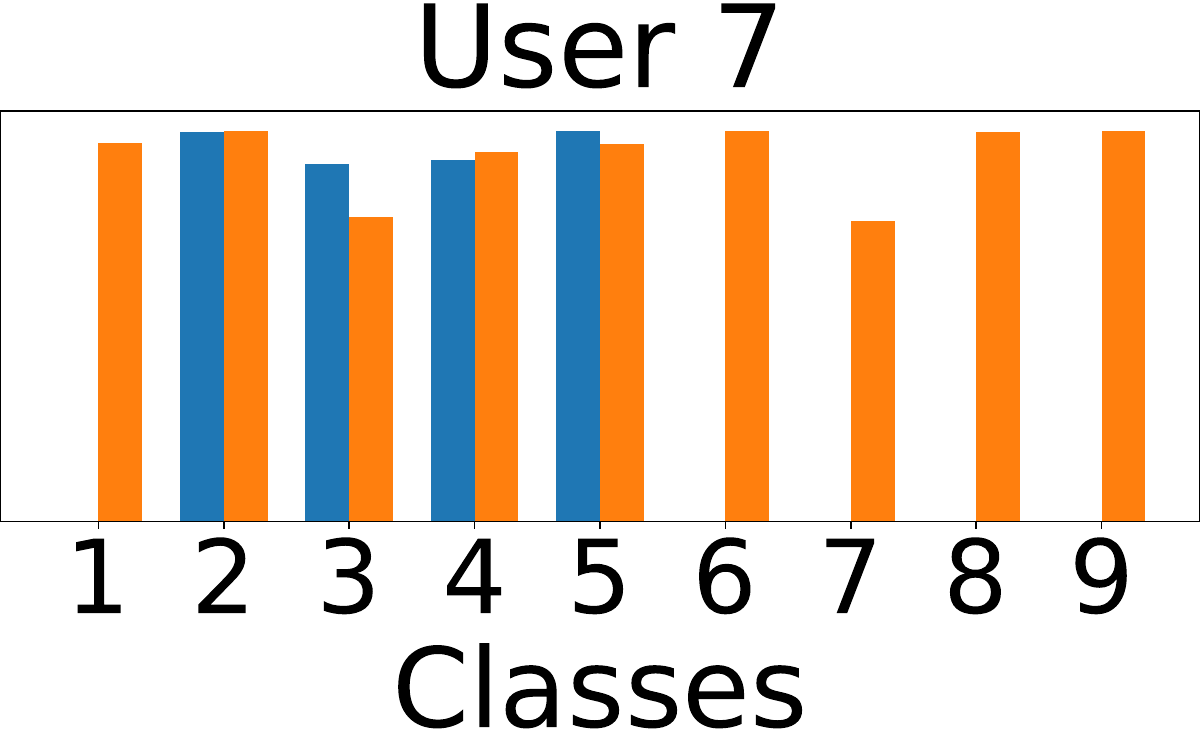}~%
    \begin{subfigure}[t]{\mywidth\linewidth}
        \includegraphics[width=\linewidth]{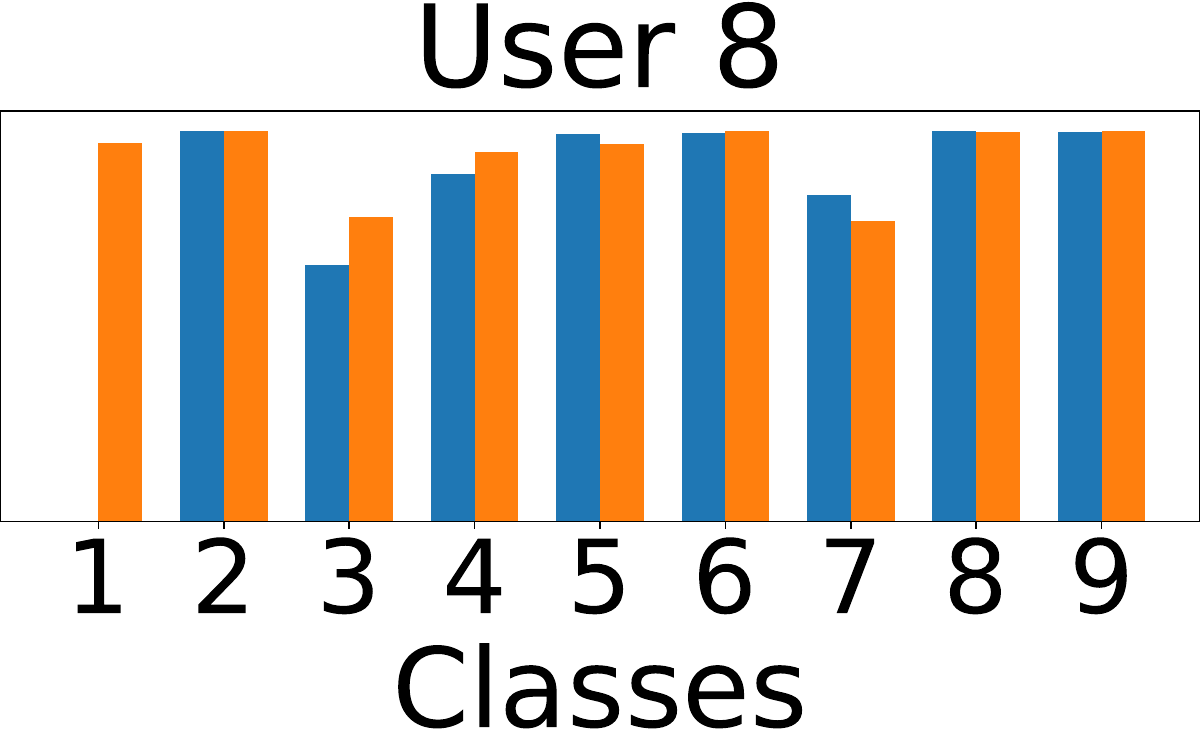}%
        \caption{$\alpha{=}$0.5}
    \end{subfigure}%
    ~\includegraphics[width=\mywidth\linewidth]{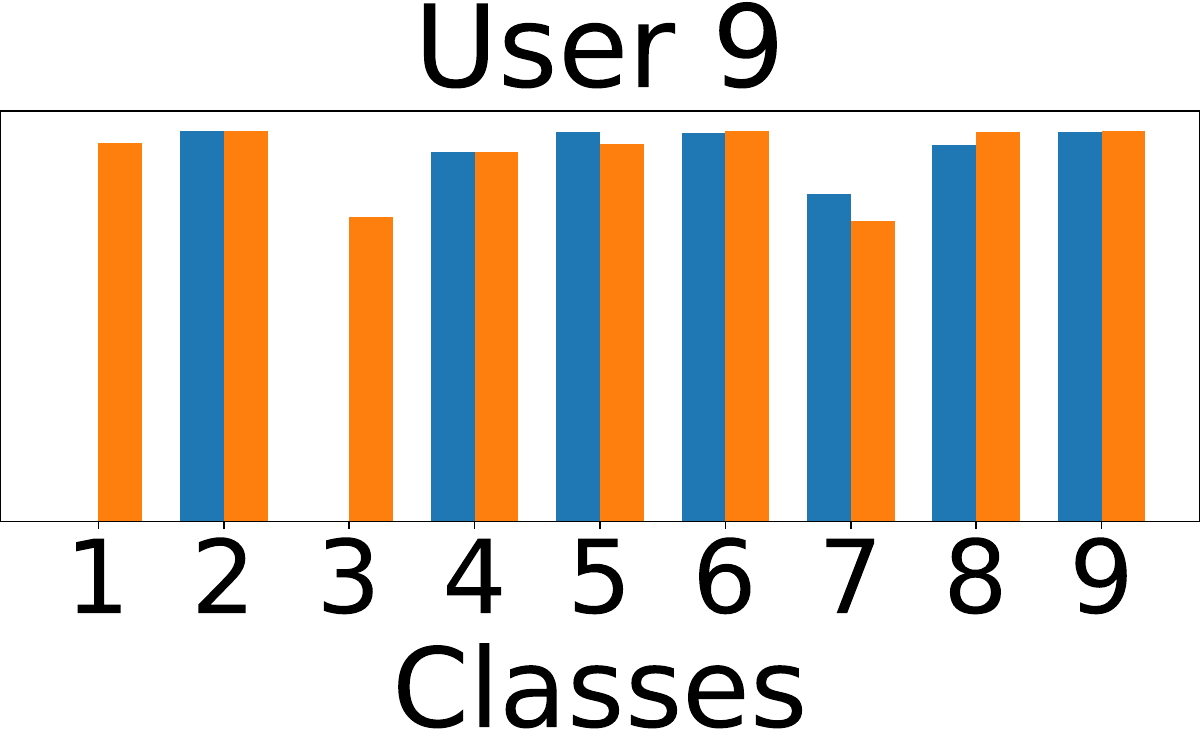}~%
    \includegraphics[width=\mywidth\linewidth]{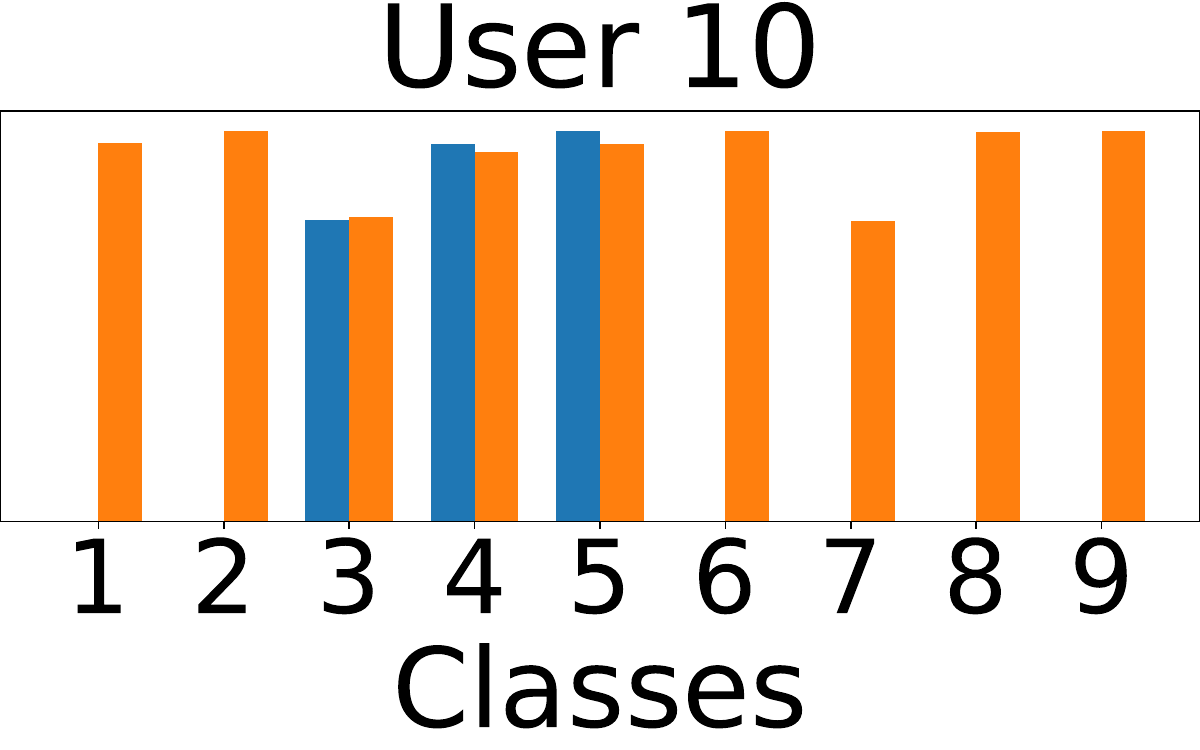}
    \includegraphics[width=\mywidth\linewidth]{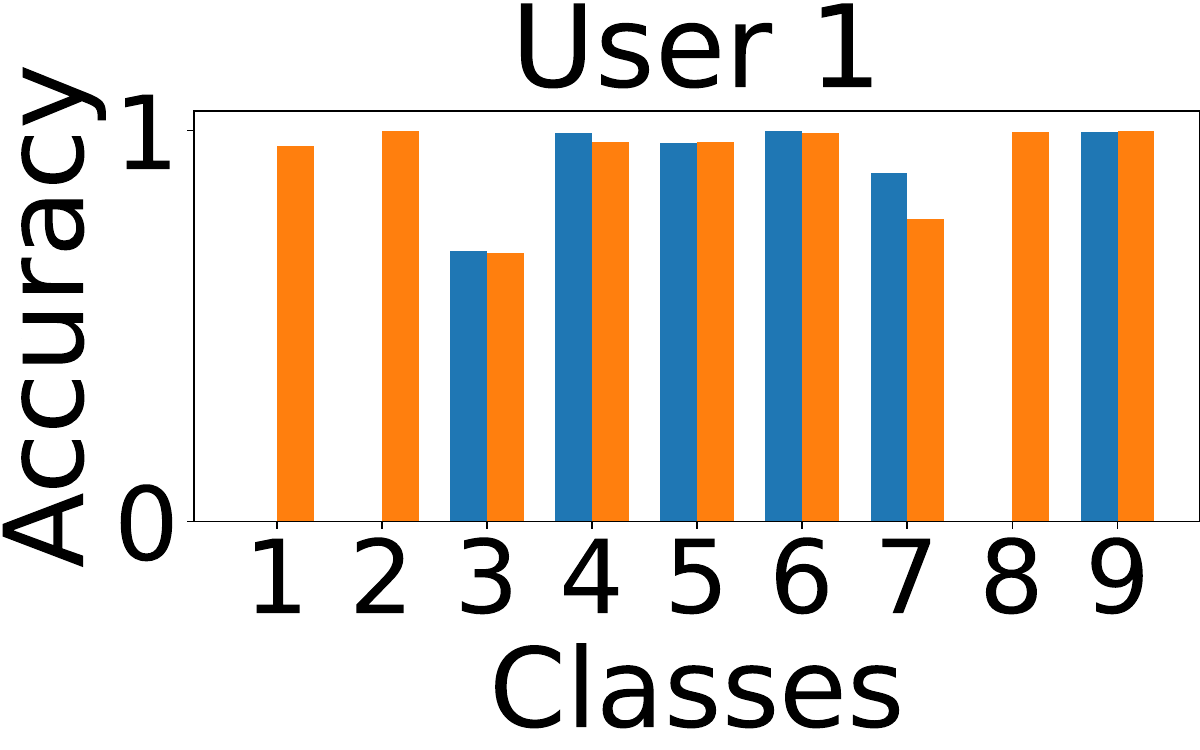}~%
    \includegraphics[width=\mywidth\linewidth]{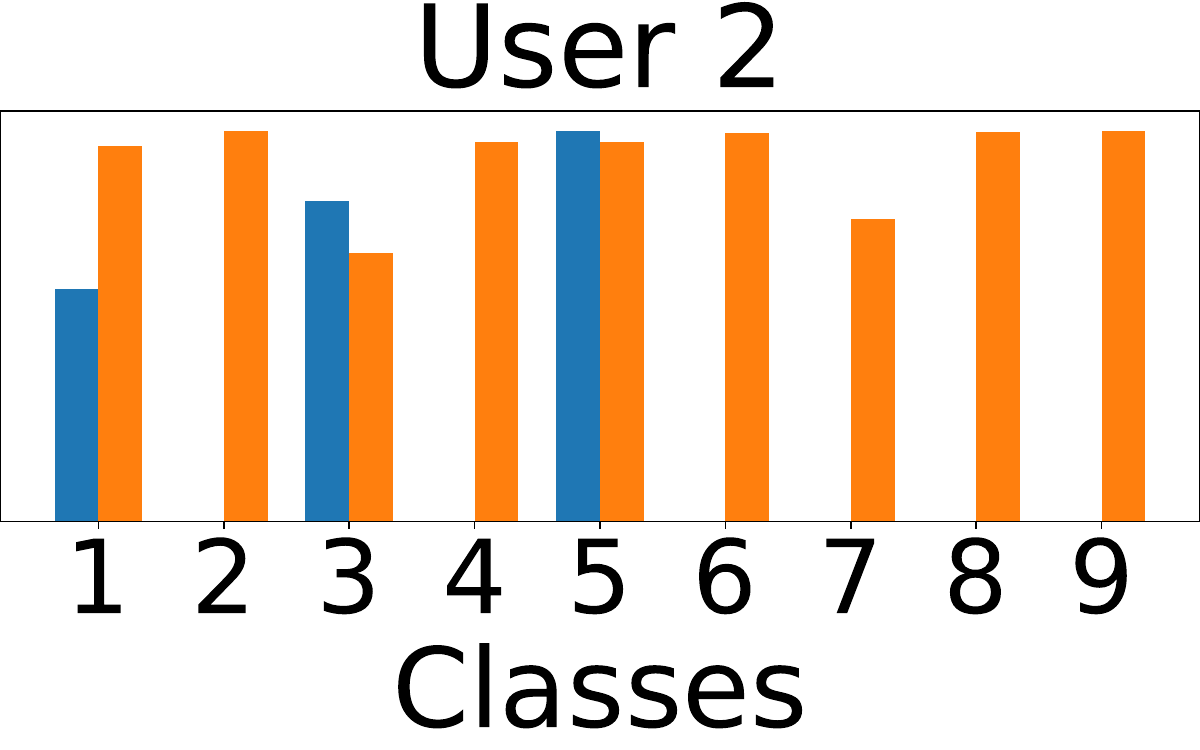}~%
    \includegraphics[width=\mywidth\linewidth]{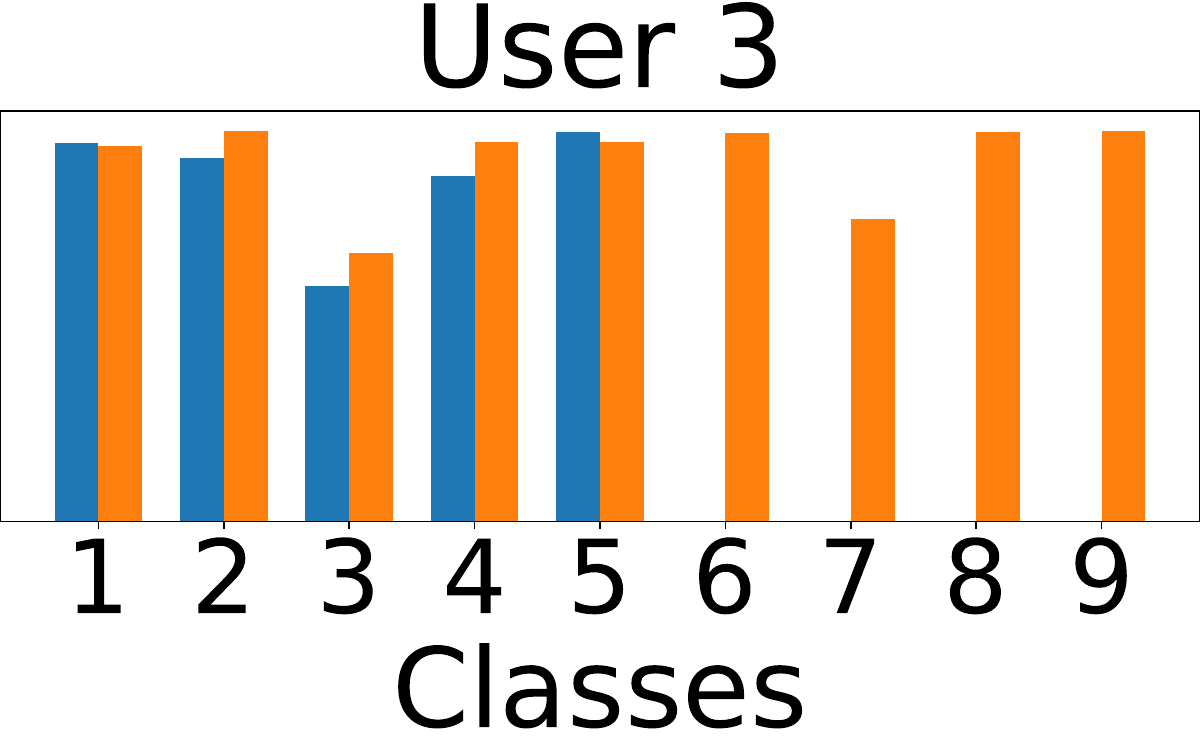}~%
    \includegraphics[width=\mywidth\linewidth]{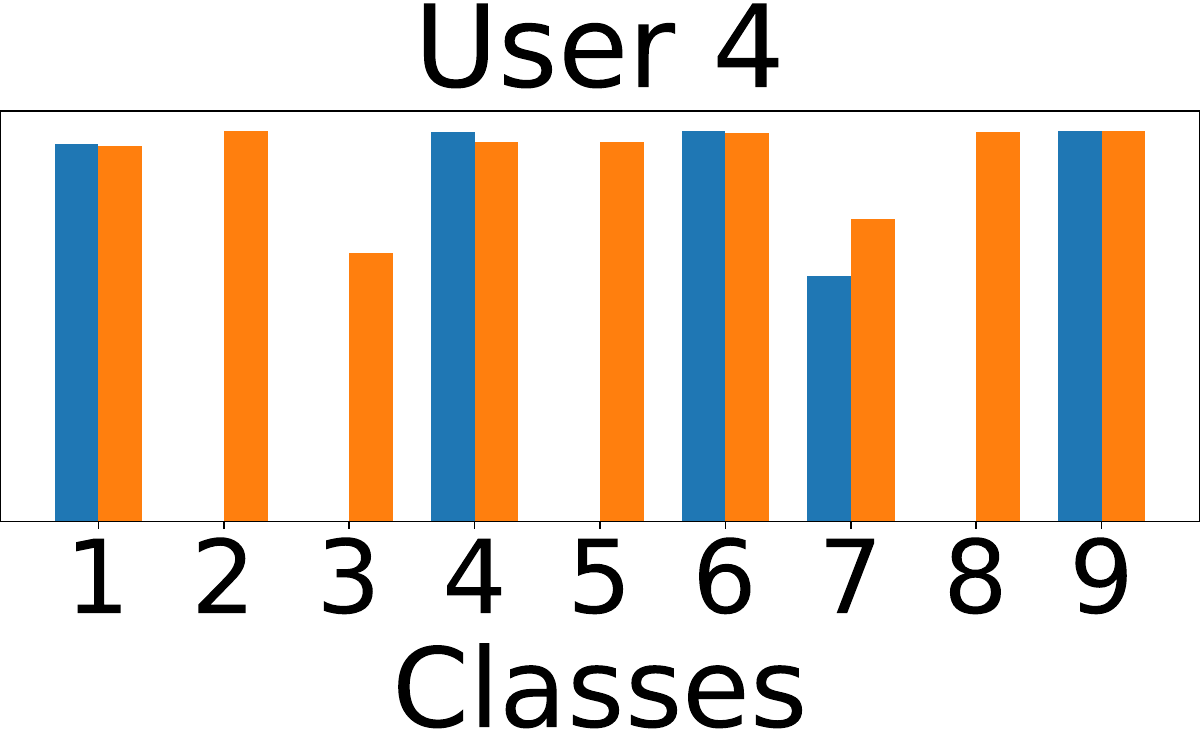}~%
    \includegraphics[width=\mywidth\linewidth]{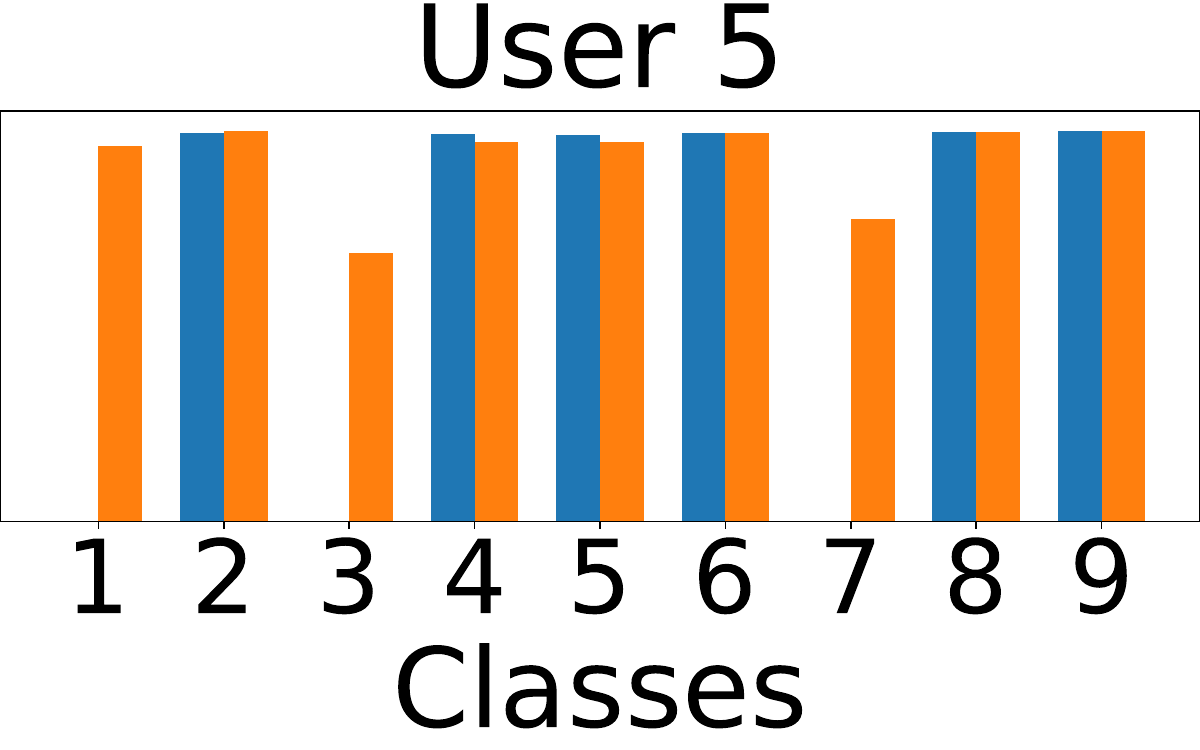}
    \includegraphics[width=\mywidth\linewidth]{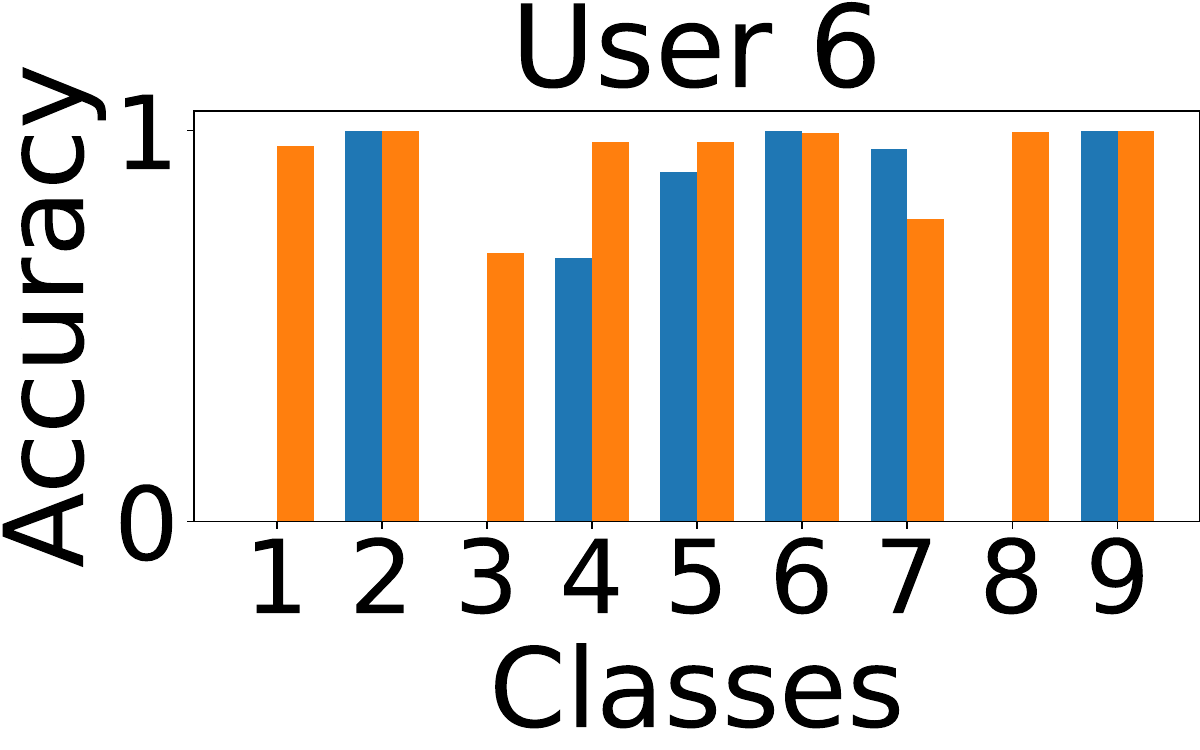}~%
    \includegraphics[width=\mywidth\linewidth]{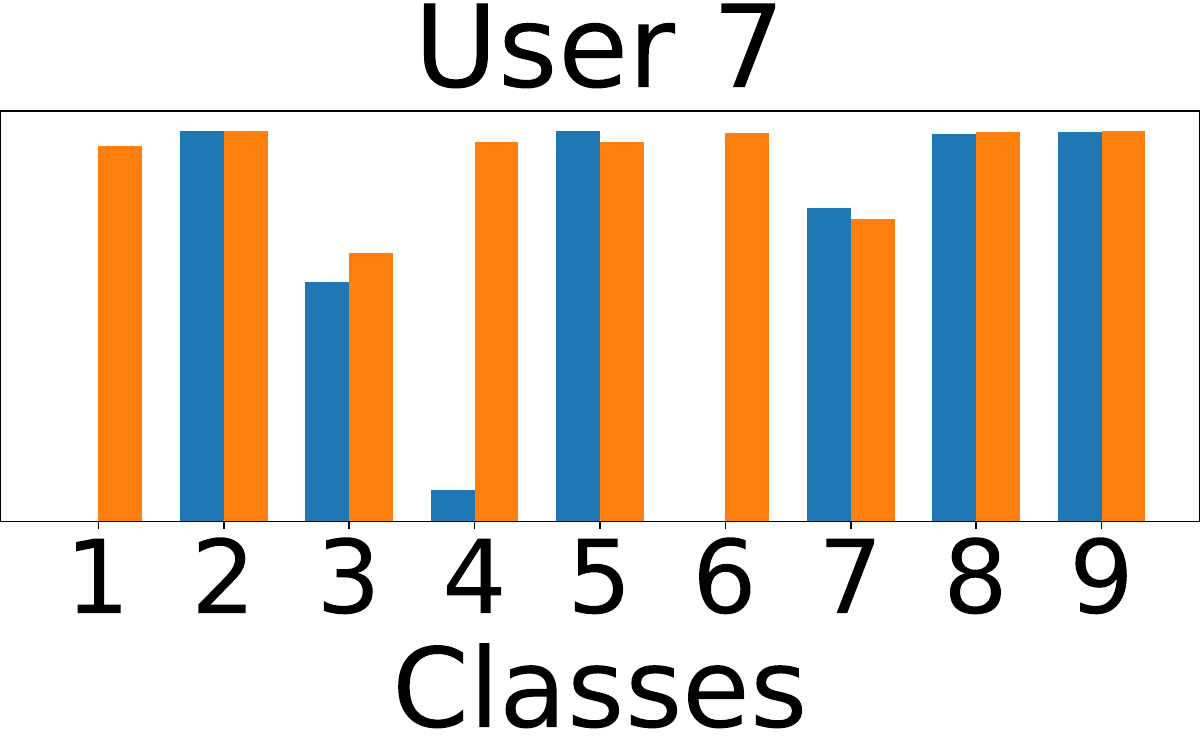}~%
    \begin{subfigure}[t]{\mywidth\linewidth}
        \includegraphics[width=\linewidth]{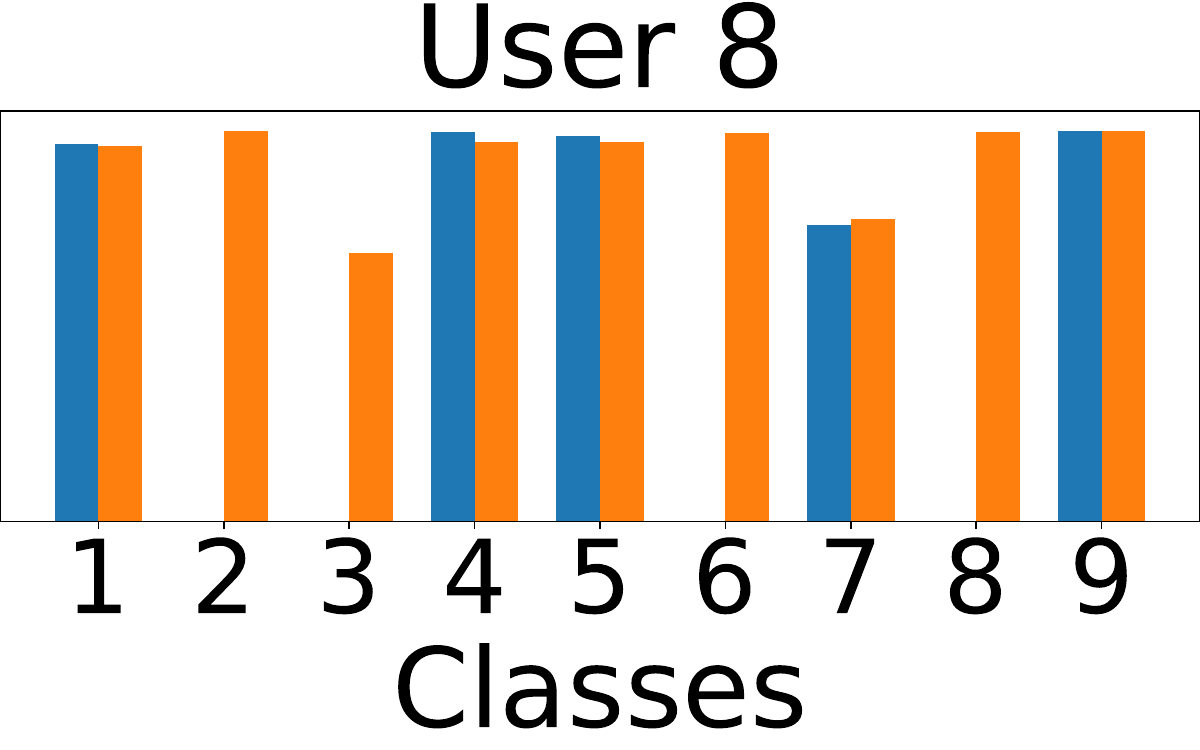}~%
        \caption{$\alpha{=}$0.25}
    \end{subfigure}%
    ~\includegraphics[width=\mywidth\linewidth]{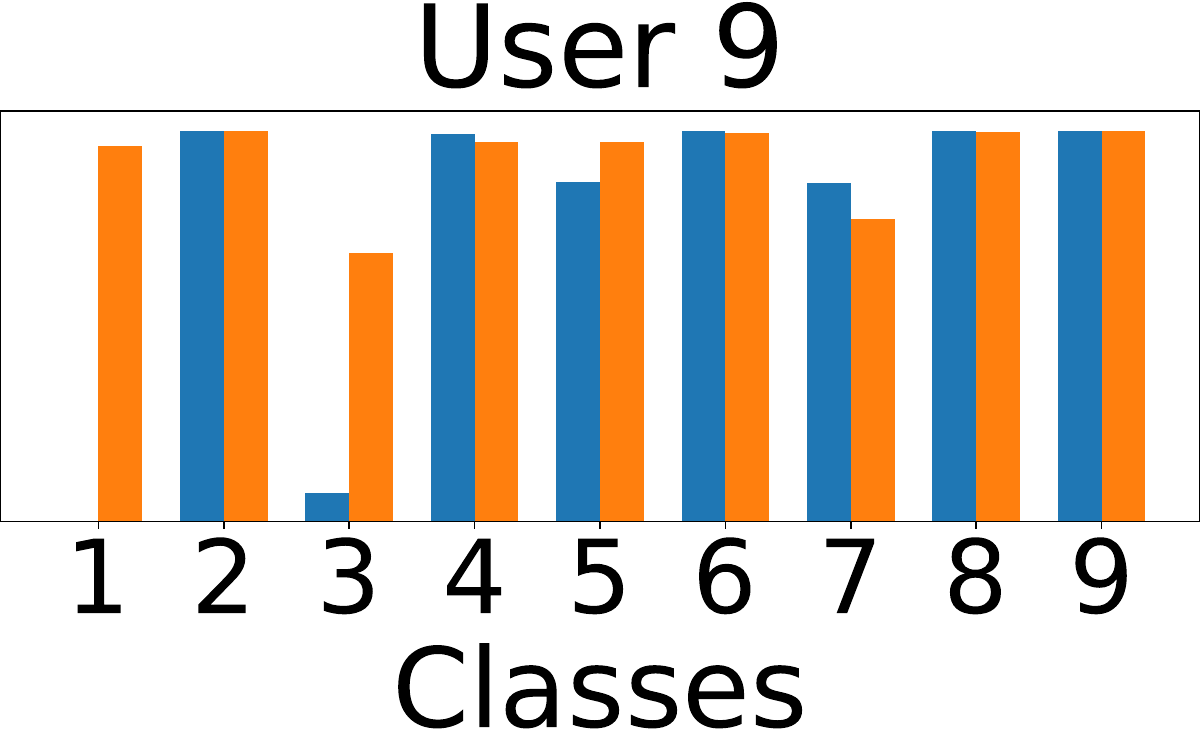}~%
    \includegraphics[width=\mywidth\linewidth]{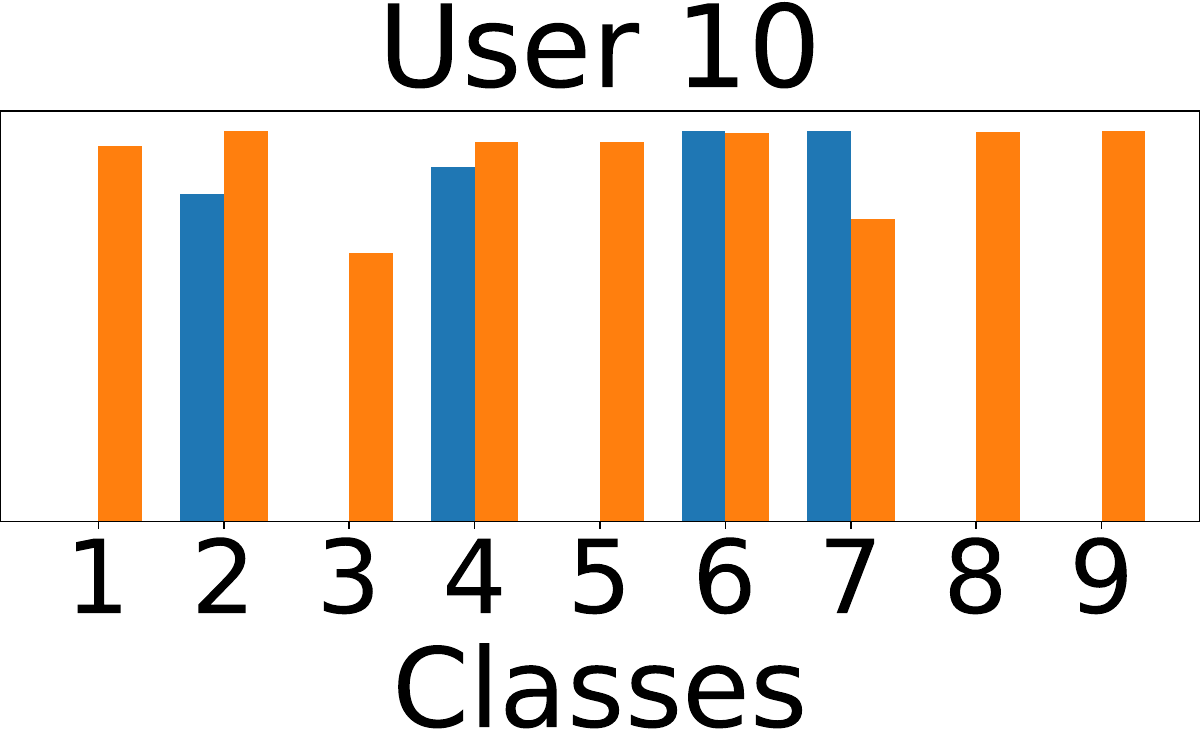}

    \caption{Per-class accuracy for all users using PROTEAN on the X-IIoTID dataset with different $\alpha$ values. Blue and orange bars represent the accuracy without FL and with PROTEAN, respectively.}
    \label{fig: before after fl xiiotid}
\end{figure*}

\subsection{Knowledge Sharing for Detecting Rare Attacks}
For each local participant, we choose the 2 attack types with the least training samples as the \textsl{rare} attack classes collected on the corresponding local participant. By setting $\alpha = 0.25, 0.5$ and $0.75$, the 2 rare attack classes amount for, on average, less than {0.39\%, 0.59\% and 0.12\% for X-IIoTID and 0.32\%, 0.59\%, 1.46\% for 5G-NIDD of training samples hosted by the other attack types across different local participants.

On each data set, we average the accuracy score of the 2 rare attack classes for each local participant with PROTEAN and Cerberus, noted as ${\mathcal{A}}_{\text{PROTEAN}}$ and ${\mathcal{A}}_{\text{Cerberus}}$. We then compare the averaged accuracy scores of PROTEAN and Cerberus in Figures~\ref{fig: ra_comparaisions_xiiotid} and~\ref{fig: ra_comparaisions_5gnidd}. %
In the analysis of the X-IIoTID dataset, PROTEAN consistently outperforms Cerberus in detecting both rare attacks across all participants. Specifically, for $\alpha$ values of 0.5 and 0.25, Cerberus failed to detect rare attacks on participants 6 and 7, and participants 4 and 5, respectively, resulting in an accuracy close to 0\%. In contrast, PROTEAN significantly improves detection accuracies to over 80\% for rare attacks on these participants.
Similarly, on 5G-NIDD, all rare attacks are being better detected for some participants for $\alpha=0.5$ but the difference between ${\mathcal{A}}_{\text{PROTEAN}}$ and ${\mathcal{A}}_{\text{Cerberus}}$ for these values is less than $6\%$. On the other hand, other participants made an improvement with PROTEAN of more than 55\%, and, for participants 0, 1, and 5, the attacks were not detected with Cerberus but they were with PROTEAN. For other $\alpha$ values, we observe that ${\mathcal{A}}_{\text{PROTEAN}}$ is higher than ${\mathcal{A}}_{\text{Cerberus}}$ for all the cases and improvement achieved 
was higher than 80\% for participant 9 when $\alpha=0.25$.
Furthermore, we perform the Mann-Whitney U hypothesis test \cite{10.1214/aoms/1177730491} over ${\mathcal{A}}_{\text{PROTEAN}}$ and ${\mathcal{A}}_{\text{Cerberus}}$  of all the participants on X-IIoTID and 5G-NIDD datasets. The hypothesis test shows that ${\mathcal{A}}_{\text{PROTEAN}}$ is significantly larger than ${\mathcal{A}}_{\text{Cerberus}}$ with a p-value less than $2\mathrm{e}{-4}$ for the X-IIoTID dataset. 
For the 5G-NIDD dataset, the p-value is less than $5\mathrm{e}{-2}$ with $\alpha=$ 0.75 and $0.25$, indicating a significantly larger value of ${\mathcal{A}}_{\text{PROTEAN}}$  than ${\mathcal{A}}_{\text{Cerberus}}$.  The only exception is at $\alpha =0.5$ on 5G-NIDD, the p-value is larger than $1\mathrm{e}{-2}$. The possible reason is that the two attack classes selected to measure ${\mathcal{A}}_{\text{PROTEAN}}$ and ${\mathcal{A}}_{\text{Cerberus}}$ on some of the participants have relatively more samples inside the 5G-NIDD dataset compared to the situation on the X-IIoTID dataset. Therefore, the accuracy gap between PROTEAN and Cerberus shrinks on these participants.  We also compute the average of ${\mathcal{A}}_{\text{PROTEAN}}$ and ${\mathcal{A}}_{\text{Cerberus}}$ scores derived on all the participants, which gives 91.32\% and 58.08\% respectively. We observe that globally, PROTEAN can reach 33.24\% higher detection accuracy than Cerberus on the rarely appearing attack types across all the local participants.

These empirical observations convey a two-fold message. First, via sharing the global attack class prototypes in PROTEAN, the local participants with the rare attacks can use the distribution of these attack data in the embedding space to enhance the training of their local models, thus mitigating the training data bias induced by the local imbalanced data. This is achieved by integrating the prototype regularization into the learning objective of PROTEAN in updating local detection models in each round, according to Equation~\ref{eq:optimization_objective} in Section~\ref{section: methodology}. Second, compared to Cerberus, we find that PROTEAN can effectively boost the detection performances over the rarely appearing attack classes. Though sharing model parameters in Cerberus facilitates driving the federated training to converge to a global detection model, local model training in Cerberus is prone to the statistical bias caused by the class imbalance between the rare attack types and other attack classes. As a result, the globally aggregated model is impacted by the bias of class imbalance of local ones, which deteriorates the detection performance over rare classes.

\begin{figure*}[t]
    \centering
     \begin{subfigure}[t]{0.375\linewidth}
    \centering
    \includegraphics[width=0.49\linewidth]{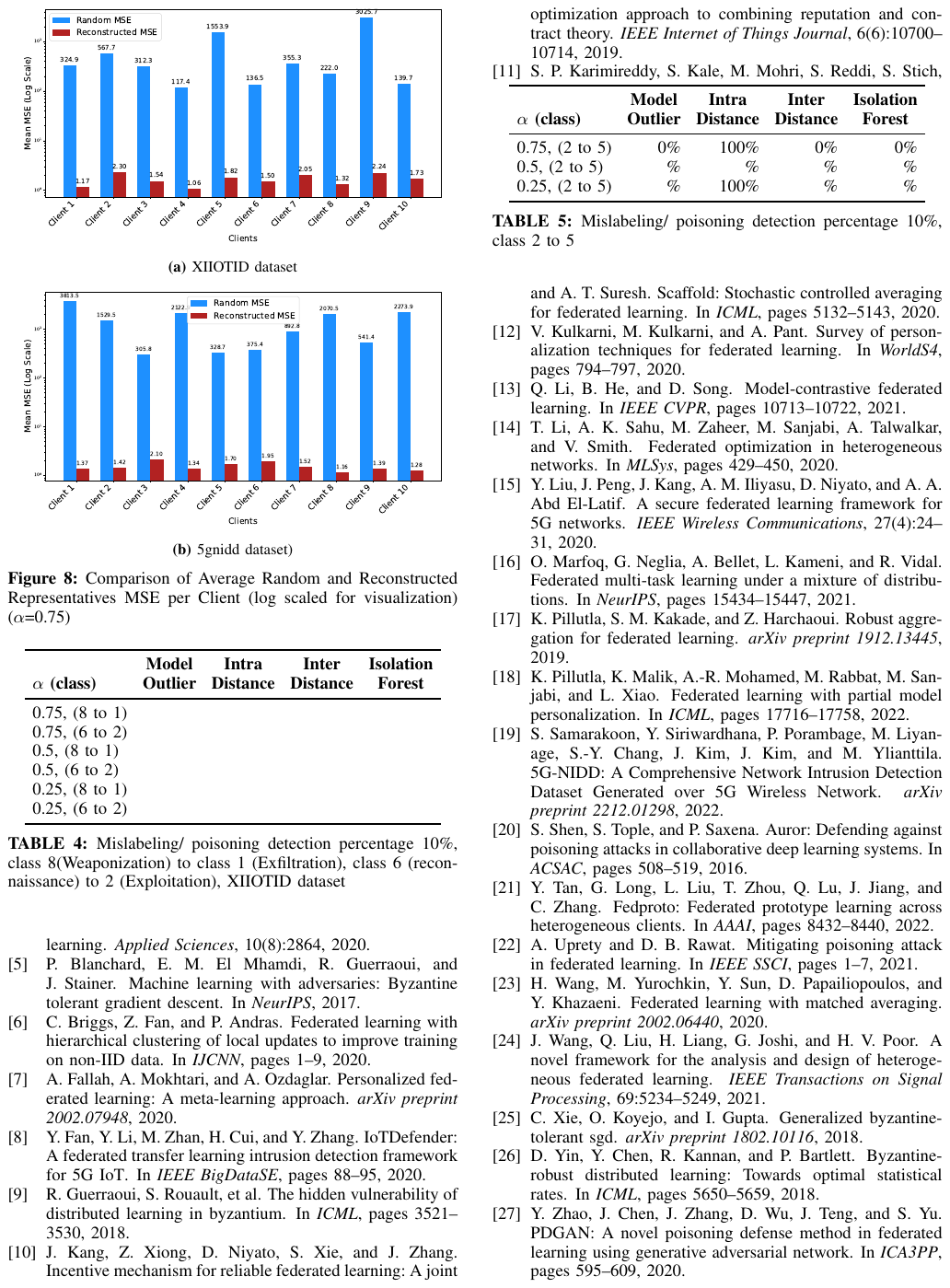}
    \includegraphics[width=0.49\linewidth]{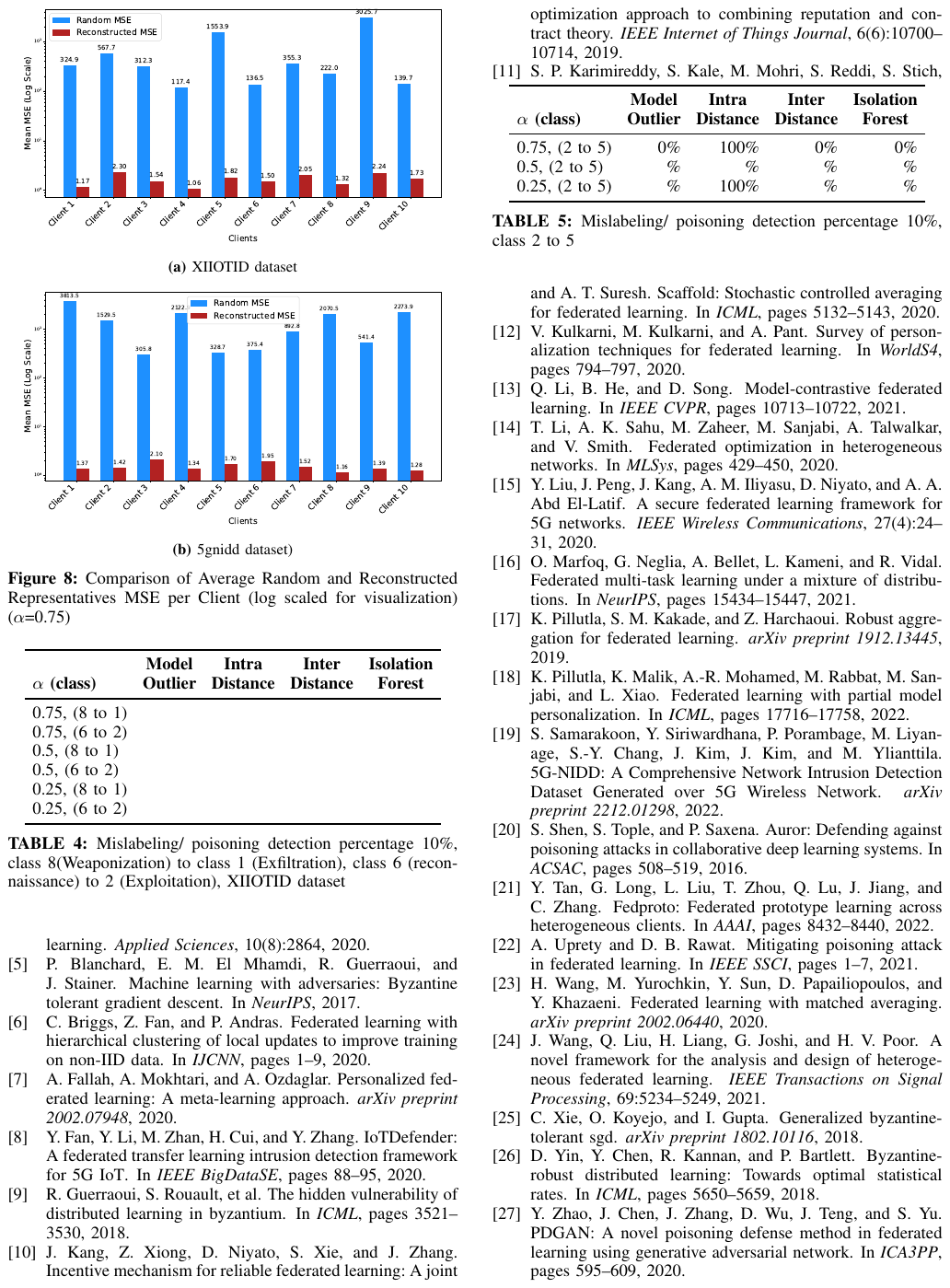}
    \end{subfigure}\\
    \begin{subfigure}[t]{0.35\textwidth}
   \centering
   \includegraphics[width=0.99\linewidth]{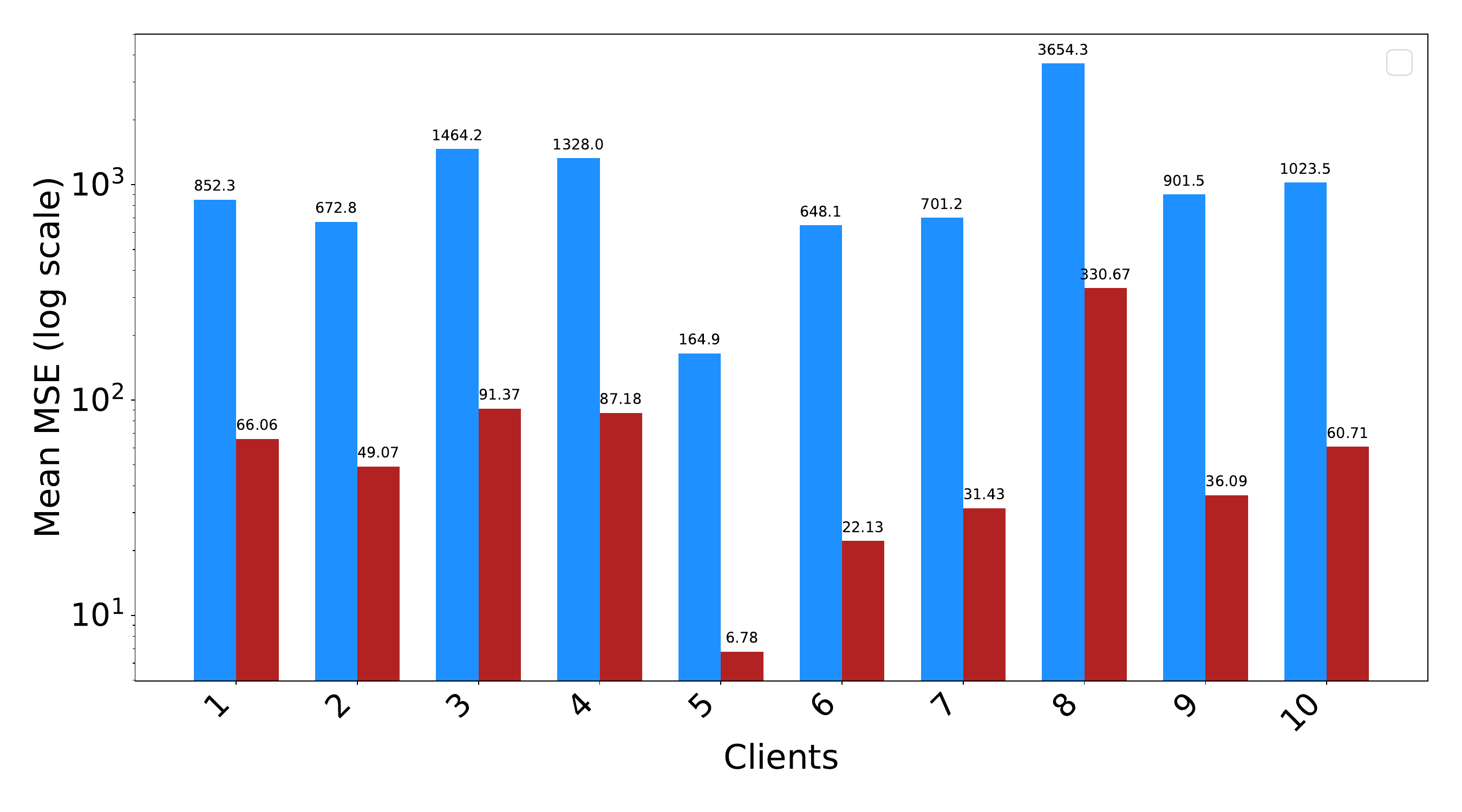}
   \caption{X-IIoTID}
   \label{fig:reconstruction_xiiotid_alpha0.75}
    \end{subfigure}
    ~
    \begin{subfigure}[t]{0.35\textwidth}
   \centering
   \includegraphics[width=0.99\linewidth]{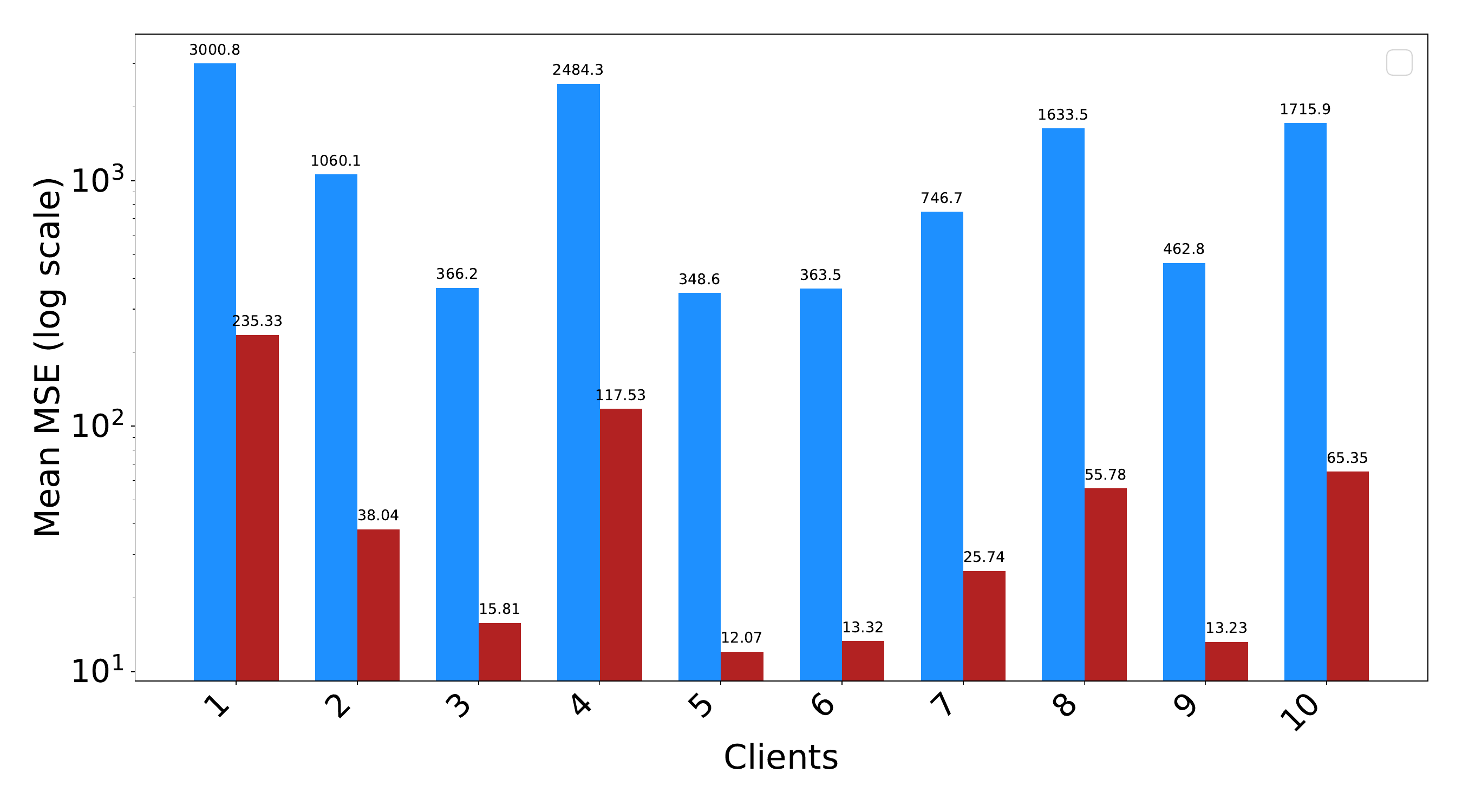}
   \caption{5G-NIDD}
   \label{fig:reconstruction_5gnidd_alpha0.75}
    \end{subfigure}
    
       \caption{Comparison between Random and Reconstructed MSEs per client (with $\alpha$=0.75). Larger/smaller gap between the two MSE values indicates more/less accurate data reconstruction.} %
       \label{fig:reconstruction0.75}
\end{figure*}

\descr{Zero-shot Learning of Unseen Cyber Intrusions via Knowledge Sharing.} %
For each participant, we further examine the detection accuracy of PROTEAN across attack classes for which the participant has no training samples (zero-shot learning \cite{8413121}). After training the IDS model using PROTEAN, we compare the accuracy score for each attack class between models trained with PROTEAN and those trained only locally (i.e., \textit{without FL}) in Figures~\ref{fig: before after fl xiiotid} and~\ref{fig: before after fl 5gnidd}  for $\alpha$ values of $0.75$, $0.5$, and $0.25$, for X-IIoTID and 5G-NIDD datasets, respectively. For the setting without FL, the detection model is locally trained with the training samples of cyber intrusions hosted by the participant. In this case, this local participant can only detect the attack types present in his own training data. Using the locally trained detection model, there are some participants limited to detecting just one or two types of cyber intrusions. By comparison, after applying PROTEAN, local participants using the shared model and prototypes were able to detect all types of attacks. 
For example, for the X-IIoTID dataset and for $\alpha$ = 0.75, participant 5 does not have any instances of classes RDOS, Reconnaissance, Tampering, and Weaponization. With the locally trained detection model, the accuracy scores of these classes are all 0\%. With PROTEAN, the accuracy of these classes are increased to 99.61\%, 77.18\%, 100\%, and 99.98\% respectively. For the same dataset with $\alpha$ = 0.5, participant 6 does not have any training instances of the same classes.
With the locally trained detection model, the accuracy scores are null for these classes for that participant. With PROTEAN, the accuracy is raised to 99.80\%, 76.91\%, 99.61\%, and 99.62\% on these classes, respectively.
For the 5G-NIDD dataset and $\alpha$ = 0.75, participant 5 does not have any instances of all classes except benign. With only the locally trained model, the accuracy of these classes is null. With PROTEAN, their accuracies are increased to 76.19\%, 91.78\%, 98.45\%, 99.59\%, 77.03\%, 99.10\%, and 99.88\%.
This indicates that local participants can share their knowledge about cyber intrusions and detect unseen attacks in their own local datasets. The empirical results demonstrate the effectiveness of PROTEAN in knowledge sharing and in effectively detecting attacks not present in the local training data.

\section{Audit of Privacy Risks in PROTEAN}

\descr{Privacy Risk.}
We start by investigating potential privacy violations in PROTEAN, more precisely, from sharing the prototypes of the attack classes and the detection model's parameters.  
To do so, we use data reconstruction attacks geared to recover the feature values of the raw network traffic flow data. 
We assume that the attacker controls a semi-honest central server. %
The attacker has full access to the parameters of the local detection models and local class-specific prototype vectors submitted by a local participant. 
The attacker's goal is to infer sensitive data (e.g., input samples) hosted by the local participant, from the shared information -- the parameters of the local detection model and local prototypes. 
In other words, we wish to assess the feasibility of reconstructing original data points from the shared elements or learning any information about them.
The data reconstruction attack is formulated as a minimization of squared error (MSE) problem. 
For a given participant $i$, the attacker attempts to recover a representative raw feature profile \( \hat{x}_{i,j} \) of the attack class $j$. By passing through the embedding layer of the detection model, 
the goal is to get an embedding vector as close as possible to the shared local class prototype ${C}_{i,j}$ of the class $j$. 
More formally:
\begin{equation}\label{eq:data_reconstruct}
\small
\hat{x}_{*i,j} =  \underset{\hat{x}_{i,j}}{\arg\min} \,\, \| \phi_{i}(\hat{x}_{i,j}) - C_{i,j} \|^2
\end{equation}
where $\phi_i(\hat{x}_{i,j})$ denotes the embedding vector produced by passing the estimated attack feature profile of the class $j$ to the local model owned by participant i, respectively. %
After estimating $\hat{x}_{*i,j}$, we compute the MSE between it and the average feature vector of all data in the class $j$ hosted by the participant $i$. 
The larger (resp., smaller) the $L_2$ distance is, the less (resp., more) successful the attacker is in recovering the representative feature profiles in the class $j$. %

\descr{Evaluating Data Privacy in the Knowledge Sharing Process of PROTEAN.}
We perform data reconstruction attacks as defined by Equation~\ref{eq:data_reconstruct} for each participant. Figure \ref{fig:reconstruction0.75} reports the MSE value averaged over all the classes hosted by each participant, noted as \textit{Reconstructed MSE}. Besides, we introduce the average MSE value derived by a random guess of the averaged feature values of each class, noted as \textit{Random MSE}. A larger gap between them %
indicates a more successful data reconstruction attack, thus the more severe privacy leak risk.

\begin{figure*}[t]
    \centering
    \begin{subfigure}[t]{0.32\linewidth}
        \includegraphics[width=\linewidth]{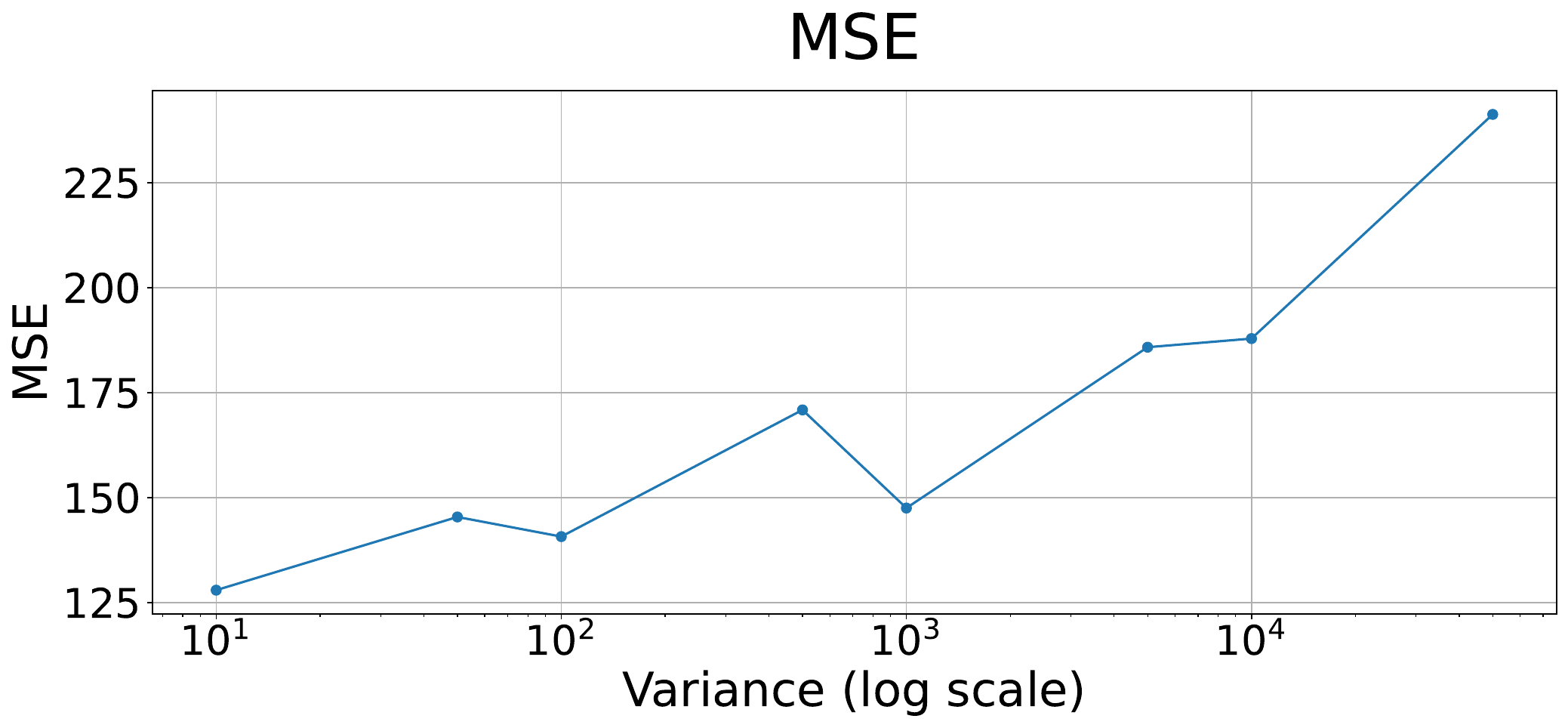}
    \end{subfigure}%
    \begin{subfigure}[t]{0.32\linewidth}
        \includegraphics[width=\linewidth]{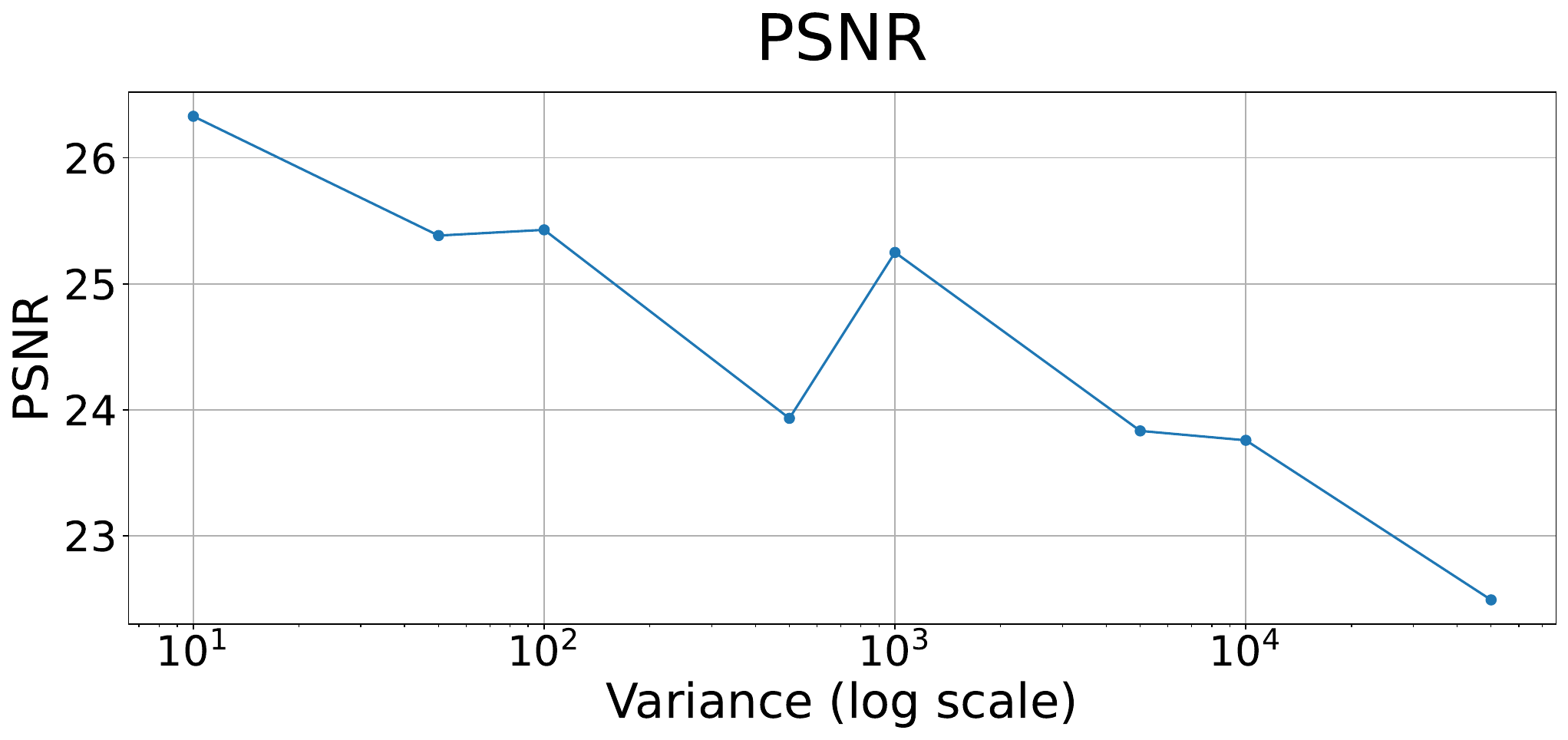}
        \caption{$\alpha=0.75$}
    \end{subfigure}%
    \begin{subfigure}[t]{0.32\linewidth}
        \includegraphics[width=\linewidth]{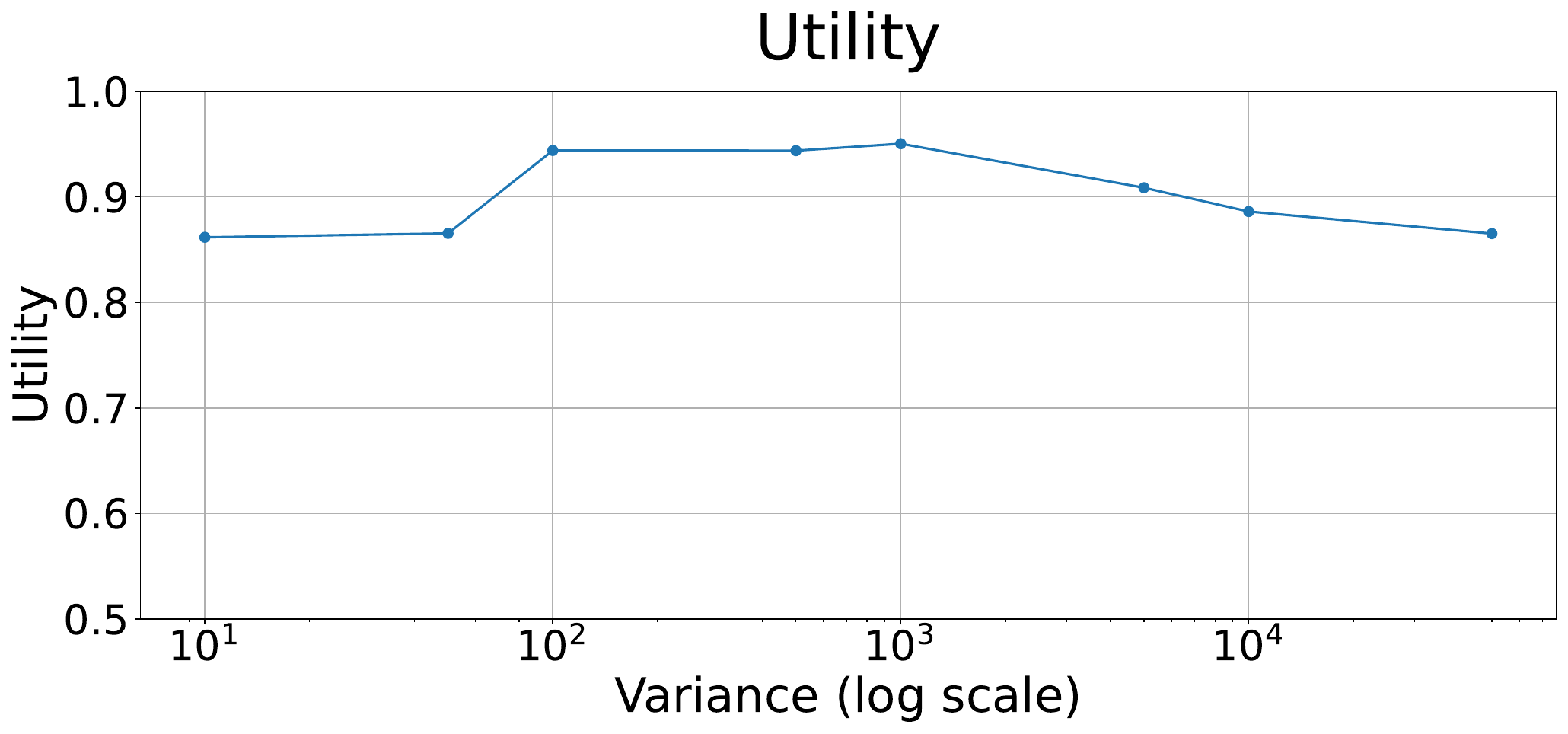}
    \end{subfigure}

    \begin{subfigure}[t]{0.32\linewidth}
        \includegraphics[width=\linewidth]{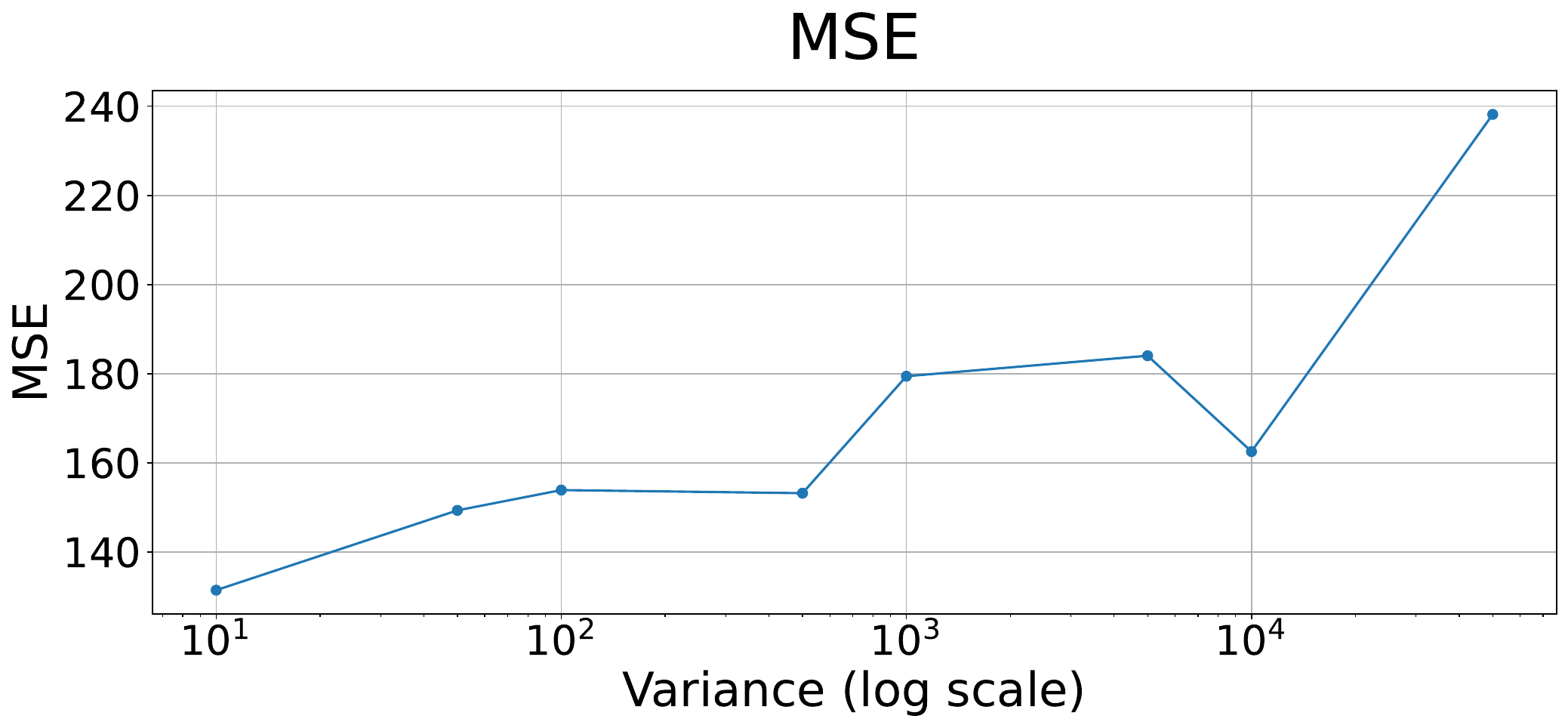}
    \end{subfigure}%
    \begin{subfigure}[t]{0.32\linewidth}
        \includegraphics[width=\linewidth]{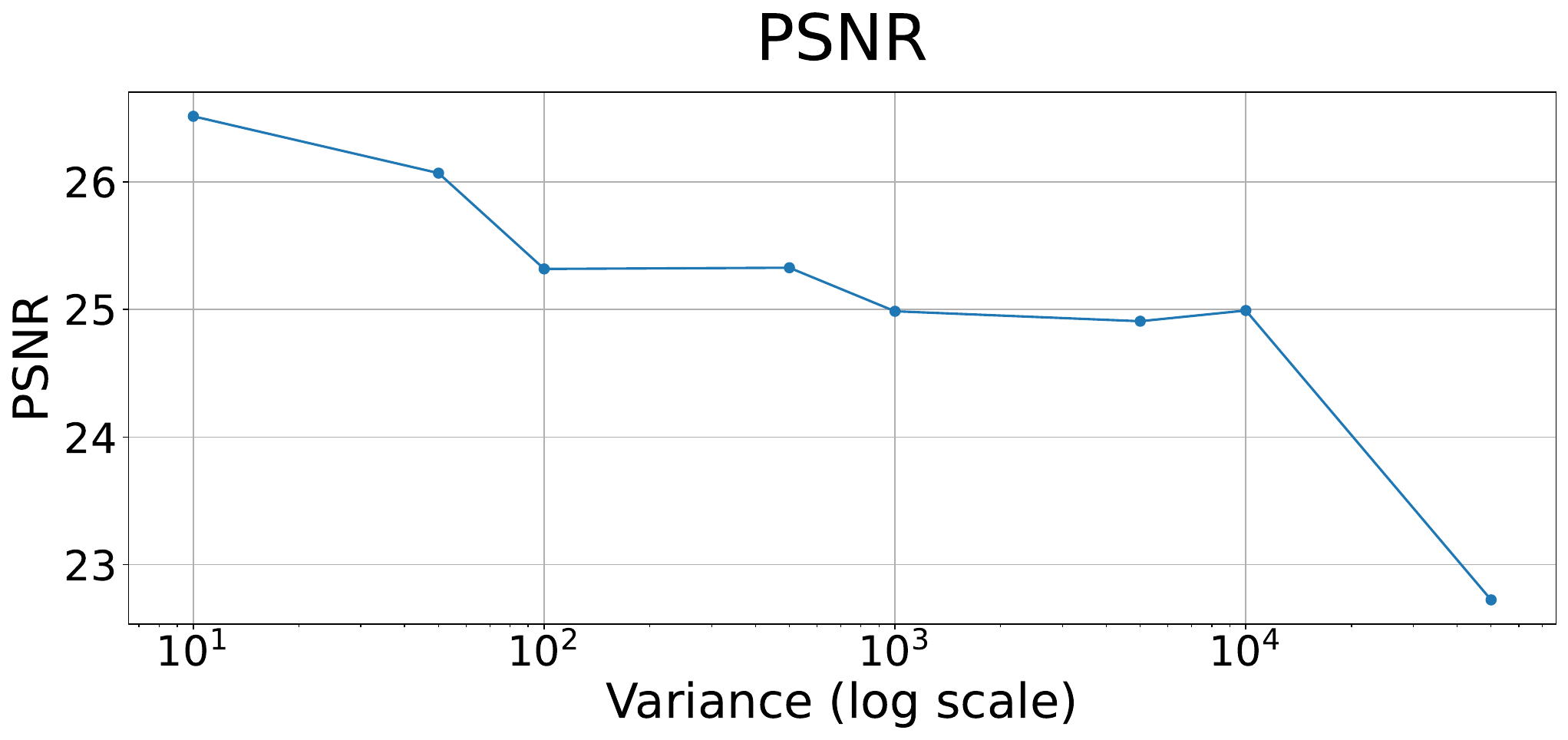}
        \caption{$\alpha=0.5$}
    \end{subfigure}%
    \begin{subfigure}[t]{0.32\linewidth}
        \includegraphics[width=\linewidth]{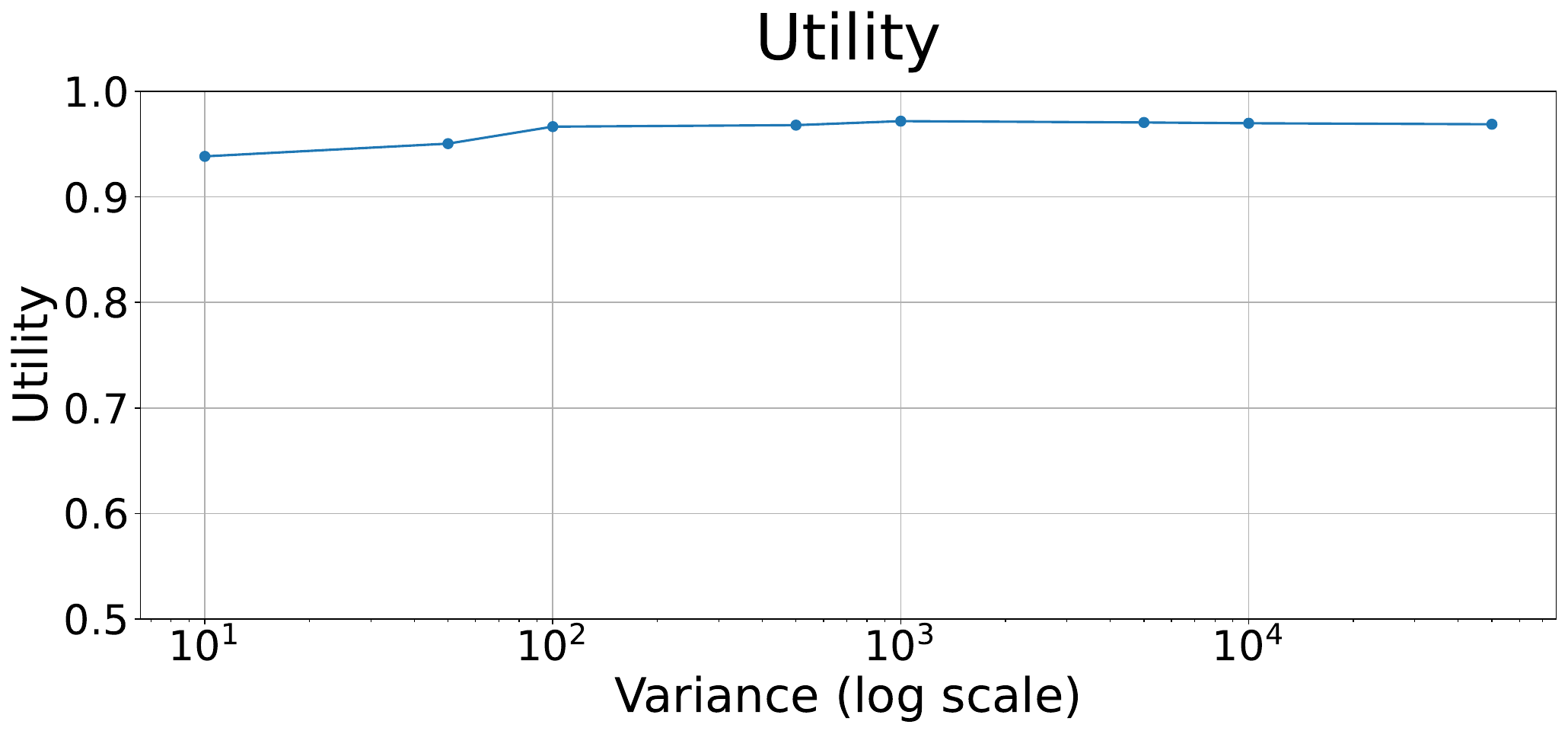}
    \end{subfigure}

    \begin{subfigure}[t]{0.32\linewidth}
        \includegraphics[width=\linewidth]{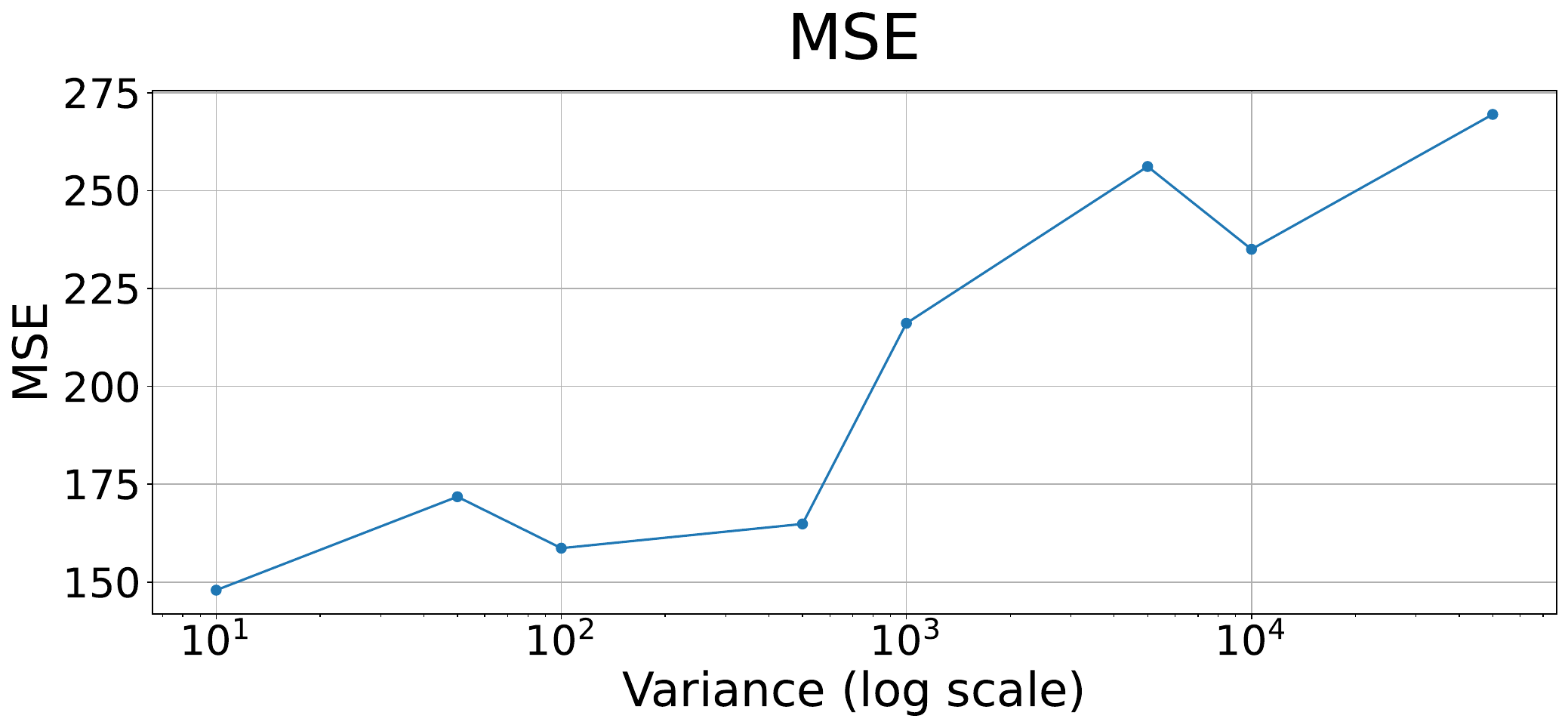}
    \end{subfigure}%
    \begin{subfigure}[t]{0.32\linewidth}
        \includegraphics[width=\linewidth]{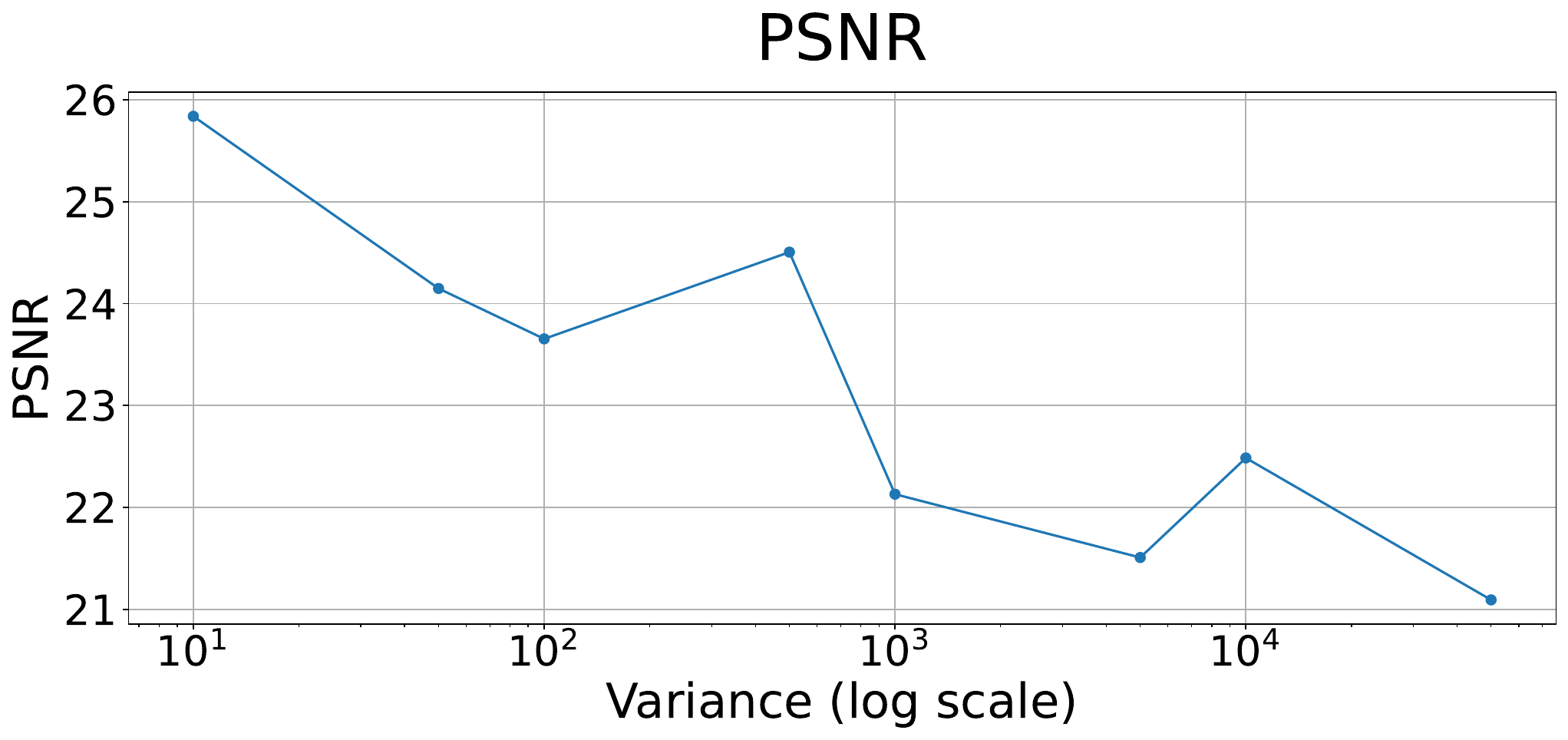}
        \caption{$\alpha=0.25$}
    \end{subfigure}%
    \begin{subfigure}[t]{0.32\linewidth}
        \includegraphics[width=\linewidth]{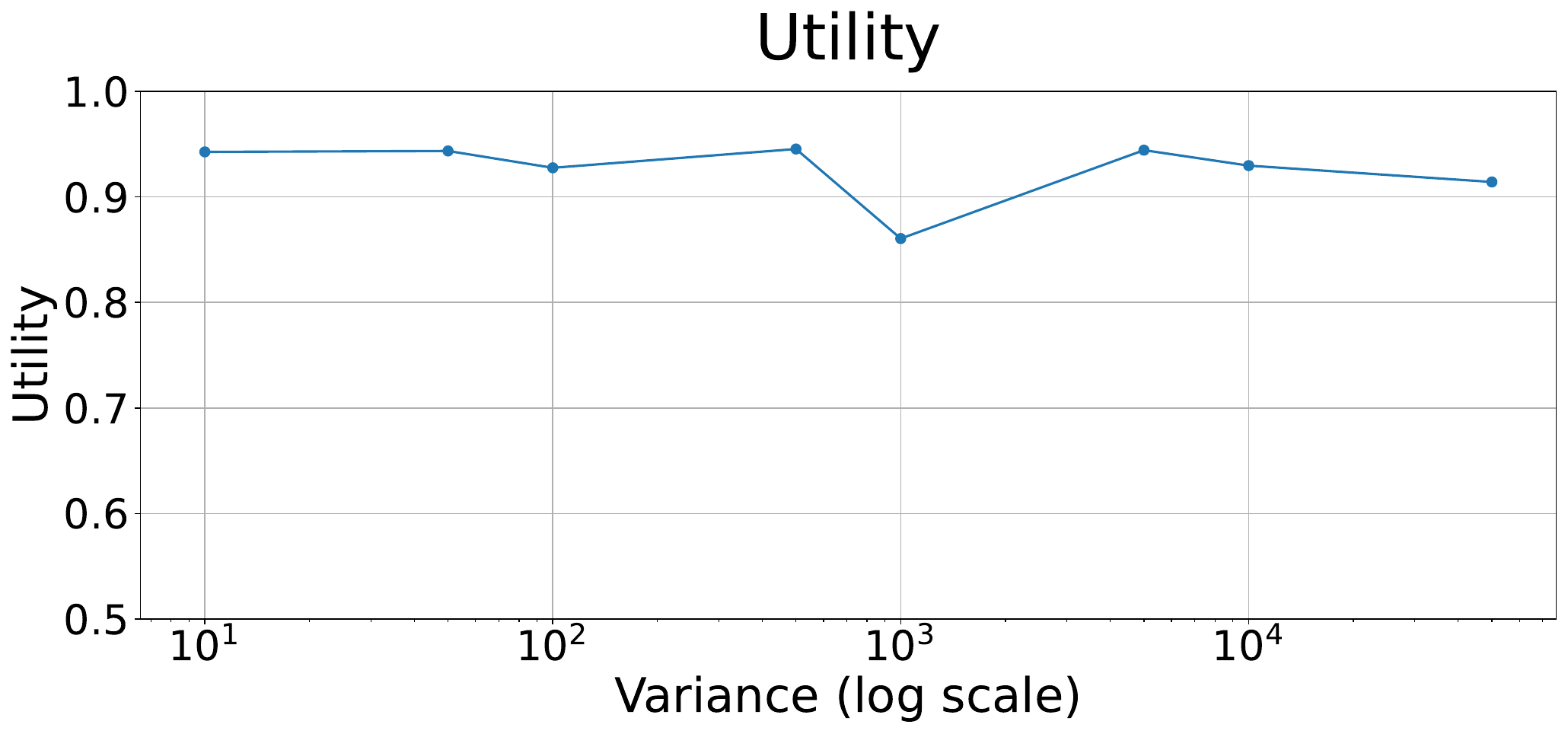}
    \end{subfigure}
    \caption{Reconstructed MSE, PSNR and Utility for $\alpha=0.75$, $0.5$ and $0.25$ on the X-IIoTID dataset.}
    \label{fig:xiiotid_combined}
\end{figure*}
 
This attack can only reconstruct class representatives for each participant but not the data itself. The leakage is restricted to a high-level summary rather than a complete reconstruction of the data. To further enhance privacy, we applied DP by adding Gaussian noise to the local prototypes, making it more difficult to reconstruct representatives: $\small C_{i,j} = C_{i,j} + \mathcal{N}(0, \sigma^2)$.

We evaluate the impact of using DP to make the reconstruction attack more challenging. We analyze the impact on utility (F1-score) and privacy loss (reconstructed MSE and Peak Signal-to-Noise Ratio (PSNR) values) by varying the noise level.
The PSNR is a widely used metric to quantify reconstruction quality \cite{xiao2023privacy}. 
In our evaluation, PSNR is computed per feature based on the per-feature MSE and the feature range of the original training samples. It is defined as:
$\text{PSNR} = 10 \log_{10} \left( \frac{(\text{Range})^2}{\text{MSE}} \right)$. The final PSNR value is obtained by averaging the per-feature PSNRs.
Higher PSNR values indicate better reconstruction.
Results are shown in Figures~\ref{fig:xiiotid_combined} and~\ref{fig:5gnidd_combined} for X-IIoTID and 5G-NIDD datasets, respectively. 
Higher noise variance improves privacy (higher MSE). 
With the addition of DP noise, we observe larger MSE values and lower PSNR values, indicating that the noise perturbation effectively reduces the success of reconstructing the averaged feature profile for each class. 
The F1-score remains largely unaffected by the noise, suggesting that the utility is preserved. %
The average of the random guess baseline MSE are 1112.75, 1380.42, and PSNR are 11.82, 14.13 for X-IIoTID and for 5G-NIDD. 
For the reconstructed profiles without DP, the average MSE are 74.10, 66.89 and PSNR are 31.78, 33.19 for X-IIoTID and for 5G-NIDD. %
However, compared to the MSE and PSNR values of the random guess baseline, the MSE and PSNR values after adding the DP noise are still high. It implies that on one hand, adding DP alleviates, yet does not completely mitigate the privacy risk in the proposed system, appealing for more effective privacy protection strategies.  On the other hand, the reconstruction attack can estimate no more than the averaged feature profile of one given class of network attacks. Compared to the data reconstruction attack scenarios described in previous privacy attack research \cite{geiping2020inverting}, the reconstructed result can not be used to uncover accurate privacy-related information in the training data.

\section{Conclusion}
\label{section: conclusion}
In this work, we present PROTEAN, an FPL-IDS designed for environments with distributed sensors and participants. The core contribution of PROTEAN is its knowledge-sharing process, where class-specific prototypes are shared across participants to enhance the detection of rare attacks in non-IID distributions. We also address key challenges such as privacy risks. Our findings show that PROTEAN maintains data privacy, as sharing prototypes does not reveal sensitive training data features. Our future work will focus on integrating secure computing techniques, such as Multi-Party Computing (MPC), to further strengthen privacy. We also aim to study PROTEAN's resilience to poisoning attacks and how to detect malicious participants. %

\descr{Acknowledgements.}
This work was carried out in the context of Beyond5G, a project funded by the French government as part of the economic recovery plan, namely ``France Relance'' and the investments for the future program.
This work has been partially supported by the French National Research Agency under the France 2030 label, SuperviZ (ANR-22-PECY-0008) and HiSec (ANR-22-PEFT-0009) projects, as well as the GRIFIN project (ANR-20-CE39-0011) and a grant from CISCO research gift on ``Privacy-Friendly Collaborative Threat Mitigation.''
The views reflected herein do not necessarily reflect the opinion of the funders.

{\small
\bibliographystyle{abbrv}
%\bibliography{mybib}

}

\appendix

\section{Convergence Analysis of PROTEAN}\label{app:proof}
We first provide the Lipschitz continuity and smoothness assumptions posed to the model trained in PROTEAN. 
Note that Lipschitz continuity and smoothness hold for most deep neural network models, as the learning objective function and non-linear transform encoded by deep neural networks inherently limit the rate of change and ensure smooth behavior.

\begin{assumption}
\textbf{Lipchitz Smoothness.} For each participant, the learning objective $\mathcal{L}$ of \textit{PROTEAN}, as given in Eq.\ref{eq:optimization_objective} is $L_{1}$-Lipschitz smooth, which means, 
\begin{equation}
\begin{split}
\small
&\|\nabla_{\omega}{\mathcal{L}} - \nabla_{\omega'}{\mathcal{L}}\|_{2} \leq L_{1}\|\omega - \omega'\|_{2} s.t. \,\,\,\,\forall{\omega,\omega'}\in{\Omega}
\end{split}
\end{equation}
where $\omega$ and $\omega'$ denote two parameter values of the detection model in the parameter value space $\Omega$. $\|\|_{2}$ denotes the Euclidean distance. With $L_1$-Lipschitz smoothness, we can derive the quadratic bound of $\mathcal{L}(\omega')$ as:
\begin{equation}
\mathcal{L}(\omega') \leq \mathcal{L}(\omega) + \langle\nabla_{\omega}\mathcal{L},(\omega'-\omega)\rangle + \frac{L_1}{2}\|\omega' - \omega\|^2_2
\end{equation}
\end{assumption}

\begin{assumption}
\textbf{Lipschitz Continuity of the embedding function.} The embedding vectors produced by the embedding function $\phi$ has the $L_{\phi}$-Lipschitz continuity. For any input $x$, we give
\begin{equation}
\small
\|\phi'(x) - \phi(x)\|_{2} \leq L_{\phi} \|\omega_{\phi} - \omega'_{\phi}\|_{2}
\end{equation}
where $\omega_{\phi}$ and $\omega'_{\phi}$ are the parameters of two embedding functions $\phi$ and $\phi'$.  
\end{assumption}

\begin{assumption}
\textbf{Lipschitz Continuity of the objective function with respect to the embedding.} For each participant, the learning objective $\mathcal{L}$ of PROTEAN is $L_{2}$-Lipschitz continuous with respect to the embeddings. It gives for any training sample $(x,y)$:
\begin{equation}
\small
|\mathcal{L}(\phi') - \mathcal{L}(\phi)| \leq L_{2}\|\omega'_{\phi} - \omega_{\phi}\|_{2}
\end{equation}
where $\omega_{\phi}$ and $\omega'_{\phi}$ are the parameters of the two different embedding functions $\phi'$ and $\phi$.   
\end{assumption}

\begin{theorem}
\textbf{The upper bound of the learning objective descent at one federated training round.} For each participant $i$, we use $\mathcal{L}^{i}_{t+1}$ and $\mathcal{L}^{i}_{t}$ to denote the learning objective values of a participant of PROTEAN, which are derived at the successive federated training rounds $t+1$ and $t$. We assume the local learning rate as $\eta$ and locally the model is trained with $T$ steps of stochastic gradient descent. Given that Assumptions 1, 2, and 3 hold, we can bound the expectation of the learning objective at the communication round $t+1$, $\mathcal{L}^{i}_{t+1}$, as given in Equation~\ref{eq:upperbound_argos}.
\begin{equation}\label{eq:upperbound_argos}
\small
\begin{split}
\mathbb{E}[\mathcal{L}^{i}_{t+1}] \leq & \mathcal{L}^{i}_{t} - (\eta - \frac{L_{1}\eta^2}{2}) \cdot \sum_{k=1}^{T}\|\nabla_{\omega^{t}_{i,k}}\mathcal{L}^{i}\|^2_2 +\\
& +  \frac{{L_1}T\eta^2}{2}\sigma^2 + (L_2 + 2\lambda{L_{\phi}})\eta{L_2}T\\
\end{split}
\end{equation}
where $\sigma$ is the upper bound of the variance of the stochastic gradient in the local training of the participant. Between the federated training rounds $t$ and $t+1$, each participant conducts local training of the detection model. 
$\omega^{t}_{i,k}$ are the locally tuned detection model parameters of the participant $i$ derived at the local training step $k$. $\nabla_{\omega^{t}_{i,k}}\mathcal{L}^{i}$ is the gradient of the learning objective of the local training process of participant $i$ with respect to the model parameters $\omega^{t}_{i,k}$.  
\end{theorem}

\begin{theorem}
\textbf{Convergence of PROTEAN.} With $\eta$ in Equation~\ref{eq:learning_rate} set as:
\begin{equation}\label{eq:learning_rate}
\small
\begin{split}
&\eta \leq \frac{\kappa-\tau}{\psi+\kappa L_{1}/2},\\
&\kappa = \sum_{i=1}^{T}\|\nabla_{\omega_{i}\mathcal{L}_{k}}\|^2_2,\quad \tau = (L_2 + 2\lambda{L_{\phi}}){L_2}T , \quad  \psi = \frac{L_{1}T\sigma^2}{2}
\end{split}
\end{equation}
The convergence of the training process of PROTEAN holds.
\end{theorem}

\section{Evaluation on the 5G-NIDD dataset}
We report the classification performances in Figures~\ref{fig: metrics_5Gnidd} and Figures~\ref{fig: ra_comparaisions_5gnidd} on the 5G-NIDD datasets. After training the IDS model using PROTEAN, we compare the accuracy score for each attack class between models trained with PROTEAN and those trained only locally (i.e., \textit{without FL}) in Figures~\ref{fig: before after fl 5gnidd}. In Figure~\ref{fig:5gnidd_combined}, we report the privacy attack results on the 5G-NIDD dataset.

\begin{figure*}[t]
    \centering
\includegraphics[width=0.7\linewidth]{Figures/num_rounds10_big/legend_fig2.pdf}
        \begin{subfigure}[t]{1\linewidth}
            \includegraphics[width=\linewidth]{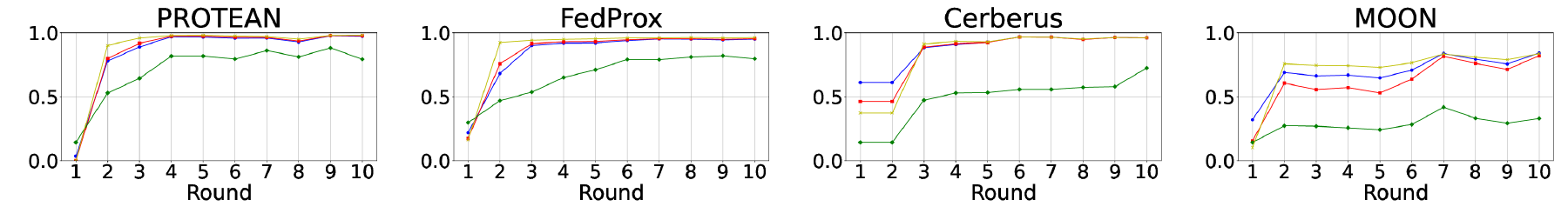}
            \caption{$\alpha=0.75$}
        \end{subfigure}
        \begin{subfigure}[t]{1\linewidth}
            \includegraphics[width=\linewidth]{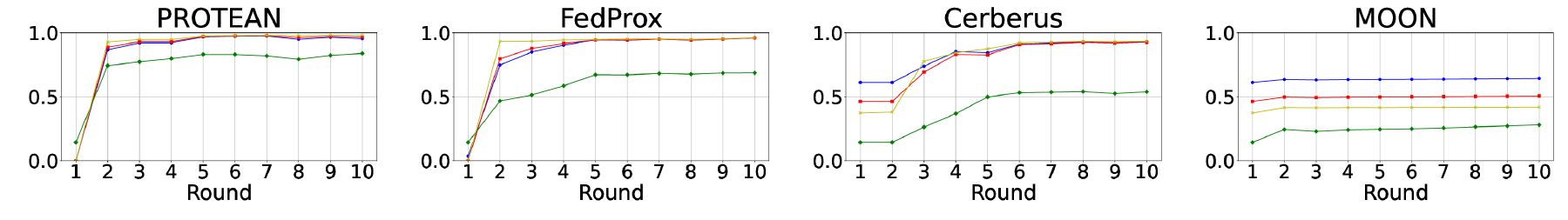}
            \caption{$\alpha=0.5$}
        \end{subfigure}
        \begin{subfigure}[t]{1\linewidth}
            \includegraphics[width=\linewidth]{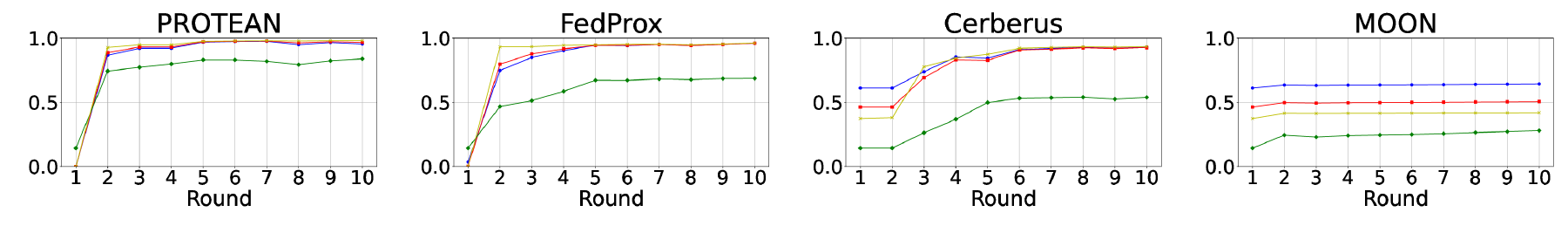}
            \caption{$\alpha=0.25$}
        \end{subfigure}

    \caption{Per-round accuracy, F1 score, macro accuracy, and precision of PROTEAN, FedProx-IDS, Cerberus, and MOON-IDS with $\alpha=$ 0.75, 0.5, and 0.25 on the 5G-NIDD dataset.}
    \label{fig: metrics_5Gnidd}
\end{figure*}

\begin{figure*}[t]
    \centering
    \begin{subfigure}[t]{0.325\linewidth}  
    \includegraphics[width=\linewidth]{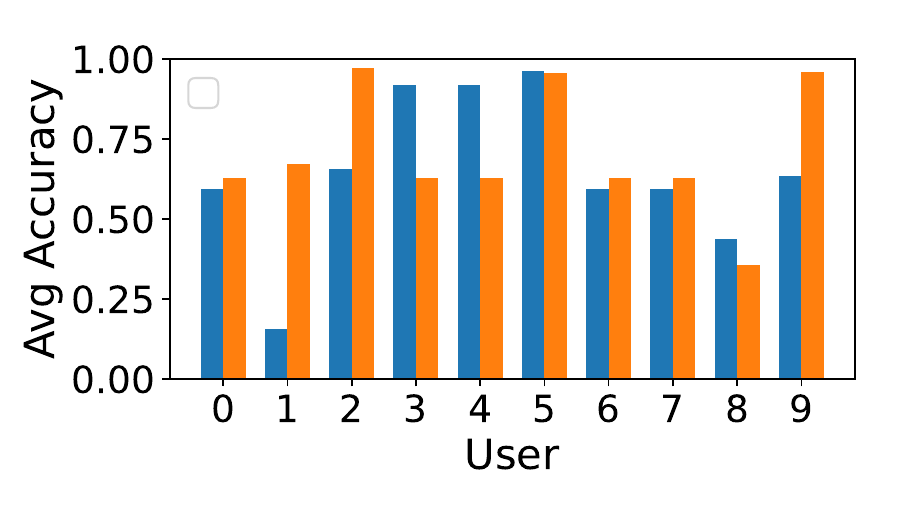}
	\caption{$\alpha{=}$0.75}
	\end{subfigure}
    \begin{subfigure}[t]{0.325\linewidth}  
    \includegraphics[width=\linewidth]{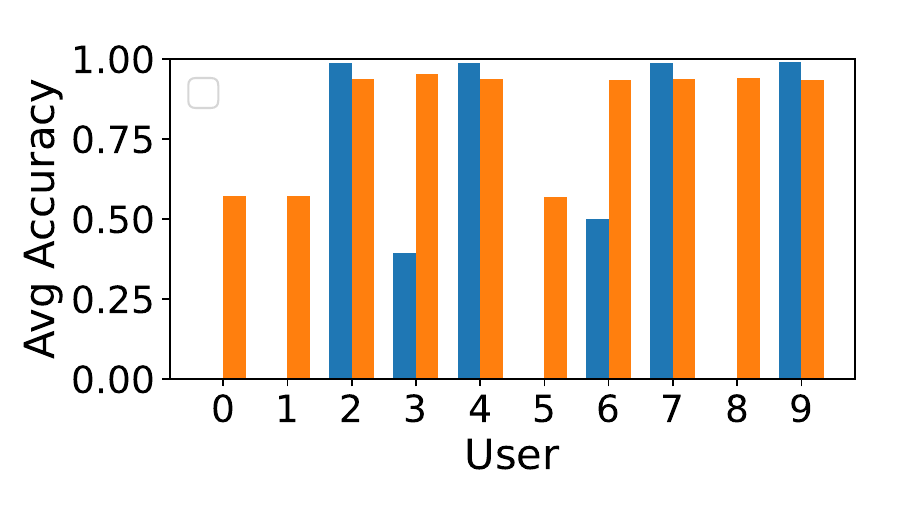}
	\caption{$\alpha{=}$0.5}
	\end{subfigure}
    \begin{subfigure}[t]{0.325\linewidth}  
    \includegraphics[width=\linewidth]{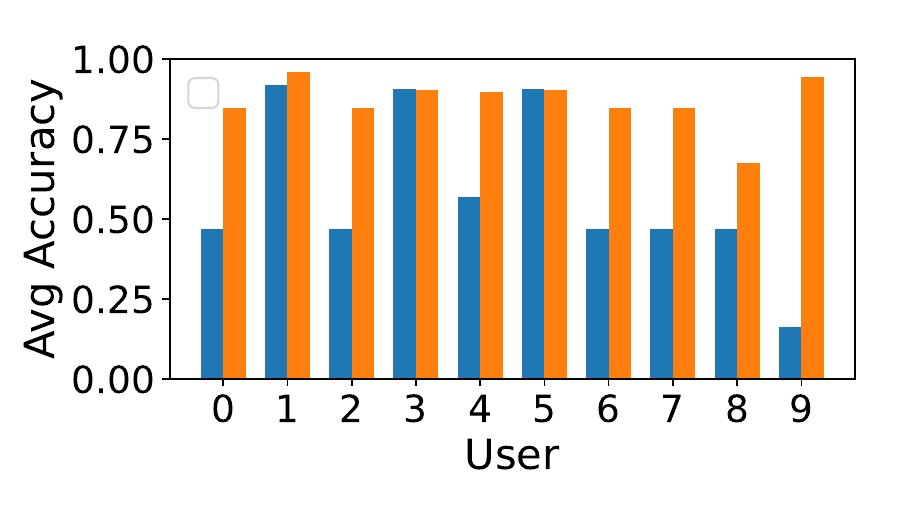}
	\caption{$\alpha{=}$0.25}
	\end{subfigure}
    \caption{Comparison of the averaged accuracy on two rare classes of each participant on the on 5G-NIDD dataset with different $\alpha$ values. Blue and orange bars represent Cerberus and PROTEAN, respectively.}
    \label{fig: ra_comparaisions_5gnidd}
\end{figure*}

\begin{figure*}[t]
    \centering

    \includegraphics[width=\mywidth\linewidth]{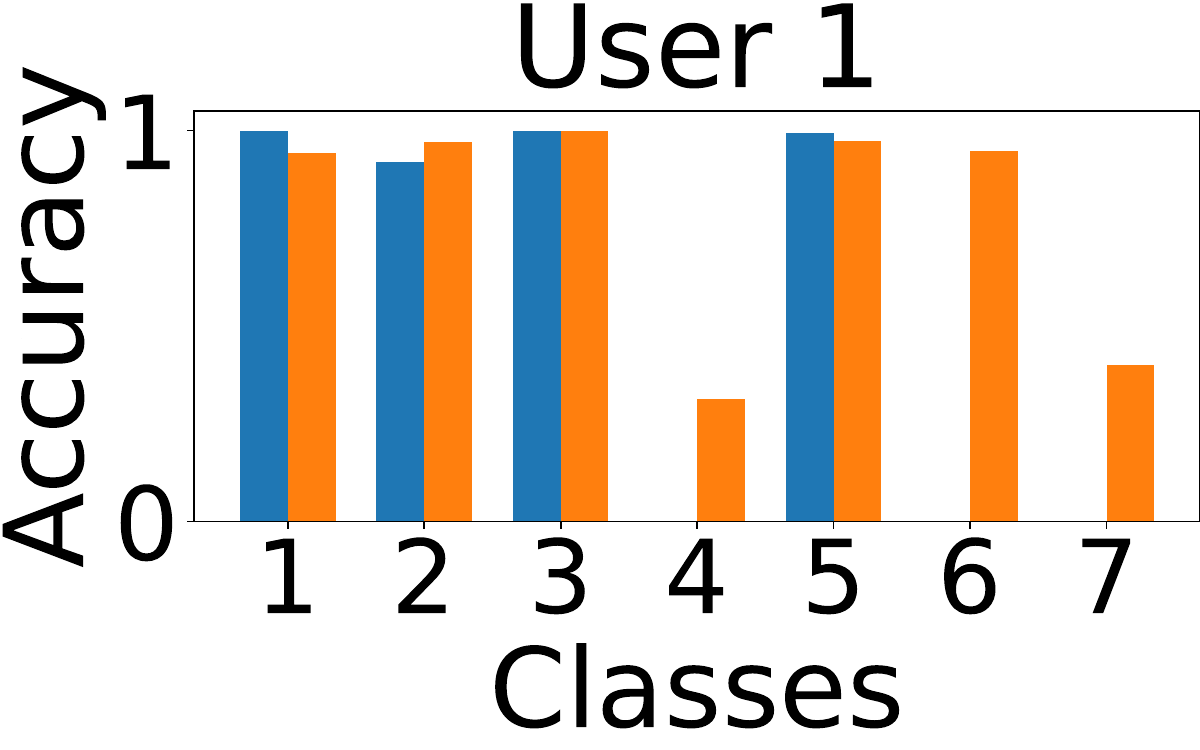}~%
    \includegraphics[width=\mywidth\linewidth]{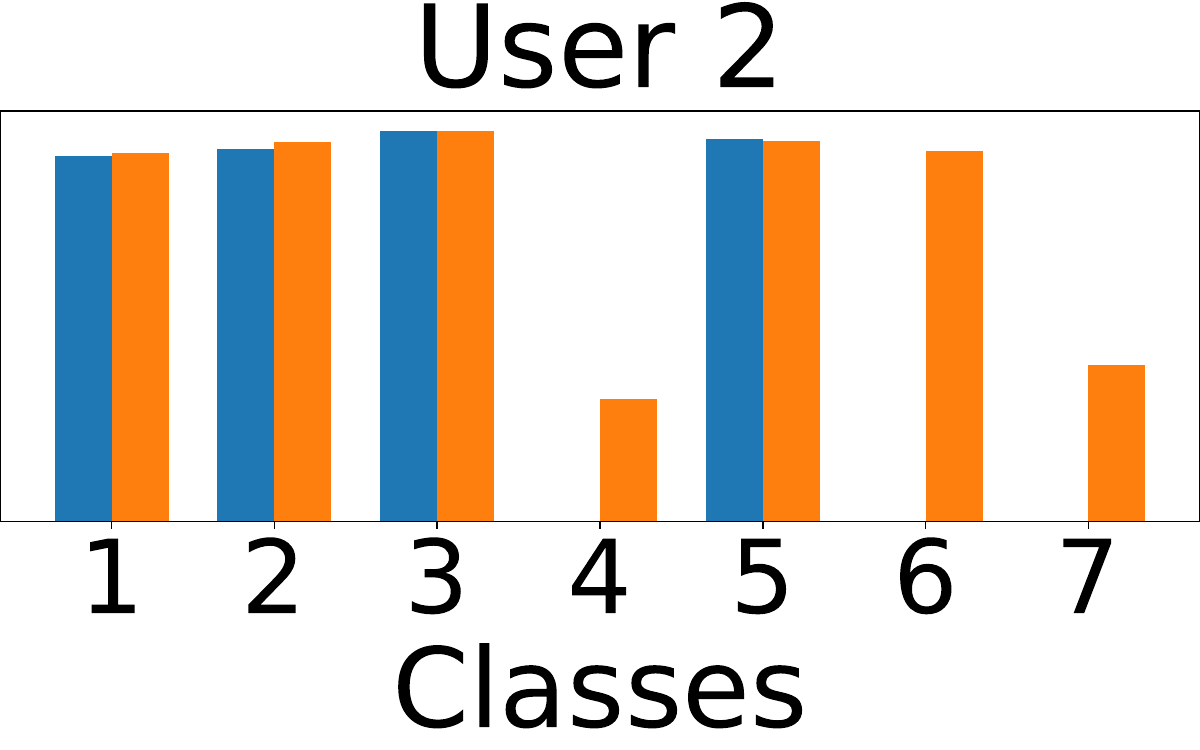}~%
    \includegraphics[width=\mywidth\linewidth]{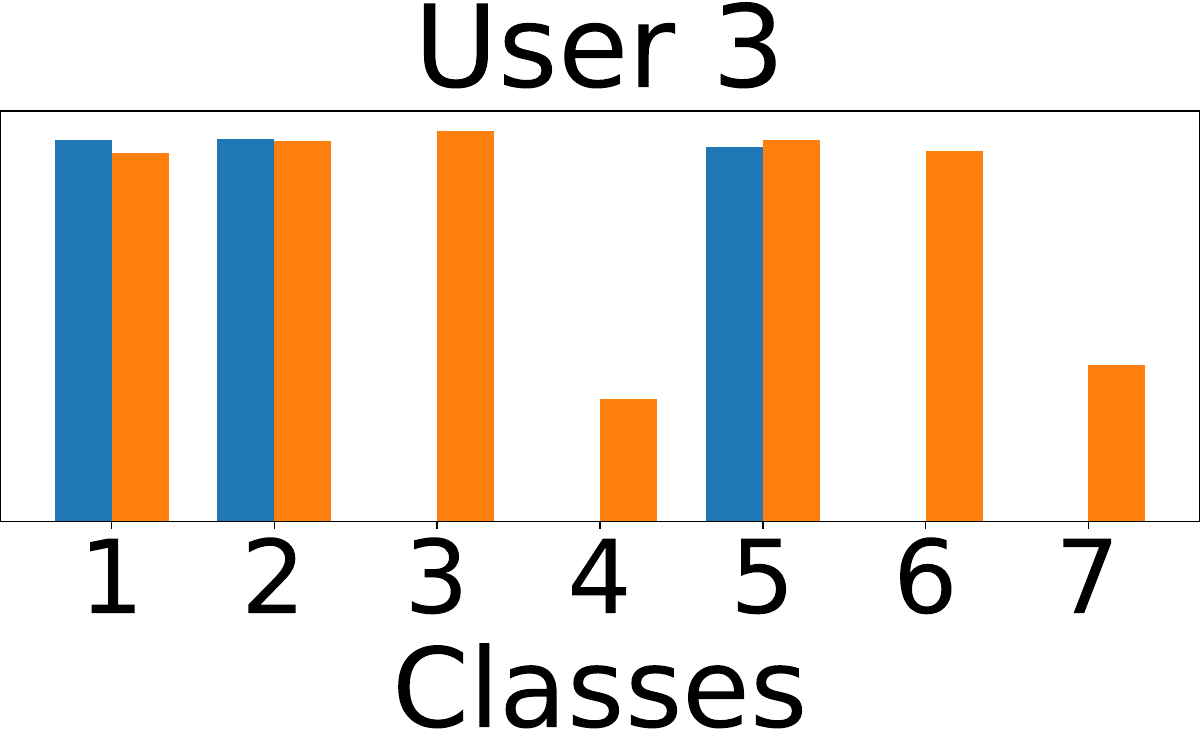}~%
    \includegraphics[width=\mywidth\linewidth]{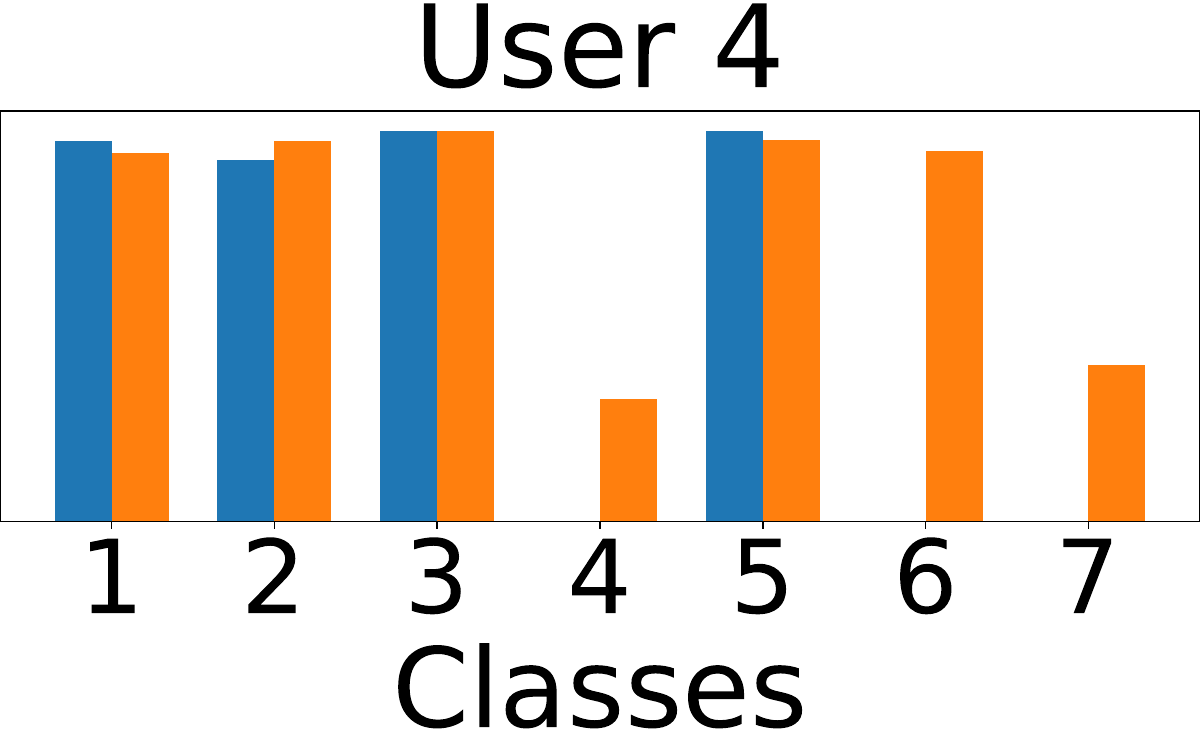}~%
    \includegraphics[width=\mywidth\linewidth]{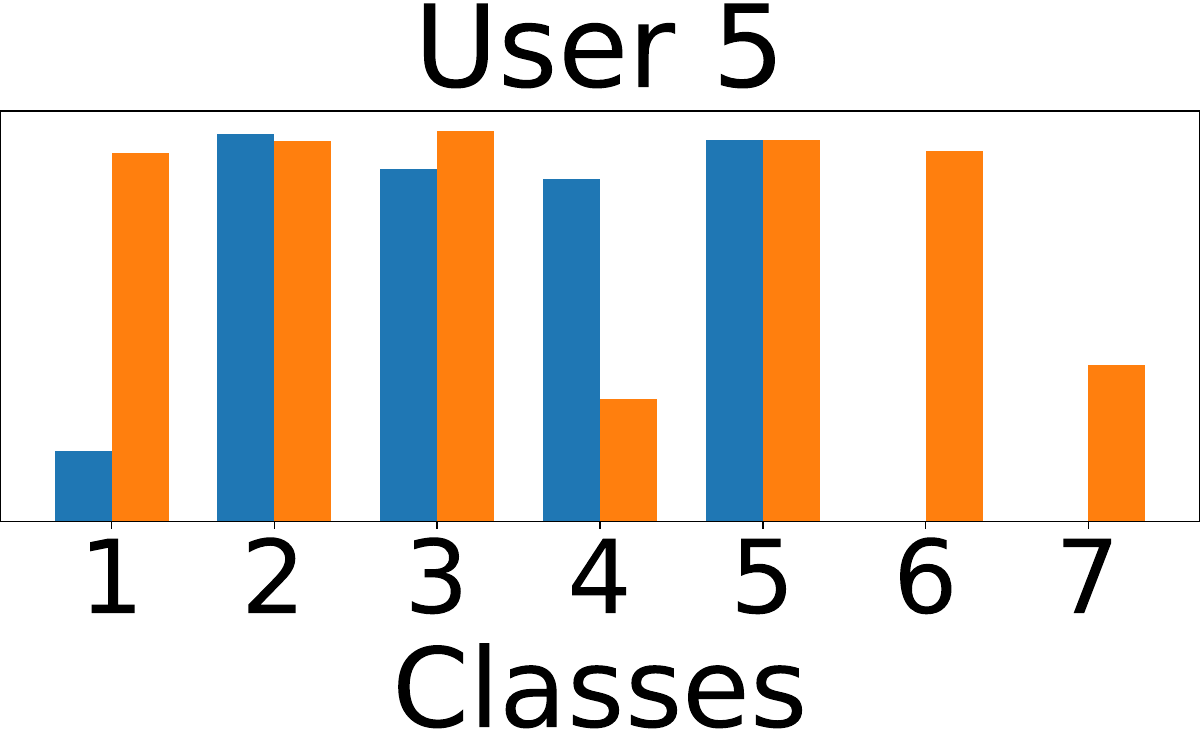}
    \includegraphics[width=\mywidth\linewidth]{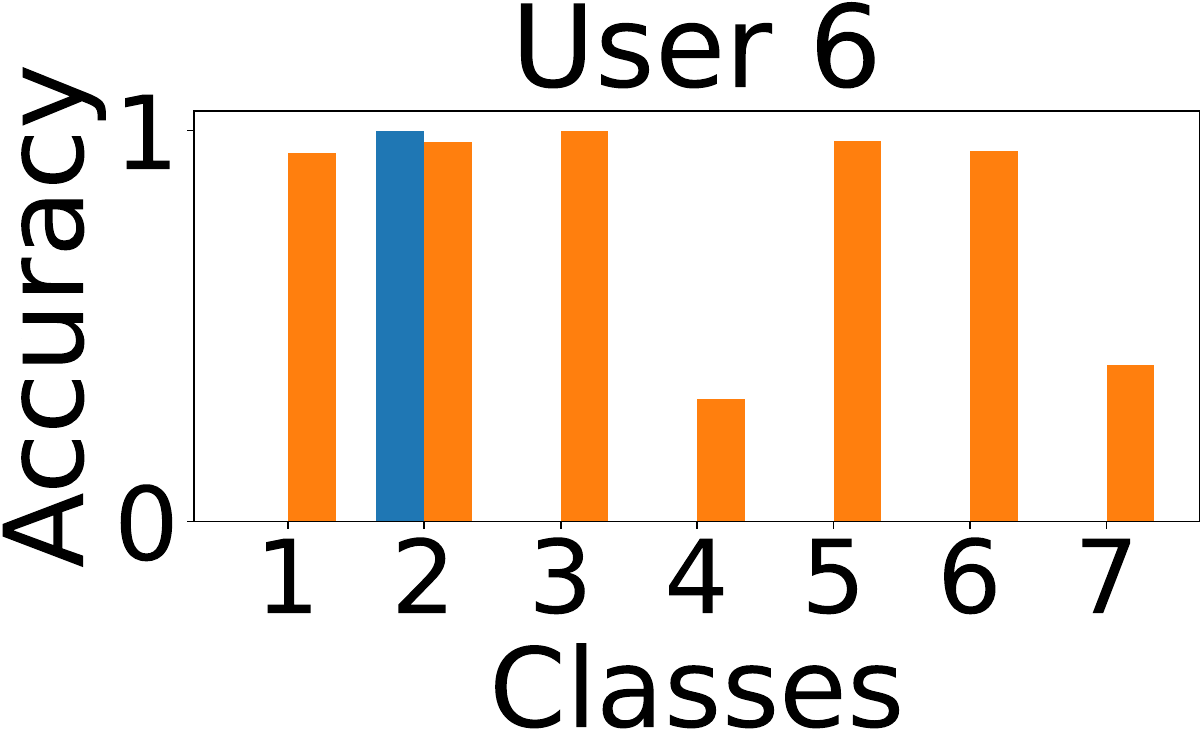}~%
    \includegraphics[width=\mywidth\linewidth]{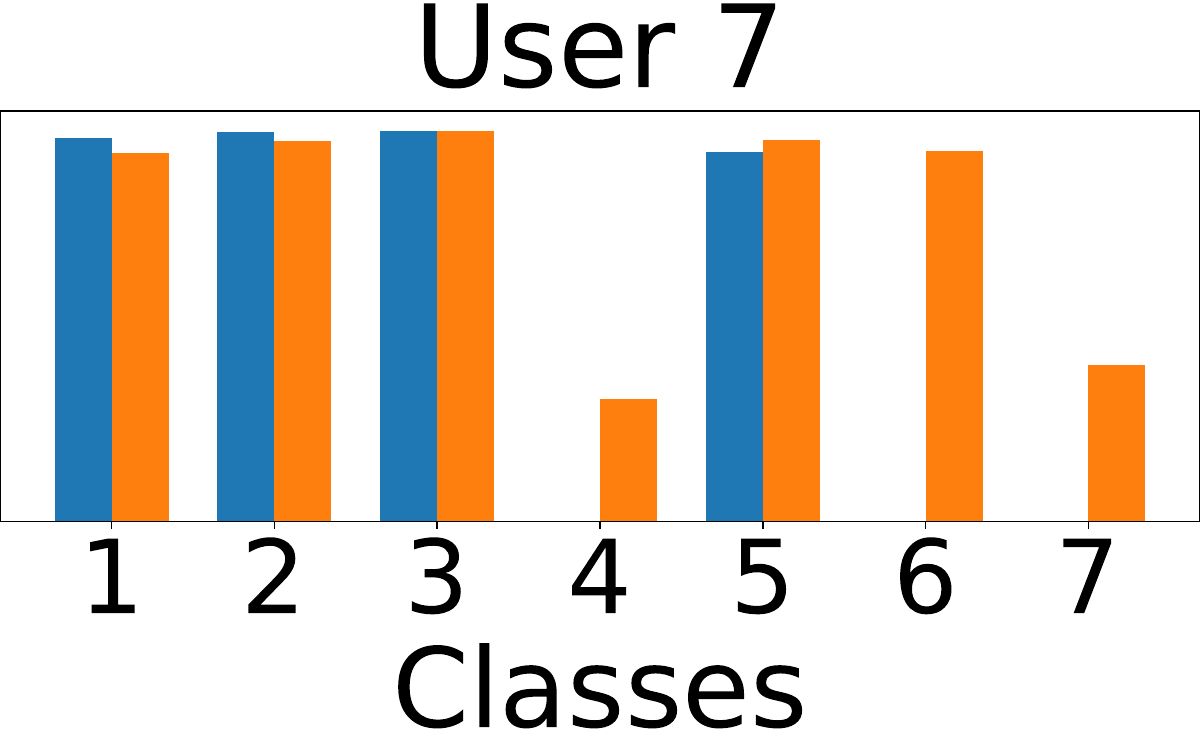}~%
    \begin{subfigure}[t]{\mywidth\linewidth}
        \includegraphics[width=\linewidth]{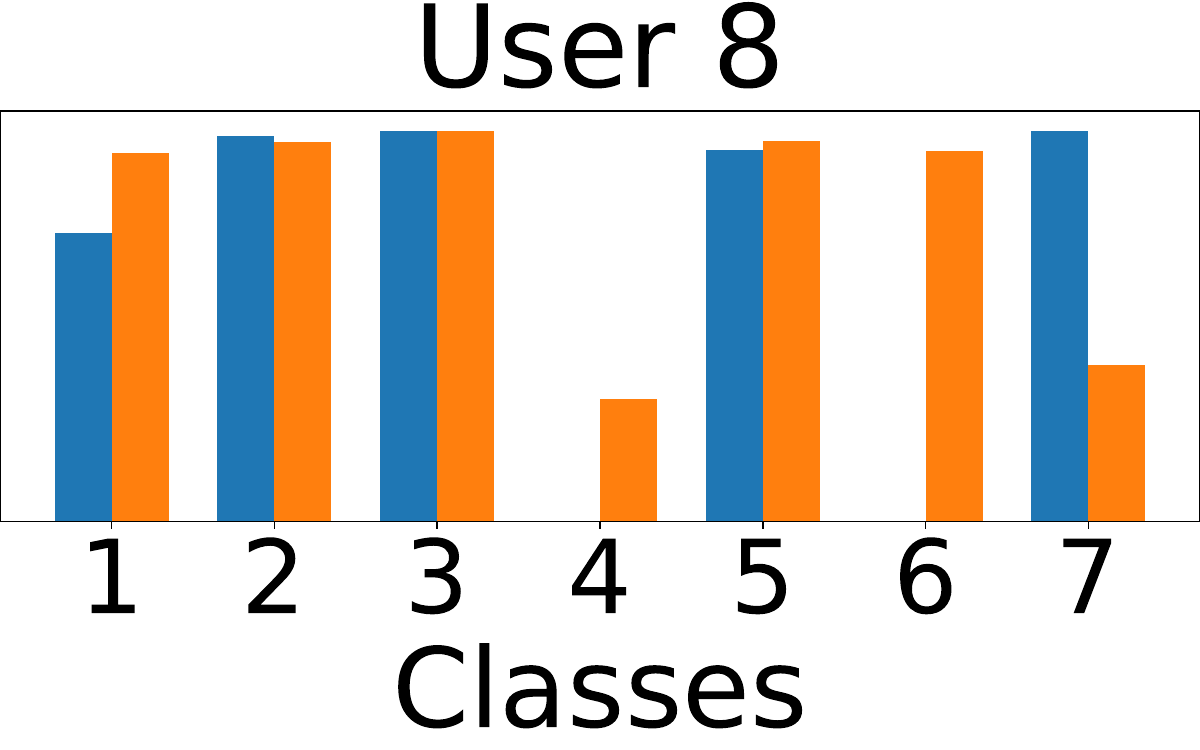}~%
        \caption{$\alpha{=}$0.75}
    \end{subfigure}%
    ~\includegraphics[width=\mywidth\linewidth]{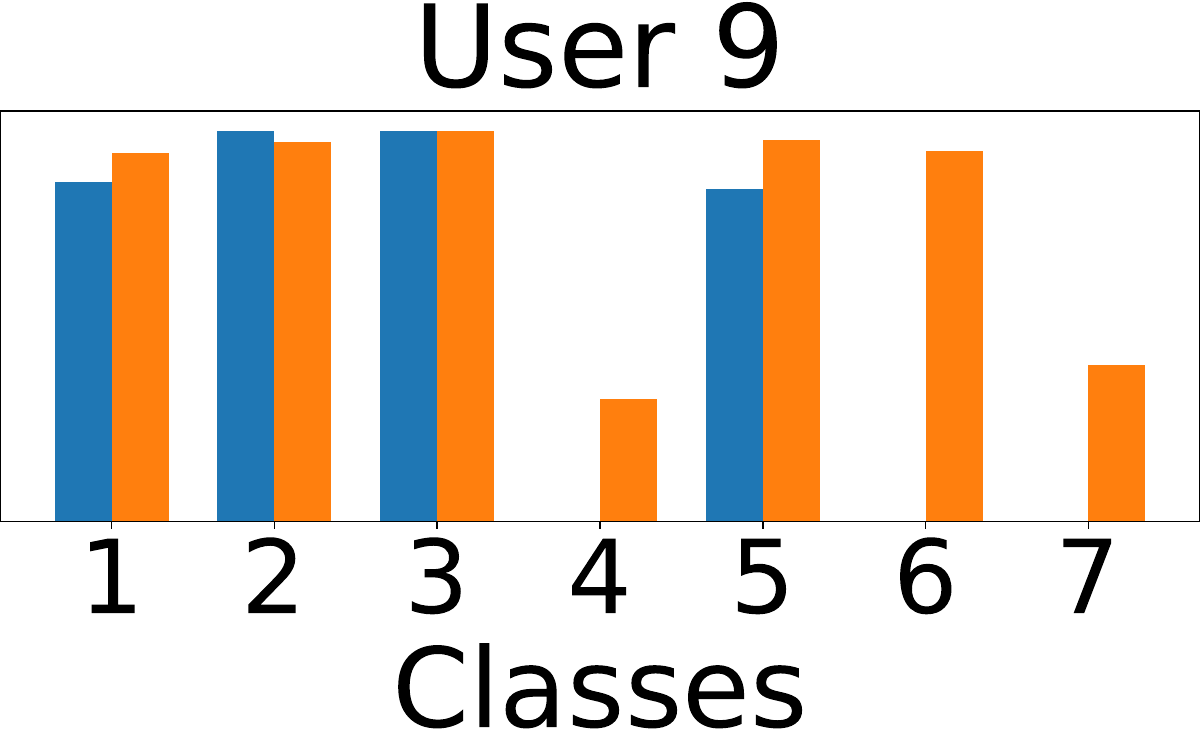}~%
    \includegraphics[width=\mywidth\linewidth]{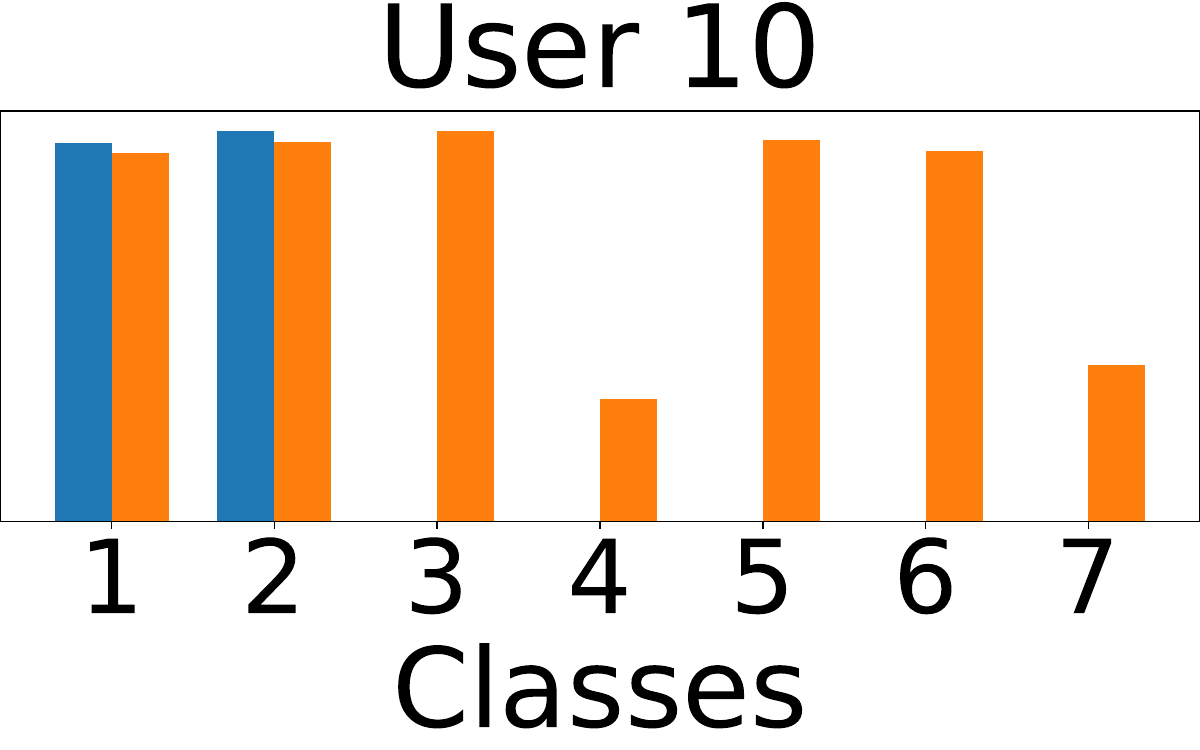}
    \includegraphics[width=\mywidth\linewidth]{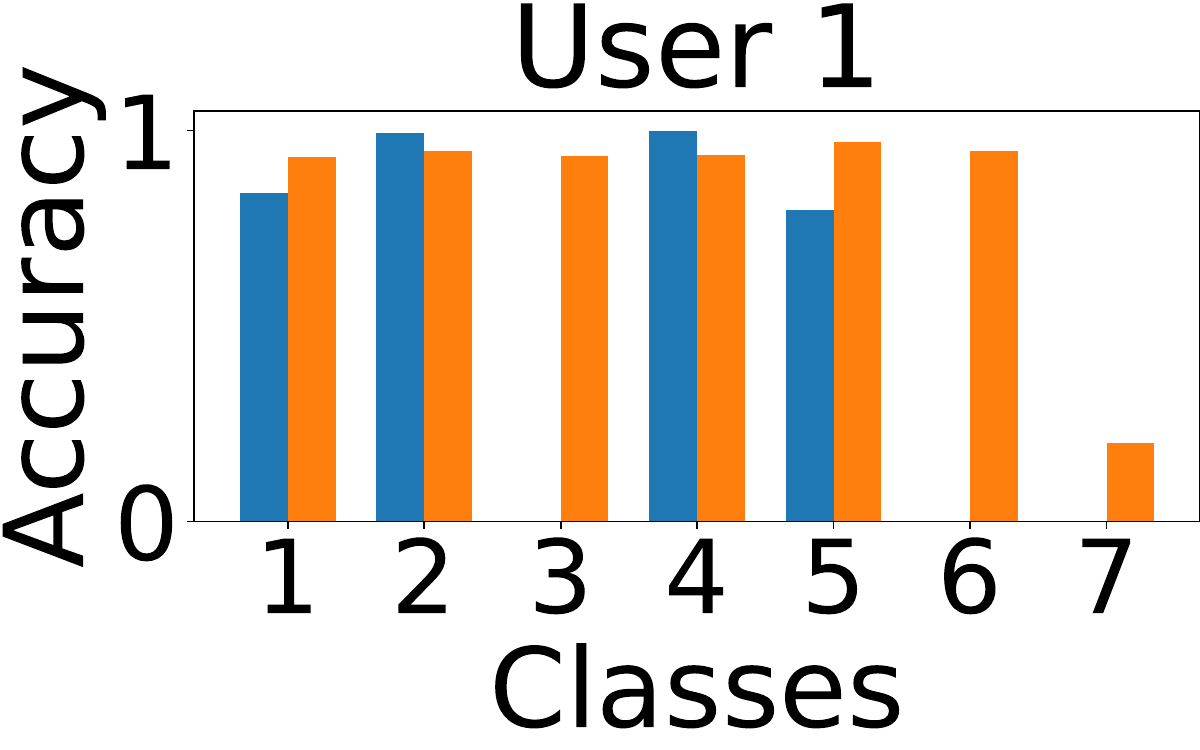}~%
    \includegraphics[width=\mywidth\linewidth]{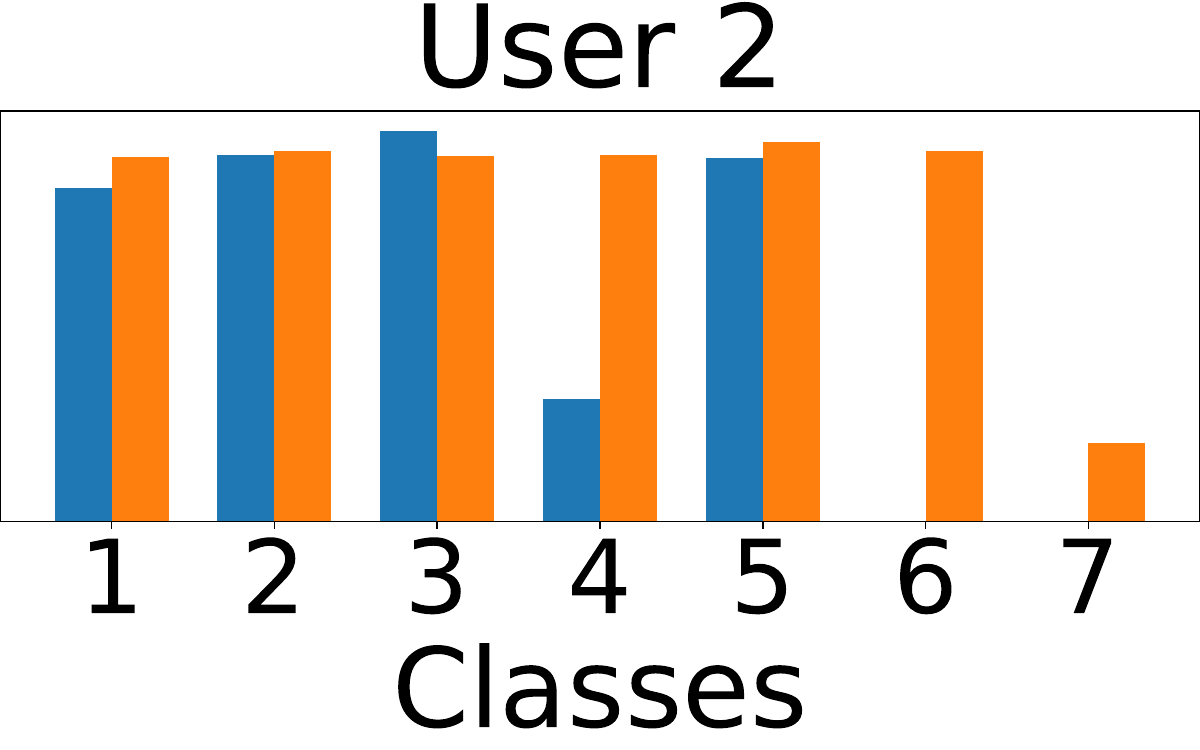}~%
    \includegraphics[width=\mywidth\linewidth]{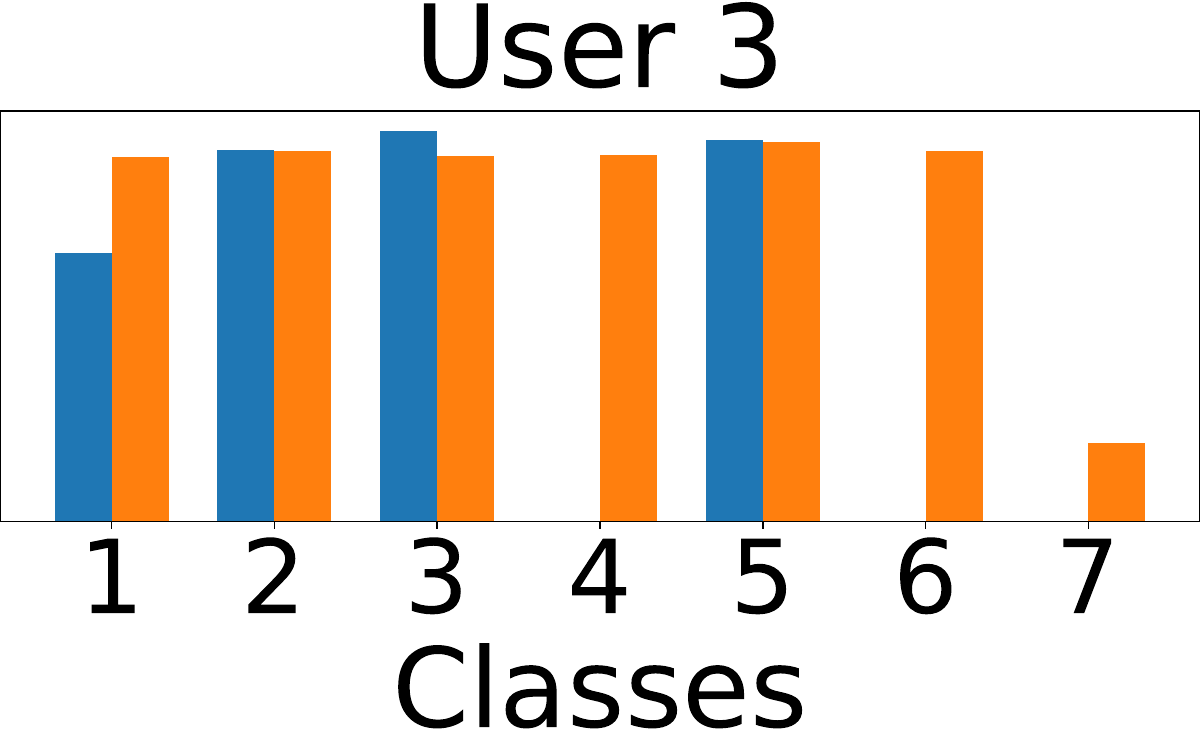}~%
    \includegraphics[width=\mywidth\linewidth]{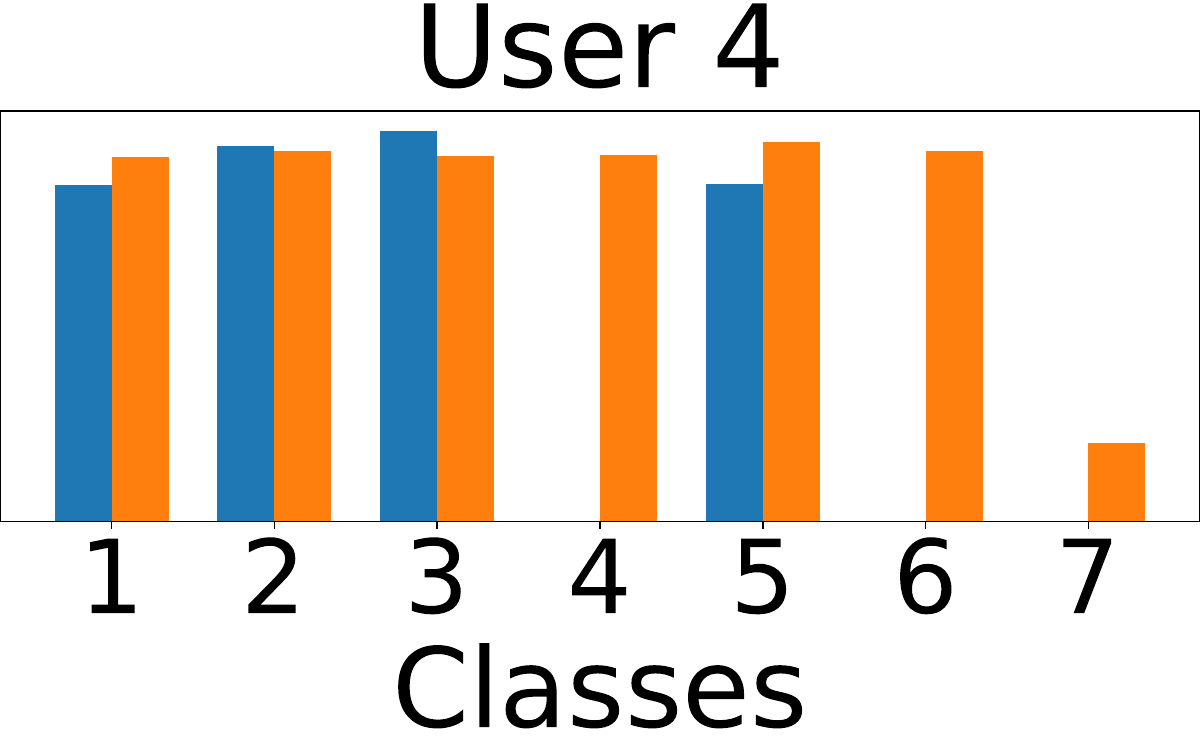}~%
    \includegraphics[width=\mywidth\linewidth]{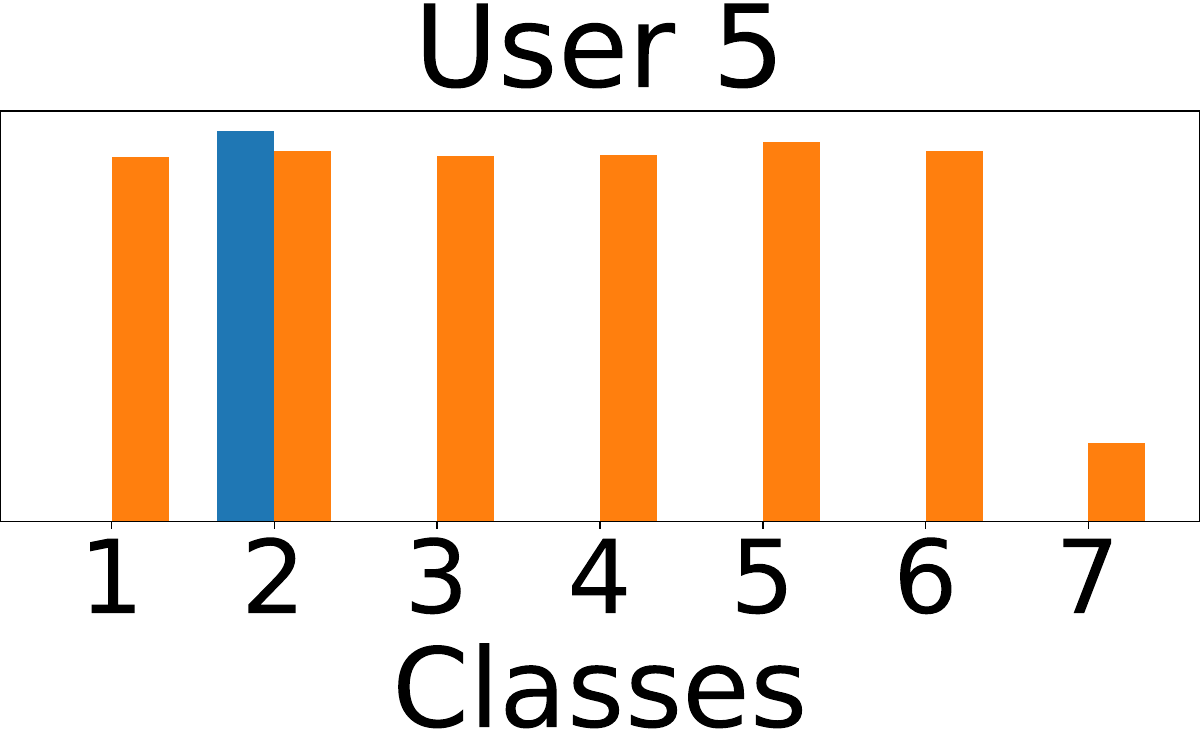}
    \includegraphics[width=\mywidth\linewidth]{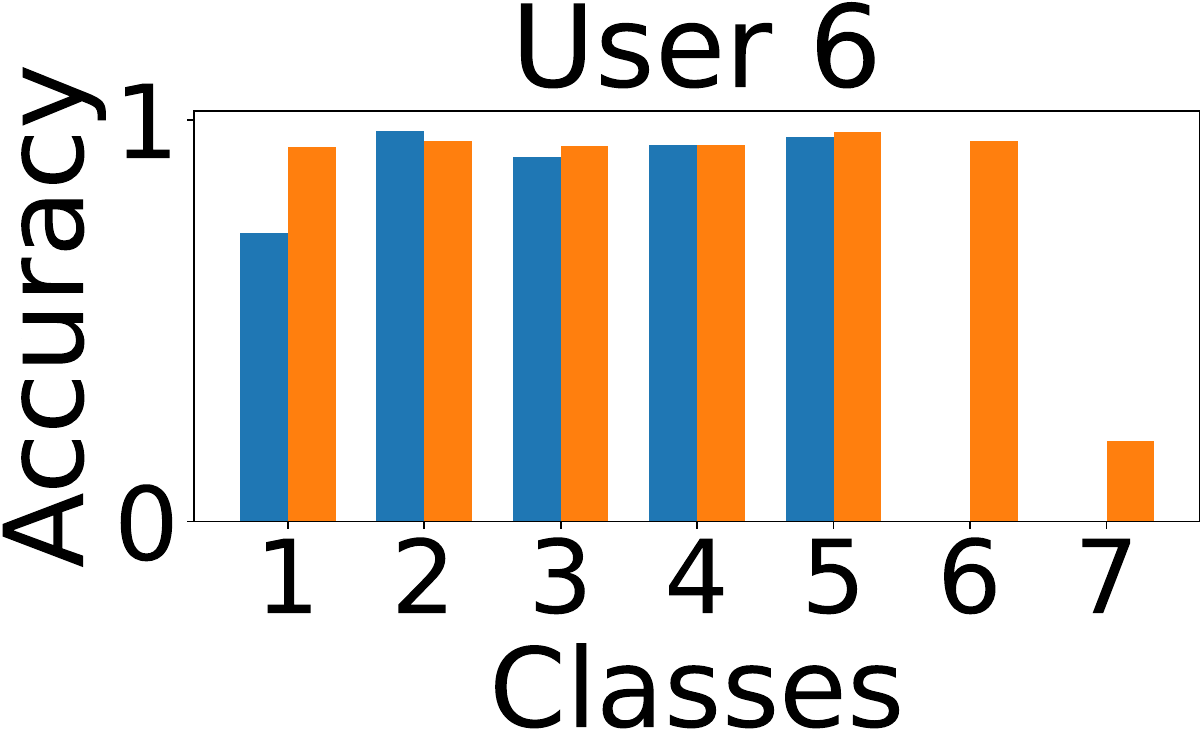}~%
    \includegraphics[width=\mywidth\linewidth]{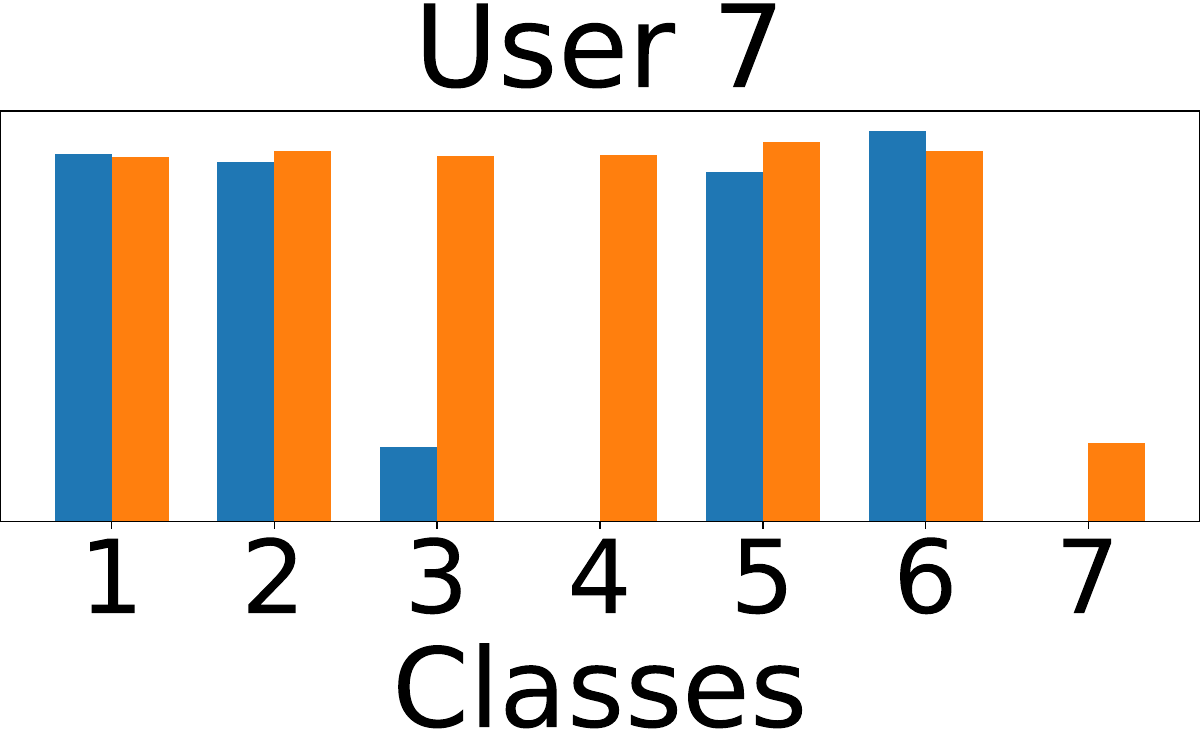}~%
    \begin{subfigure}[t]{\mywidth\linewidth}
        \includegraphics[width=\linewidth]{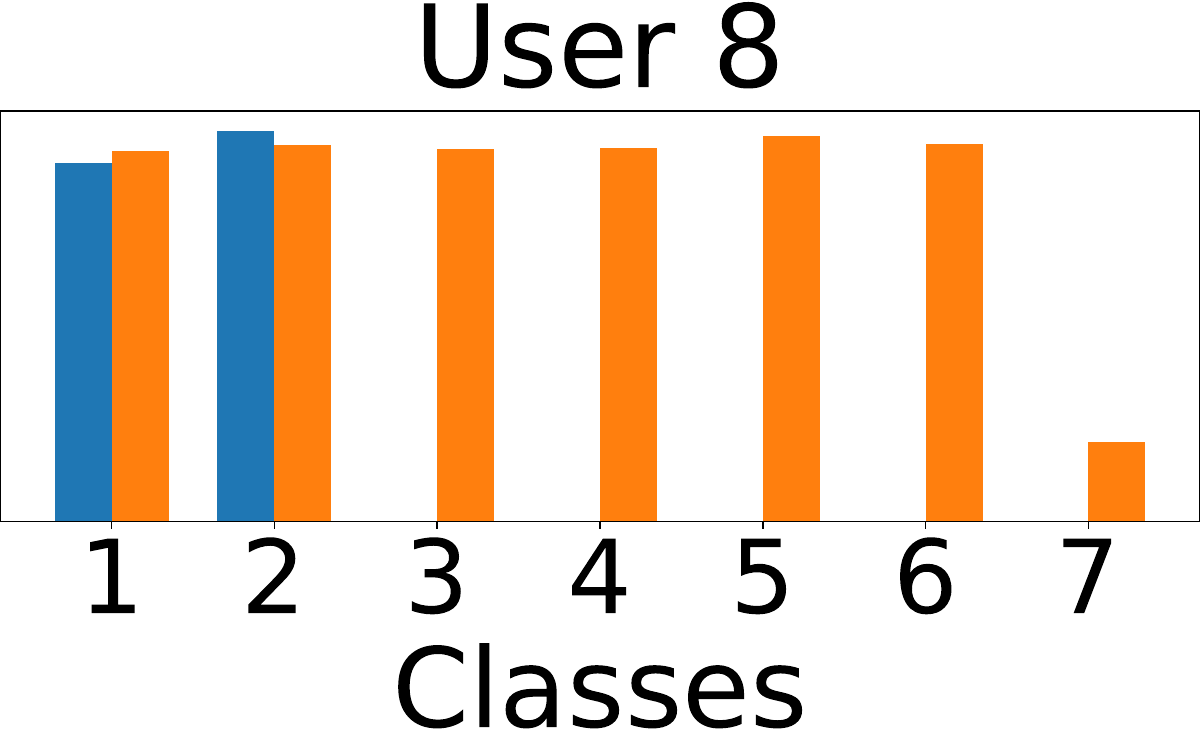}~%
        \caption{$\alpha{=}$0.5}
    \end{subfigure}%
    ~\includegraphics[width=\mywidth\linewidth]{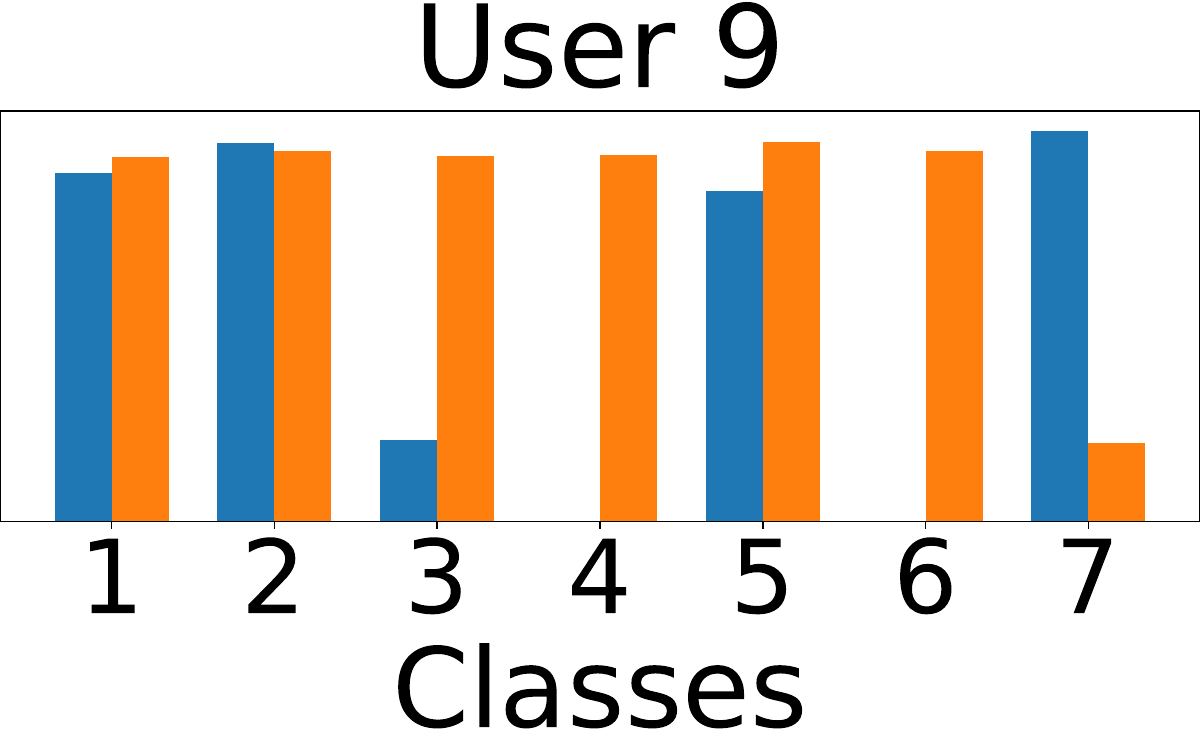}~%
    \includegraphics[width=\mywidth\linewidth]{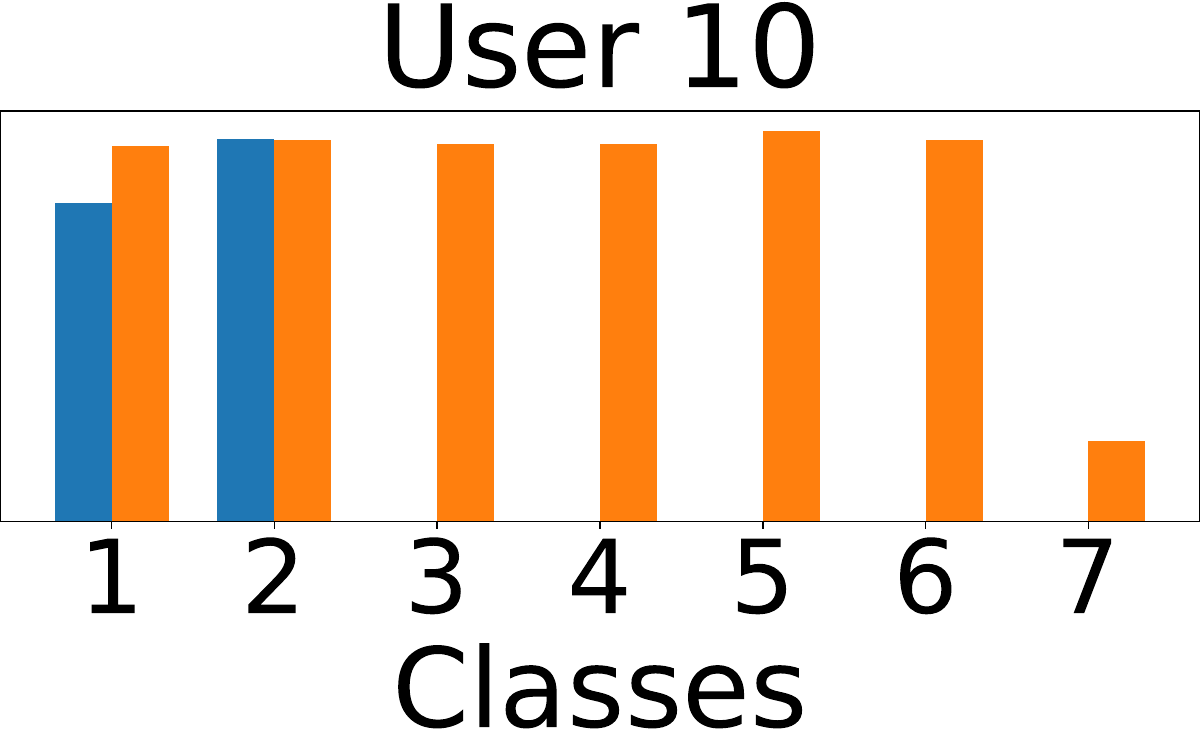}

    \includegraphics[width=\mywidth\linewidth]{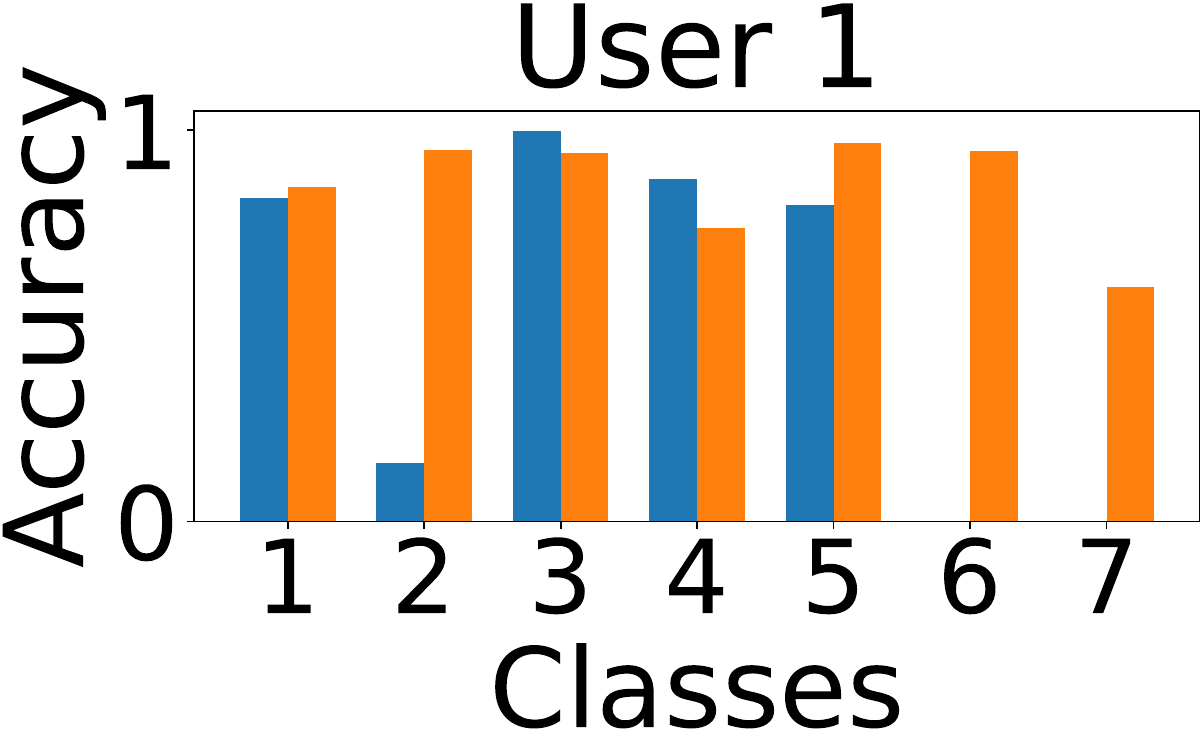}~%
    \includegraphics[width=\mywidth\linewidth]{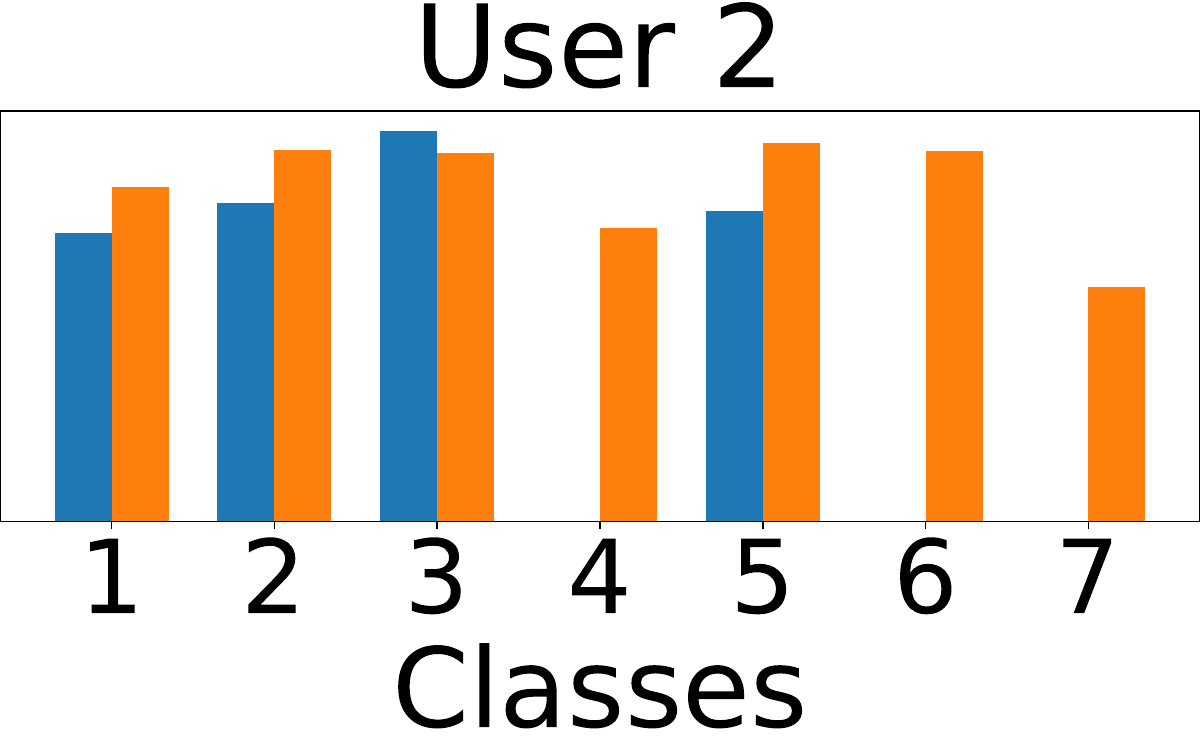}~%
    \includegraphics[width=\mywidth\linewidth]{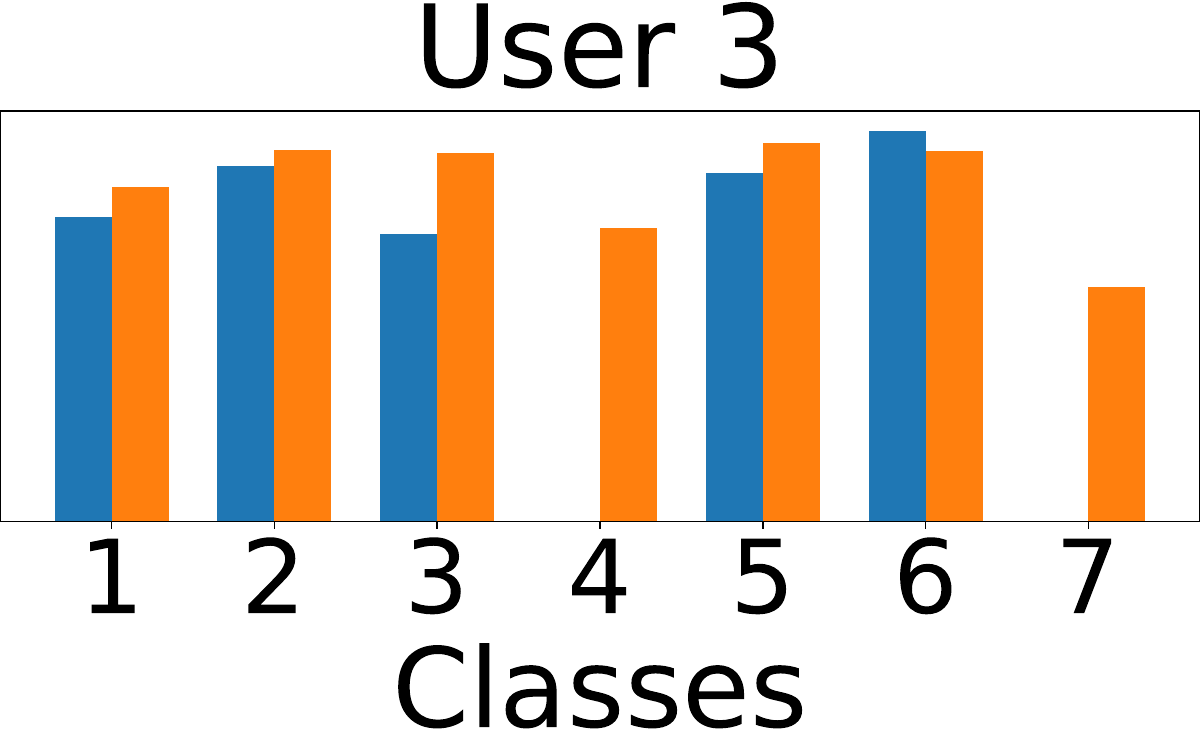}~%
    \includegraphics[width=\mywidth\linewidth]{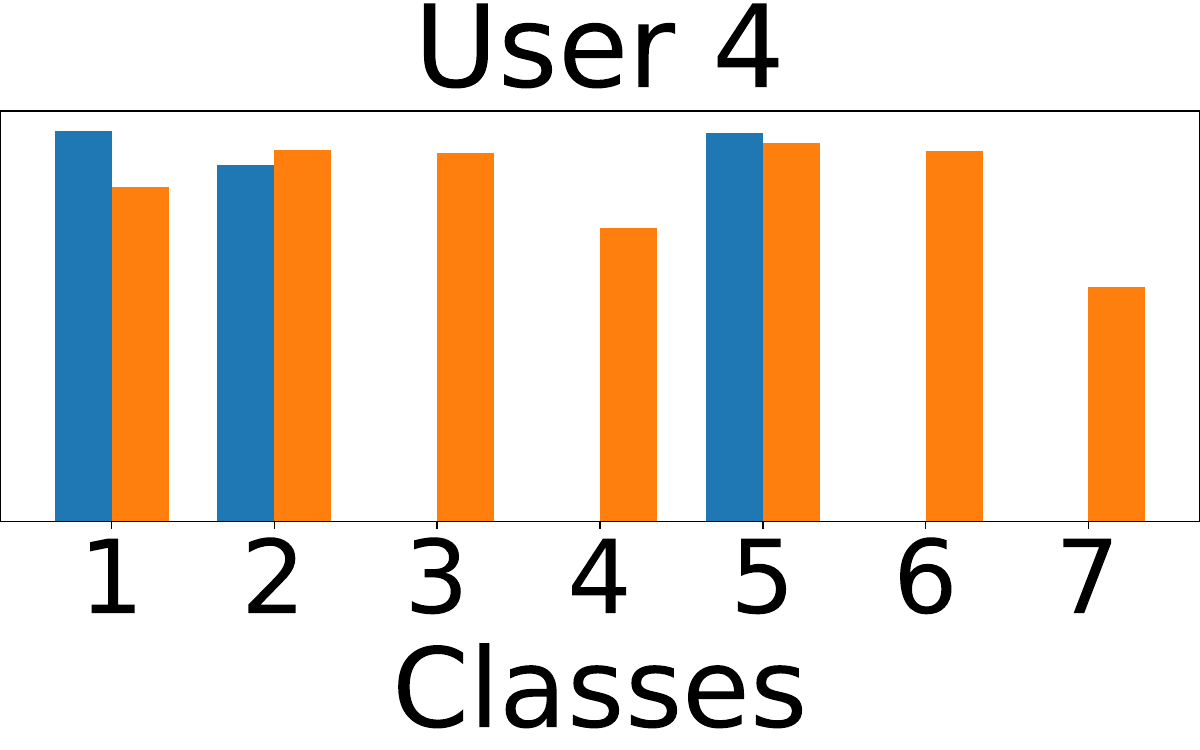}~%
    \includegraphics[width=\mywidth\linewidth]{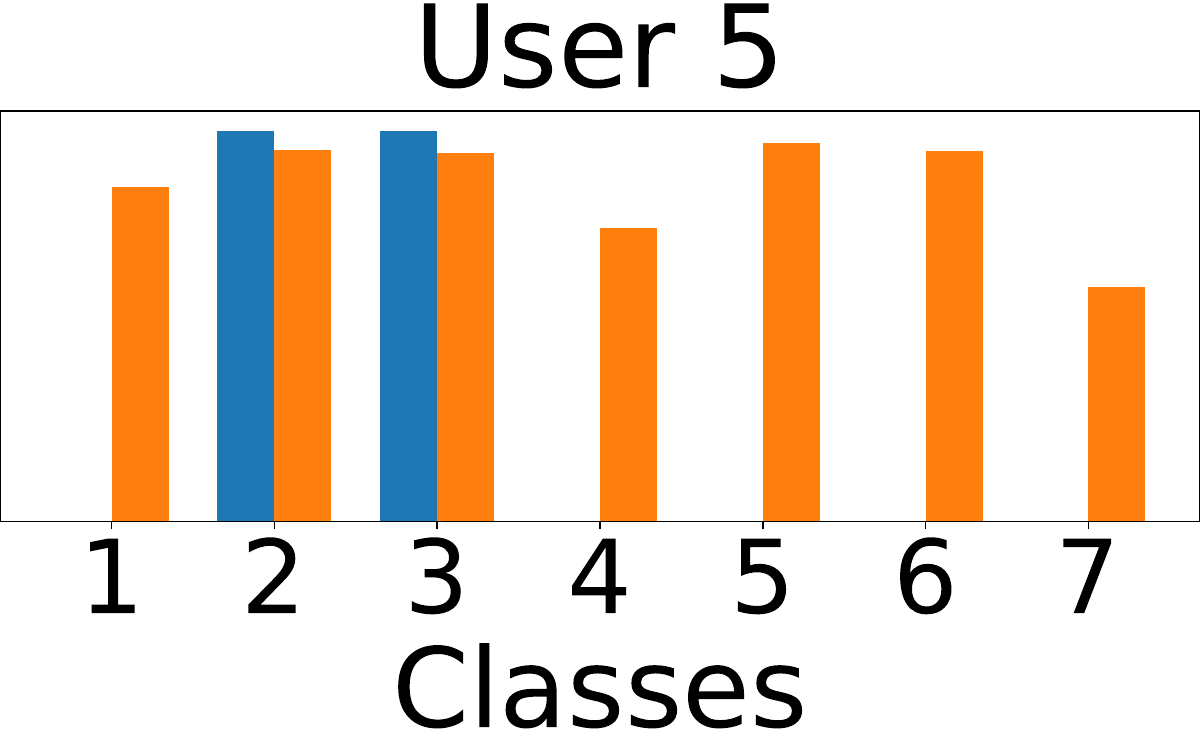}
    \includegraphics[width=\mywidth\linewidth]{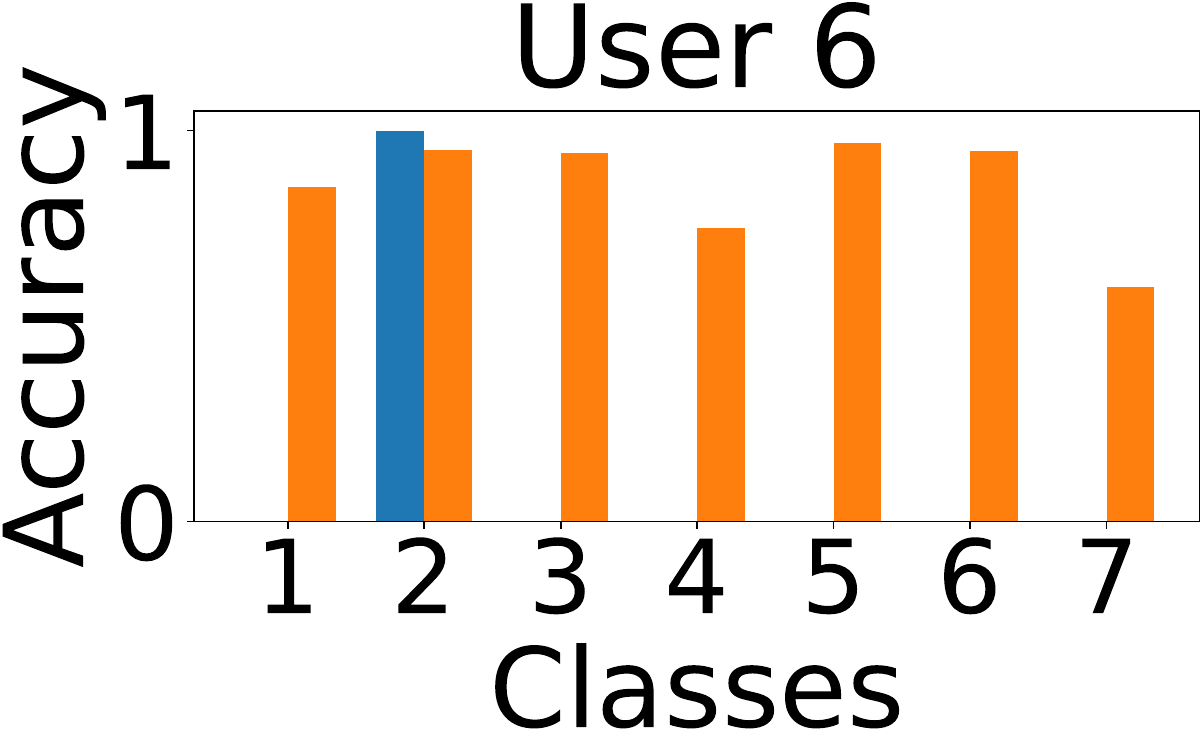}~%
    \includegraphics[width=\mywidth\linewidth]{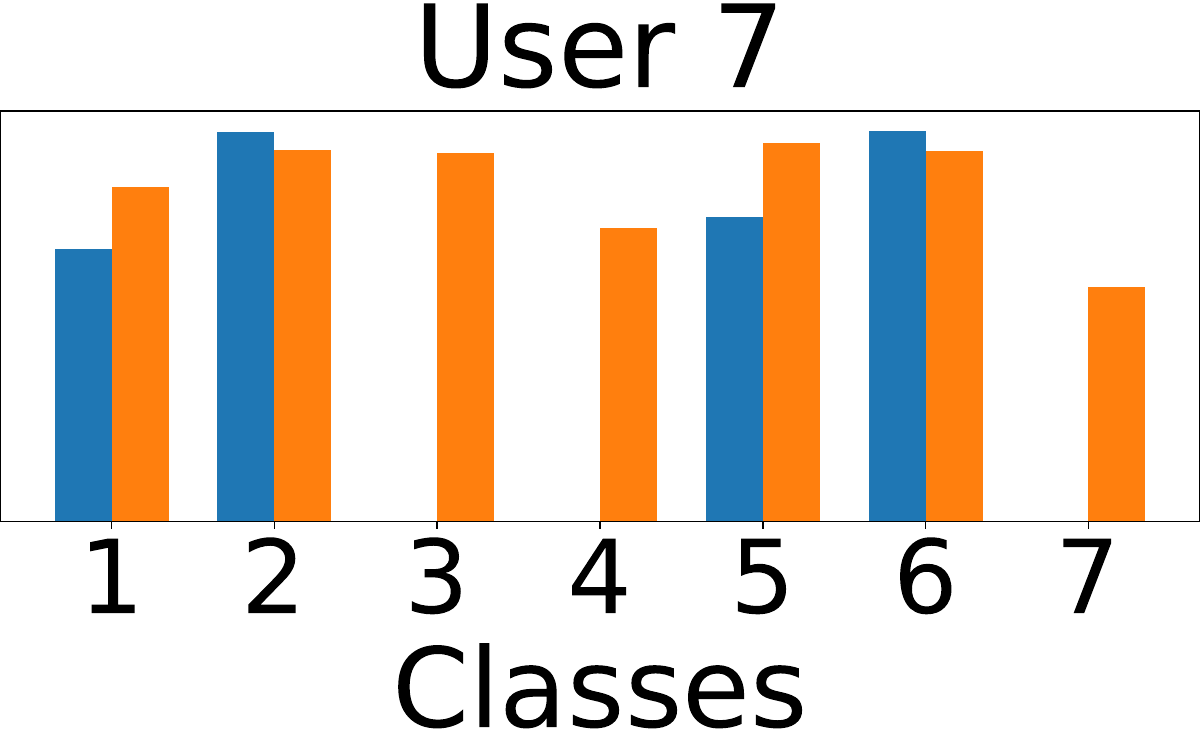}~%
    \begin{subfigure}[t]{\mywidth\linewidth}
        \includegraphics[width=\linewidth]{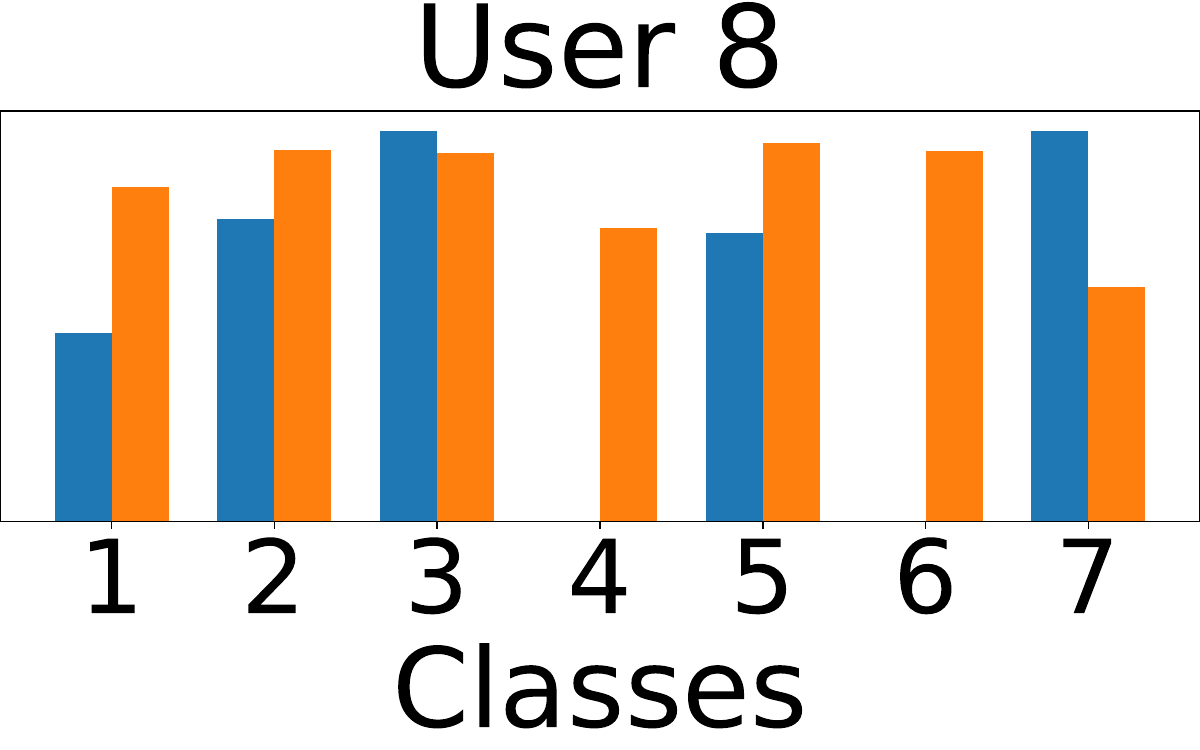}~%
        \caption{$\alpha{=}$0.25}
    \end{subfigure}%
    ~\includegraphics[width=\mywidth\linewidth]{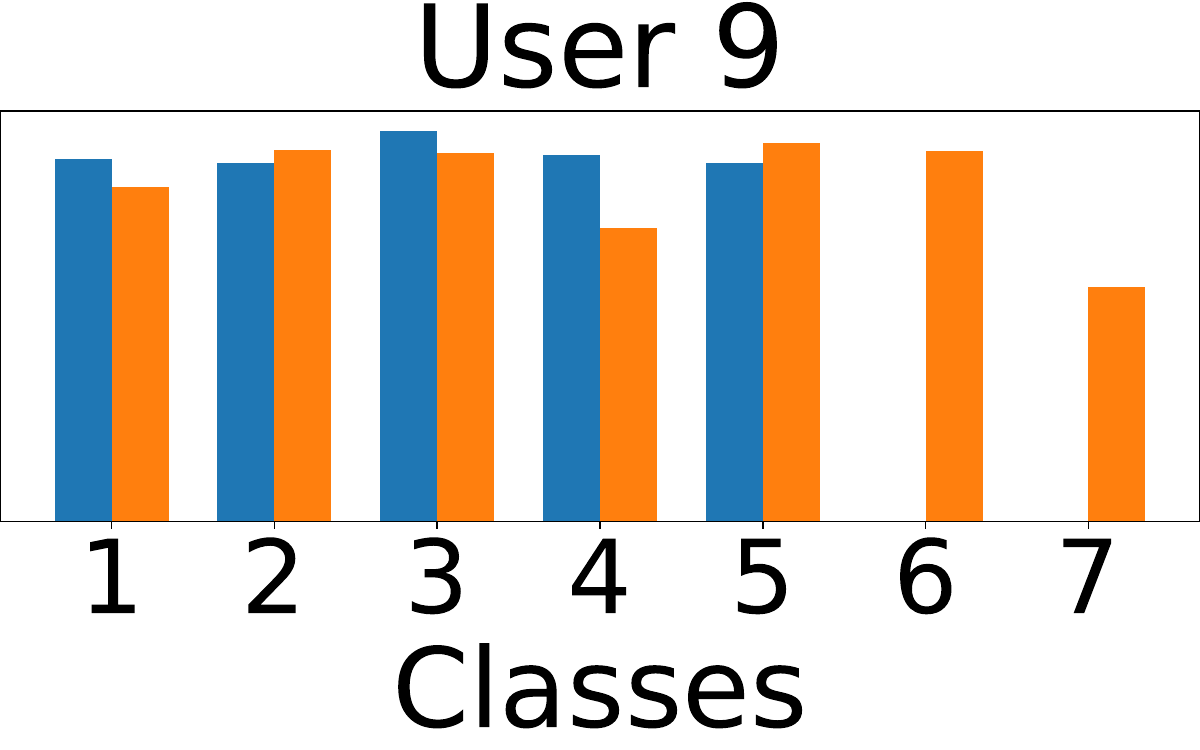}~%
    \includegraphics[width=\mywidth\linewidth]{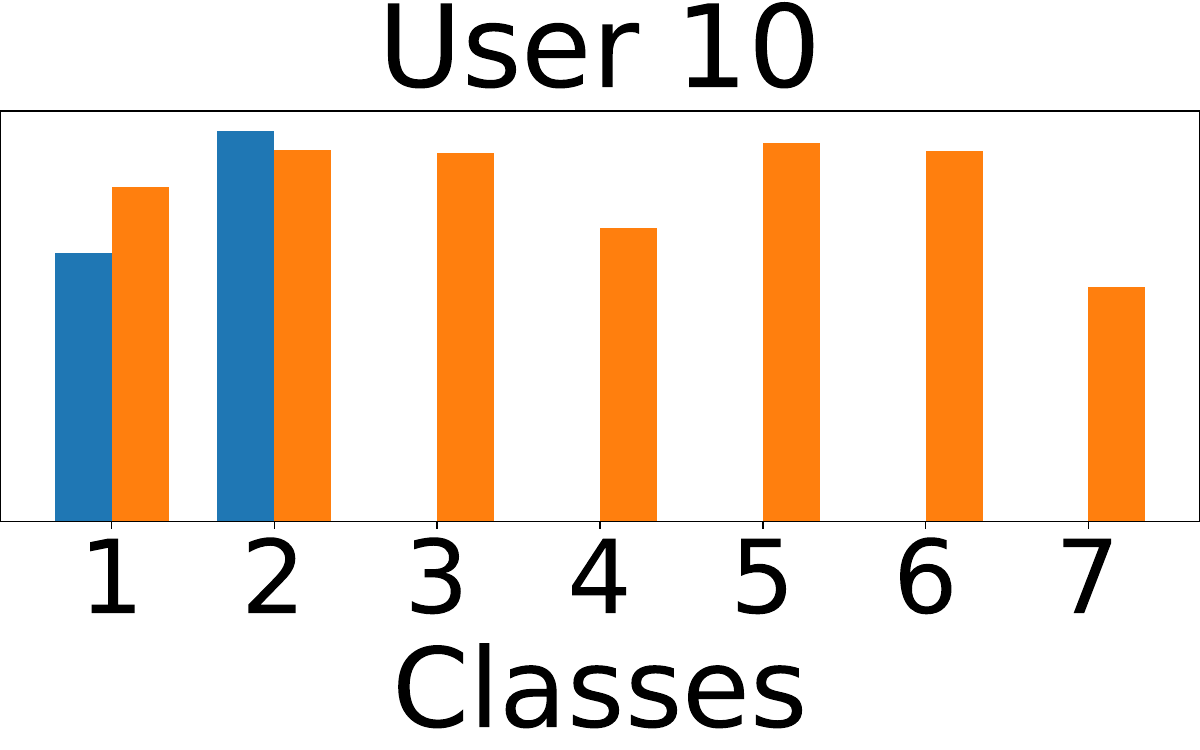}

    \caption{Per-class accuracy for all users using PROTEAN on the 5G-NIDD dataset with different $\alpha$ values. Blue and orange bars represent the accuracy without FL and with PROTEAN, respectively.}
    \label{fig: before after fl 5gnidd}
\end{figure*}

\begin{figure*}[t]
    \centering
    \begin{subfigure}[t]{0.325\linewidth}
        \includegraphics[width=\linewidth]{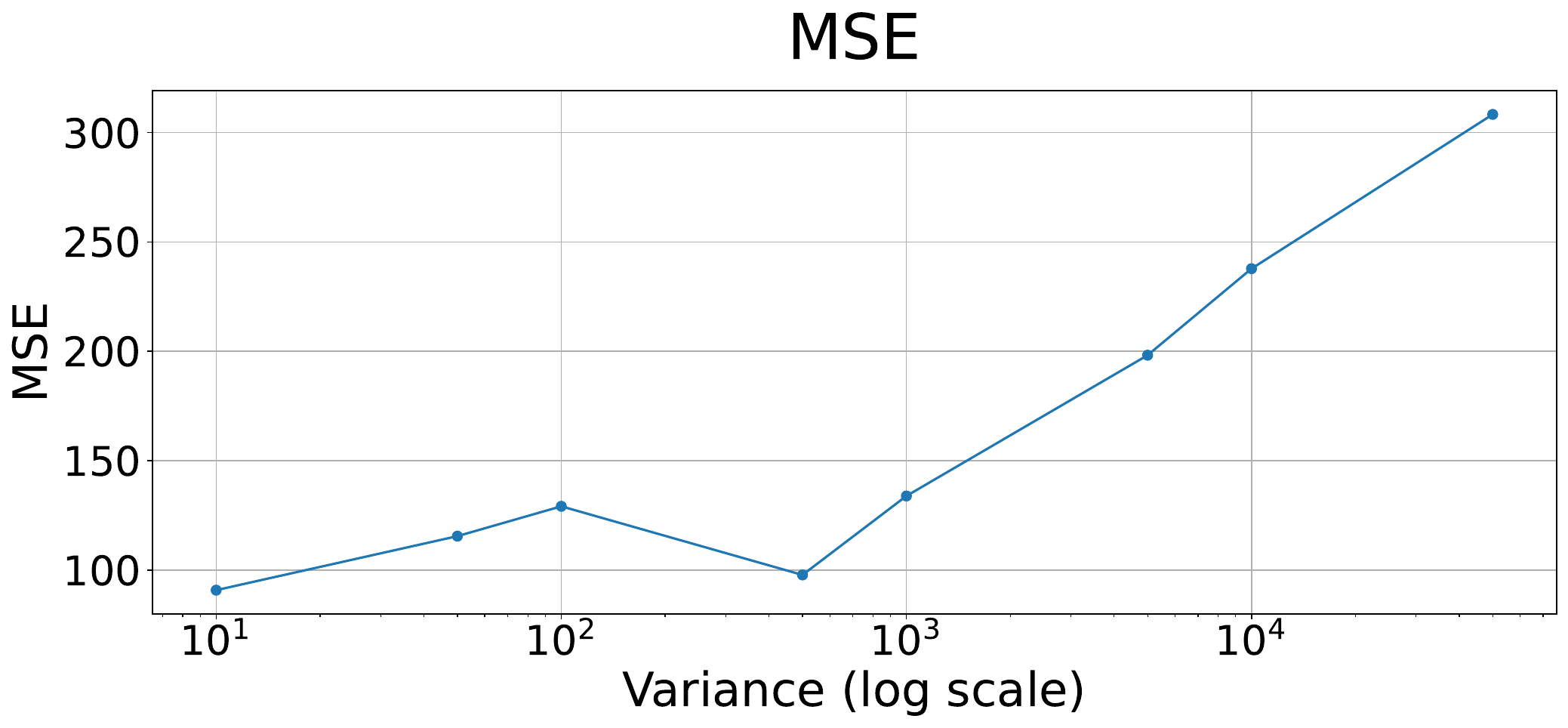}
    \end{subfigure}%
    \begin{subfigure}[t]{0.325\linewidth}
        \includegraphics[width=\linewidth]{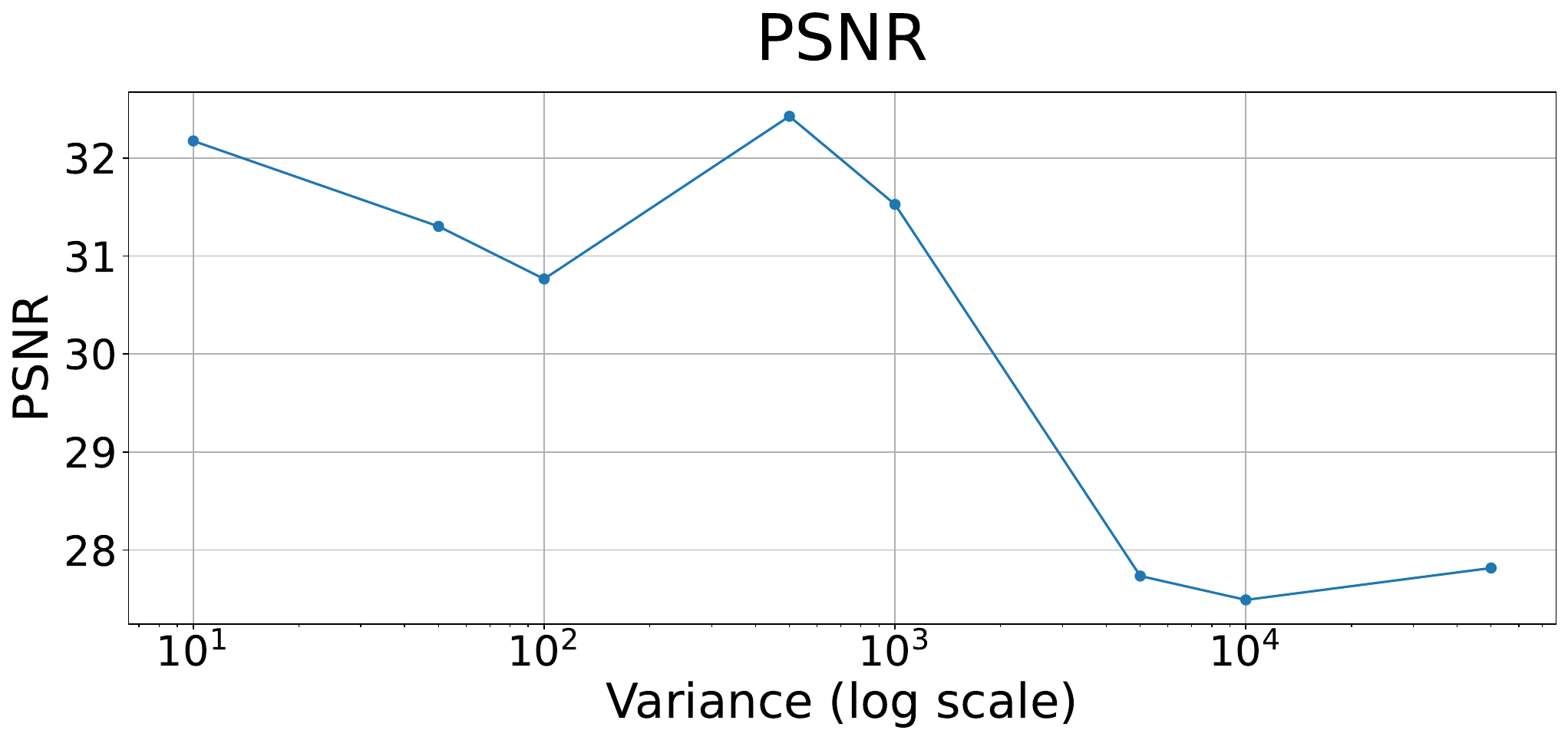}
        \caption{$\alpha=0.75$}
    \end{subfigure}%
    \begin{subfigure}[t]{0.325\linewidth}
        \includegraphics[width=\linewidth]{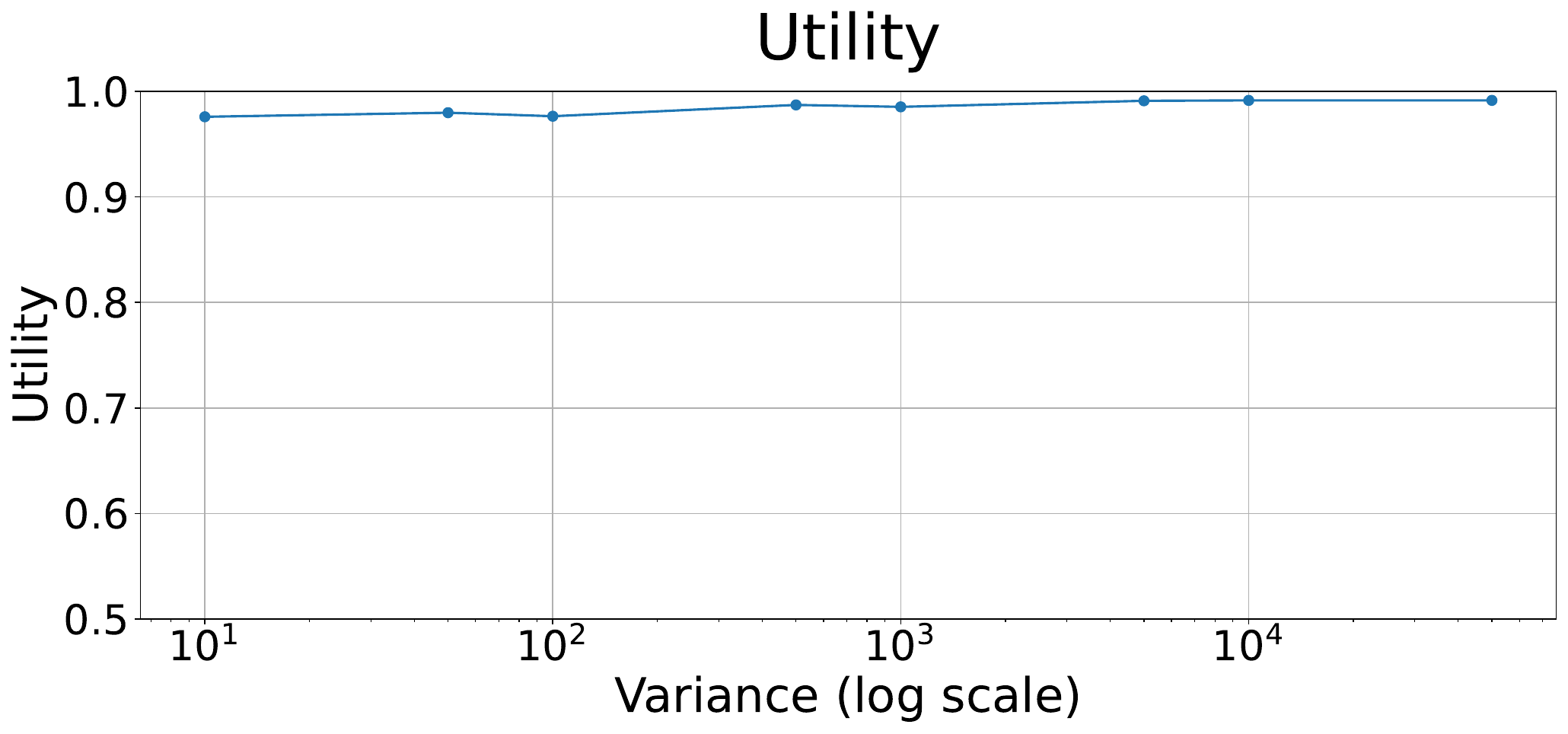}
    \end{subfigure}

    \begin{subfigure}[t]{0.325\linewidth}
        \includegraphics[width=\linewidth]{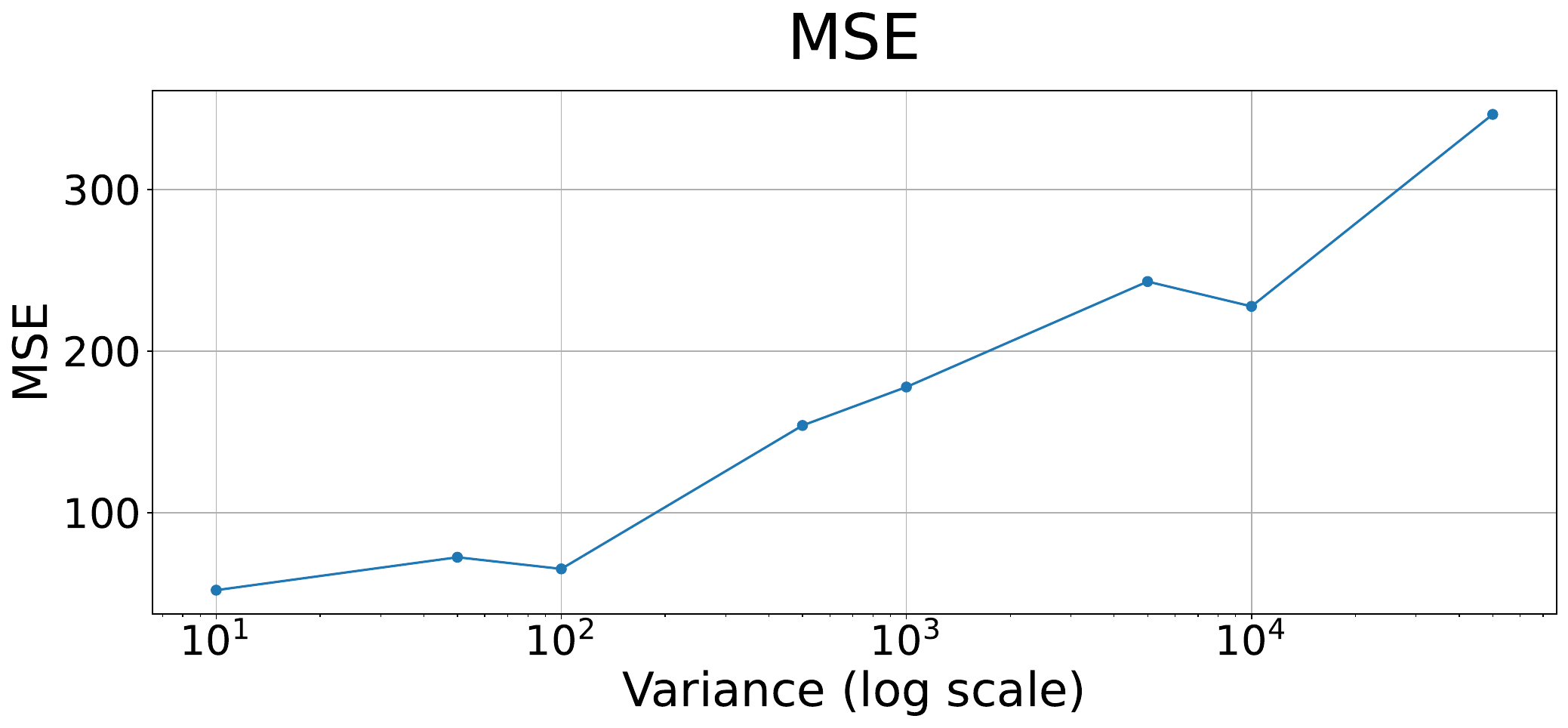}
    \end{subfigure}%
    \begin{subfigure}[t]{0.325\linewidth}
        \includegraphics[width=\linewidth]{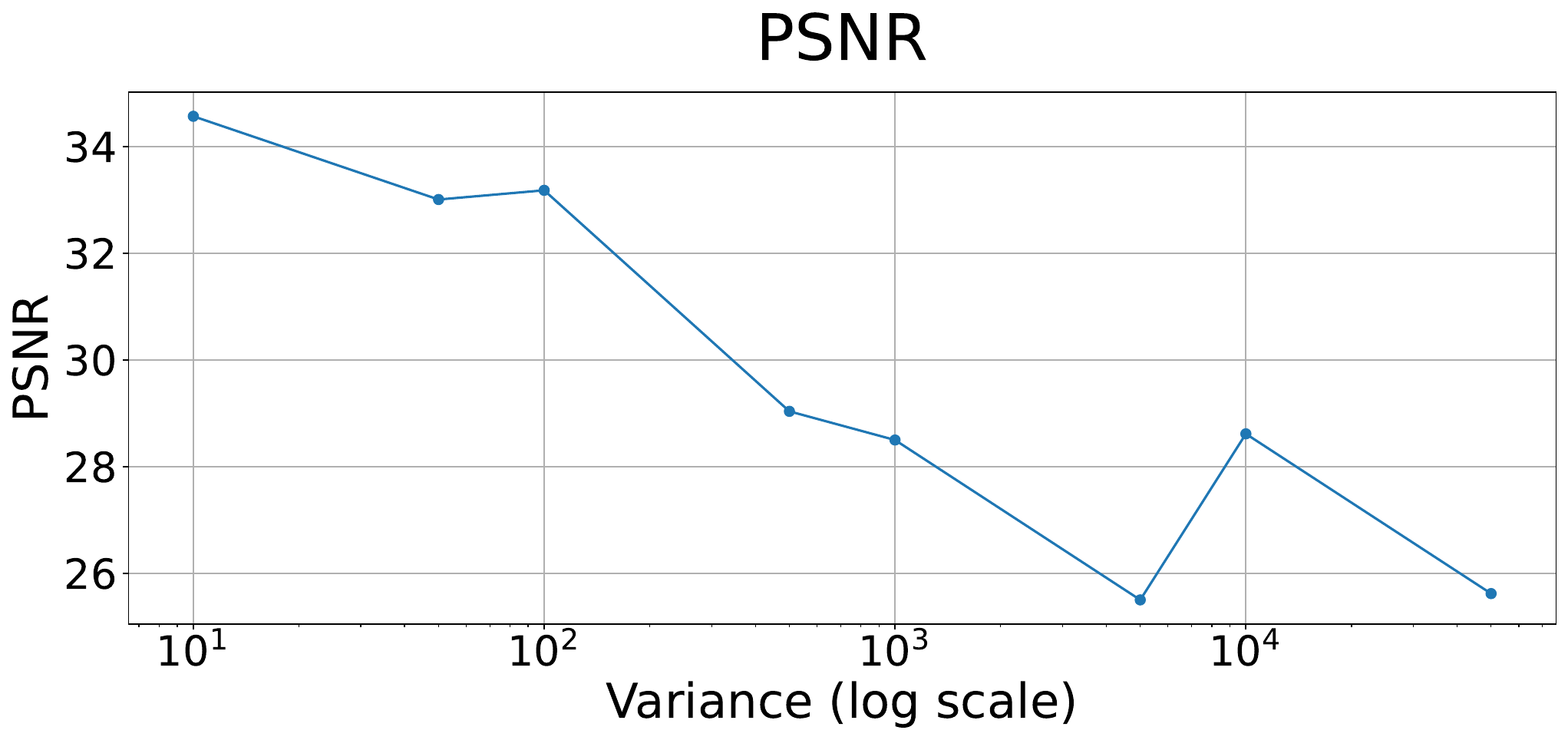}
        \caption{$\alpha=0.5$}
    \end{subfigure}%
    \begin{subfigure}[t]{0.325\linewidth}
        \includegraphics[width=\linewidth]{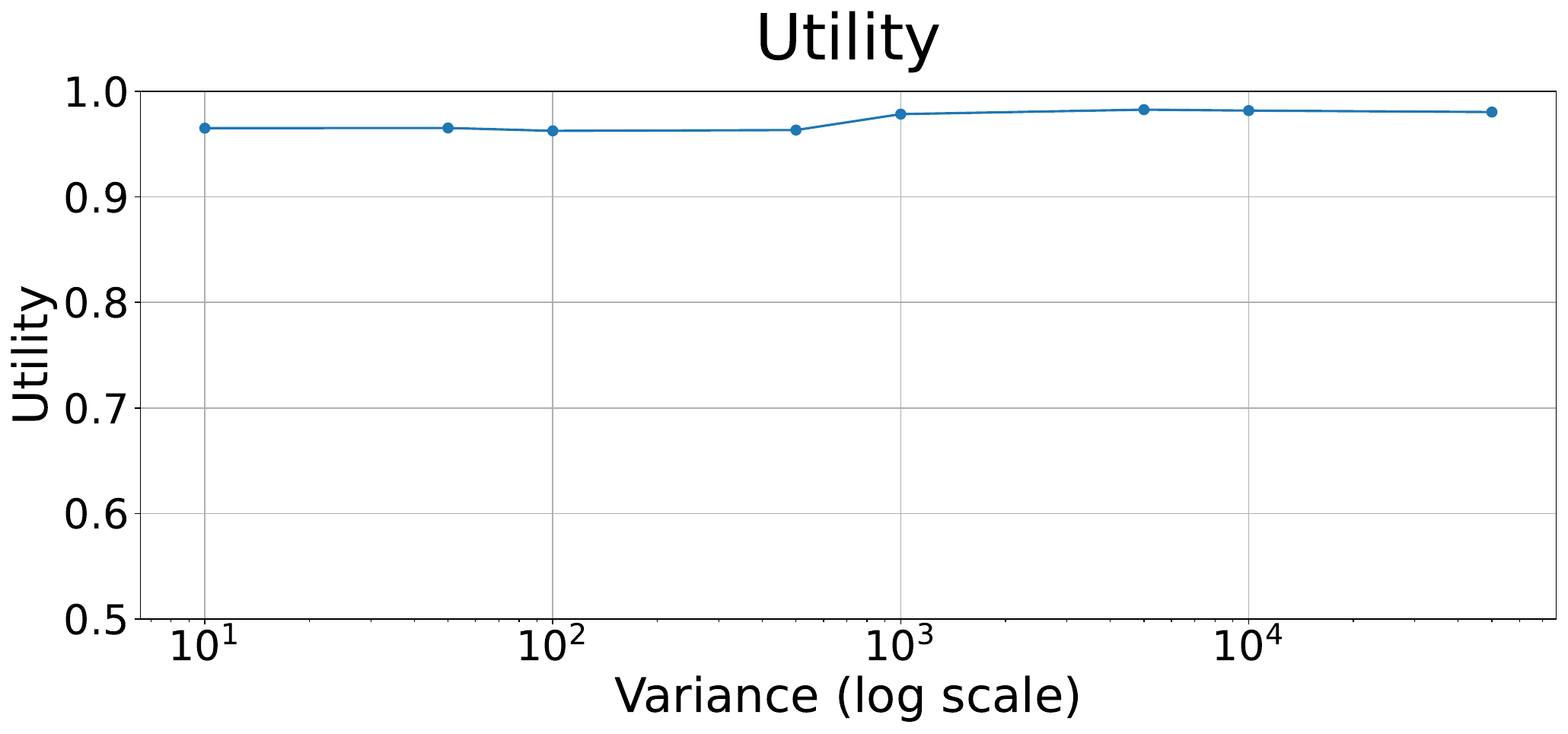}
    \end{subfigure}

    \begin{subfigure}[t]{0.325\linewidth}
        \includegraphics[width=\linewidth]{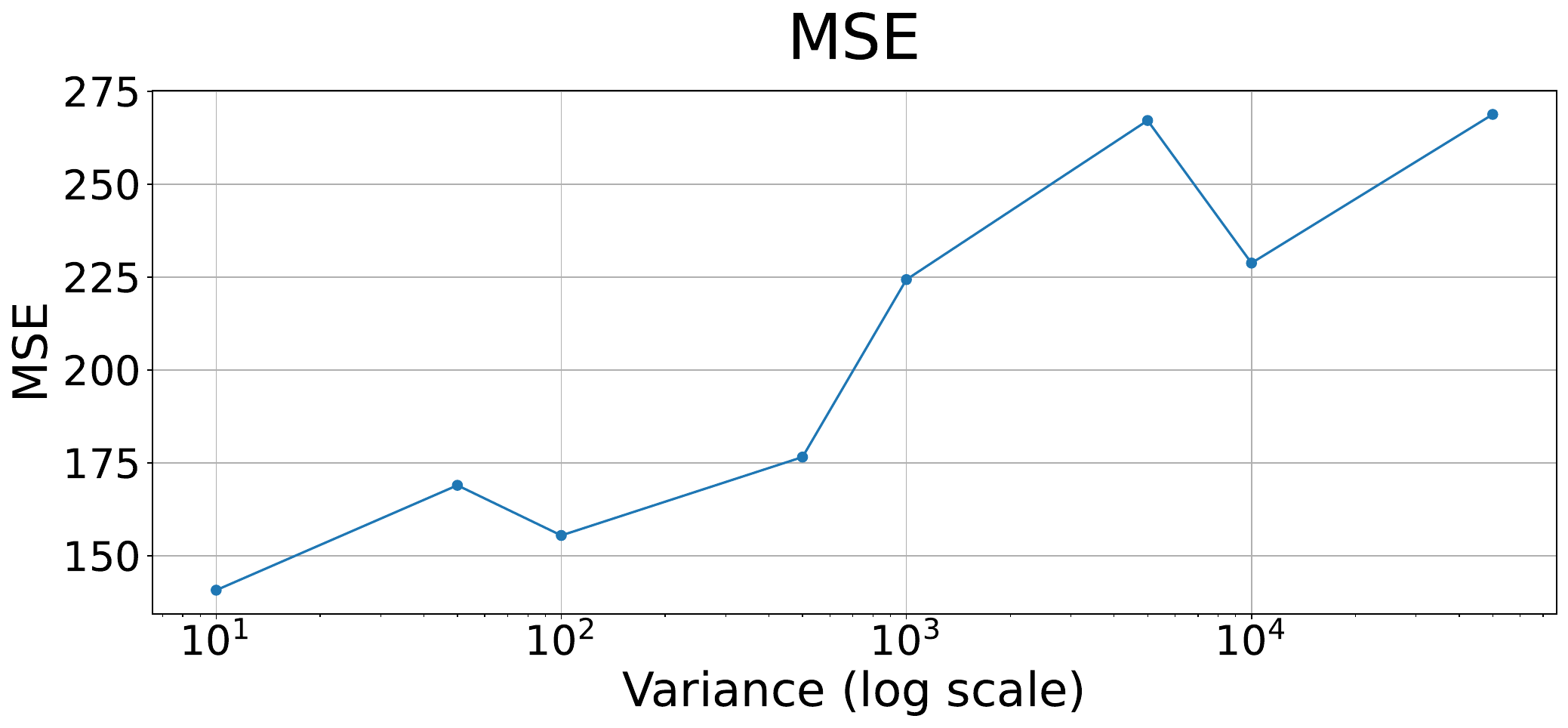}
    \end{subfigure}%
    \begin{subfigure}[t]{0.325\linewidth}
        \includegraphics[width=\linewidth]{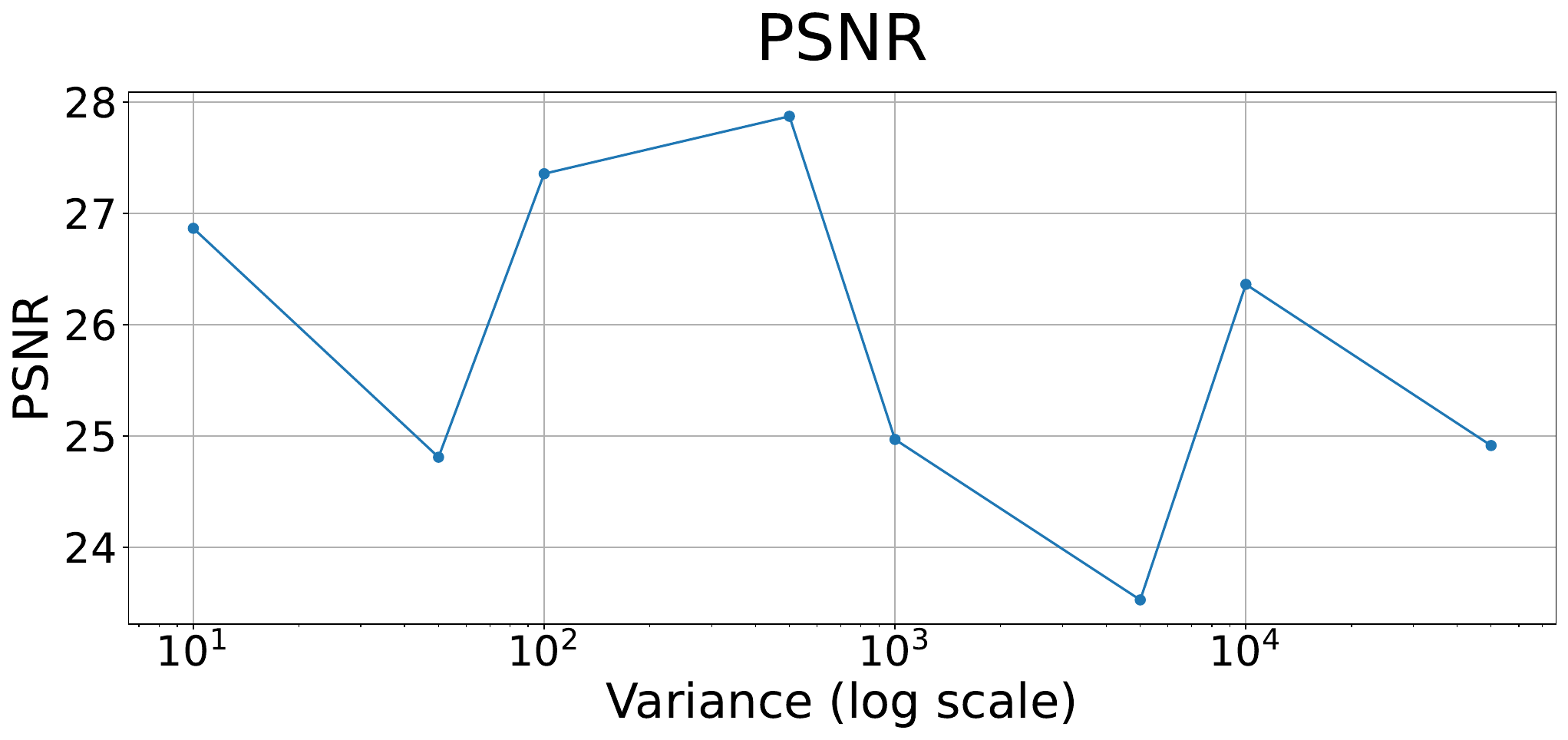}
        \caption{$\alpha=0.25$}
    \end{subfigure}%
    \begin{subfigure}[t]{0.325\linewidth}
        \includegraphics[width=\linewidth]{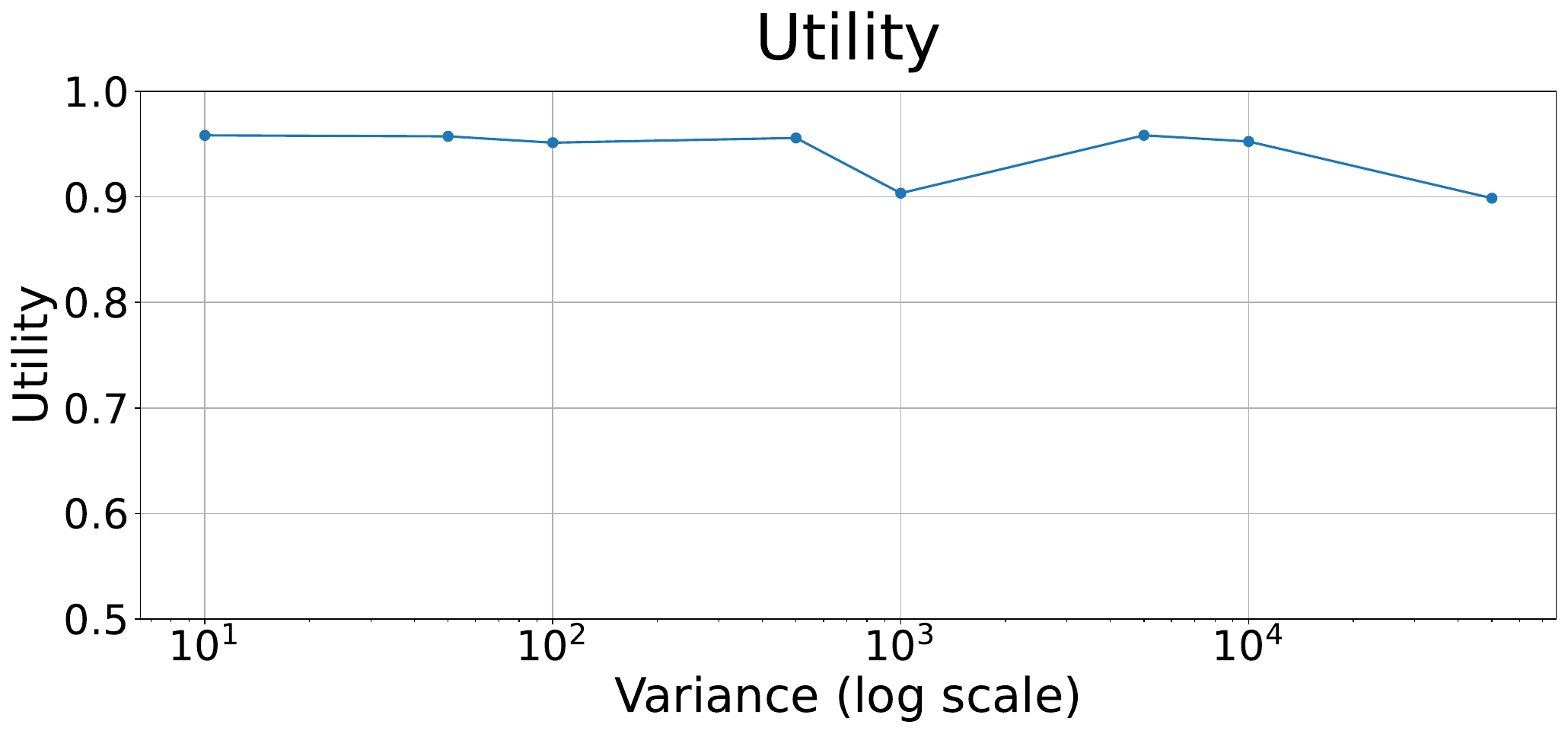}
    \end{subfigure}

    \caption{Reconstructed MSE, PSNR and Utility for $\alpha=0.75$, $0.5$ and $0.25$ on the 5G-NIDD dataset.}
    \label{fig:5gnidd_combined}
\end{figure*}

\end{document}